%% file: main.tex
\documentclass[12pt]{article}
\usepackage[margin=1in]{geometry}
\usepackage[authoryear,longnamesfirst]{natbib}
\usepackage{amsmath,amssymb,amsthm}
\usepackage{mathtools}
\usepackage{bm}
\usepackage{graphicx}
\usepackage{booktabs}
\usepackage[table]{xcolor}
\usepackage{threeparttable}
\usepackage{tabularx}
\usepackage{array}
\usepackage{enumitem}
\usepackage{needspace}
\usepackage{placeins}
\usepackage{setspace}
\usepackage{etoolbox}
\usepackage{microtype}
\usepackage{caption}
\usepackage[colorlinks,
linkcolor=blue, anchorcolor=blue, citecolor=blue]{hyperref}
\renewcommand*{\HyperDestNameFilter}[1]{\jobname-#1}
\hypersetup{
  pdftitle={Convolution Skew-t Distributions},
  pdfauthor={Xinxian Chen, Peter Reinhard Hansen, Chen Tong}
}

\AtBeginEnvironment{table}{\setstretch{1.18}}
\AtBeginEnvironment{figure}{\setstretch{1.18}}
\newcolumntype{Y}{>{\centering\arraybackslash}X}
\definecolor{blockgray}{gray}{0.92}
\newcommand{\bc}[1]{\cellcolor{blockgray}#1}

\newtheorem{theorem}{Theorem}
\newtheorem{proposition}{Proposition}
\newtheorem{lemma}{Lemma}

\newtheorem{assumption}{Assumption}
\theoremstyle{definition}
\newtheorem{definition}{Definition}
\theoremstyle{remark}
\newtheorem{remark}{Remark}

\newcommand{\E}{\mathbb{E}}
\newcommand{\Var}{\operatorname{var}}

\newcommand{\diag}{\operatorname{diag}}
\newcommand{\vech}{\operatorname{vech}}
\newcommand{\vecop}{\operatorname{vec}}
\newcommand{\rank}{\operatorname{rank}}
\newcommand{\CST}{\operatorname{CST}}
\newcommand{\ST}{\operatorname{ST}}

\newcommand{\Avar}{\operatorname{Avar}}

\begin{document}

\title{\textbf{Convolution Skew-$t$ Distributions}\thanks{Chen Tong acknowledges financial support from the National Natural Science Foundation of China (72301227, 72673150) and the Fujian Provincial Natural Science Foundation of China (2025J08008).}}
\author{\parbox{0.95\textwidth}{\centering
  Xinxian Chen$^{\ddagger}$ \quad Peter Reinhard Hansen$^{\mathsection}$\thanks{Corresponding author: Peter Reinhard Hansen. Email: \texttt{hansen@unc.edu}.} \quad Chen Tong$^{\ddagger}$\\[0.70cm]
  \small $^{\ddagger}$Department of Finance, School of Economics,\\[0.15cm]
  Xiamen University, China\\[0.3cm]
  $^{\mathsection}$Department of Economics,\\[0.15cm]
  University of North Carolina at Chapel Hill, United States}}
\date{October 25, 2026}
\maketitle

\begin{abstract} \setstretch{1.3}
We introduce the convolution skew-$t$ (CST) distribution, which combines
independent standardized skew-$t$ subvectors through a general loading matrix
to accommodate heterogeneous tails, directional asymmetry, and non-elliptical
dependence.  We establish identification and develop likelihood inference
using analytical scores and a one-dimensional reduction of Fisher information.
For the finite-variance model, we prove consistency and asymptotic normality
of maximum likelihood and characterize the consequences of imposing
false symmetry.
A raw representation extends the family to all positive tail indices.
Simulations assess finite-sample performance, and a realized-volatility
application illustrates the empirical value of allowing tail thickness and
asymmetry to vary across latent subvectors.
\end{abstract}


\newpage

\onehalfspacing

\setlength{\abovedisplayskip}{10pt}
\setlength{\belowdisplayskip}{10pt}
\setlength{\abovedisplayshortskip}{10pt}
\setlength{\belowdisplayshortskip}{10pt}

\section{Introduction}\label{sec:introduction}

Heavy tails and asymmetry are persistent features of financial and economic
data and are central to the measurement of extreme risk
\citep{Cont2001,Embrechts1997,McNeil2015}.  They also motivate skewed
conditional densities and time-varying higher moments in financial
econometrics \citep{HarveySiddique1999,Jondeau2003}.  In multivariate systems,
assets or economically meaningful groups may differ in both tail thickness
and the direction of asymmetry.  Shared exposure to a latent shock can then
generate patterns of joint extremes that covariance alone does not capture.
Evidence that tail risk can have granular, rather than exclusively
market-wide, origins further motivates models with heterogeneous tail
sources \citep{Andersen2025}.
For example, market-wide and sector-specific shocks may differ in their
tail behavior even when they affect some of the same assets.  A useful
multivariate model should allow such exposures to overlap and distinguish
their tail properties from their contributions to means and covariances.

The multivariate Student-$t$ distribution accommodates heavy tails but
assigns one tail index to the entire vector and preserves central symmetry.
Skew-$t$ and related families relax symmetry
\citep{Azzalini1996,Branco2001a,Azzalini2003a,Sahu2003}, with applications to
multivariate volatility \citep{Bauwens2005}.  Centered and unified
parameterizations clarify their moment implications and relationships
\citep{ArellanoValleAzzalini2013,Wang2024}, but standard specifications retain
a common degrees-of-freedom parameter.  Alternative multivariate $t$ models
allow scale mixing to vary across coordinates or directions
\citep{Finegold2011,Forbes2014}.  Our interest is in combining heterogeneous
tails with directional asymmetry at the level of latent groups.

The convolution-$t$ (CT) distribution of \citet{Hansen2026} provides a
natural starting point.  It combines independent Student-$t$ subvectors
through a general loading matrix, allowing each subvector to have its own
tail index.  Observed variables can share exposures to several subvectors,
producing non-elliptical dependence without imposing block-diagonal
covariance.  The latent subvectors are symmetric, however, so CT cannot
capture distinct directions of asymmetry across these tail sources.
This makes CT a useful benchmark for studying whether heterogeneous tail
sources also have different patterns of right--left imbalance.  Retaining
its convolution structure allows the effect of adding asymmetry to be
examined within a common framework for dependence.

We introduce the convolution skew-$t$ (CST) distribution by replacing the
symmetric latent subvectors with standardized multivariate skew-$t$
subvectors of \citet{Azzalini2003a}.  Its representation is
\begin{equation}\label{eq:intro-model}
  Y=\mu+\Xi Z,
  \qquad
  Z=(Z_1^{\prime},\ldots,Z_K^{\prime})^{\prime},
\end{equation}
where the independent $Z_k$ have zero mean, identity covariance, and their
own tail and skewness parameters.  The nonsingular loading matrix $\Xi$
maps the latent sources into observed coordinates, giving
$\E(Y)=\mu$ and $\Var(Y)=\Xi\Xi^{\prime}$.  This separates the first two
moments from higher-order shape while allowing overlapping exposures to
asymmetric tail shocks.  CT and the conventional multivariate skew-$t$ are
special cases, while the multivariate Student-$t$ and Gaussian distributions
are included as special or limiting cases.  Within a subvector, coordinates
share a scale-mixture shock; across subvectors, tail indices and skewness
directions can differ.  Consequently, observed variables need not belong
to disjoint clusters: the same variable can be exposed to several latent
sources with different tail properties.  We use ``subvector'' for the stochastic partition of $Z$,
``latent subspace'' for the span of the corresponding loading columns, and
``block'' for matrix partitions.

Extending CT to asymmetric subvectors raises additional construction and
identification questions.  The location parameter of a raw skew-$t$
subvector is not its mean, and its covariance depends on its shape.
Moreover, rotating a latent subvector changes both its loading columns and
its skewness coordinates.  We address these features jointly, so that the
resulting parameters have a clear interpretation and likelihood inference
respects the model's observational equivalences.

CST is related to affine transformations of independent univariate skewed
variables \citep{FerreiraSteel2007}, which are closest to its scalar-subvector
case.  Allowing multivariate subvectors adds shared tail shocks within each
latent group and requires identification at the subvector level.
Independent-factor conditional-density models likewise combine flexible
non-Gaussian factors, with an emphasis on scalar factors and their dynamics
\citep{GhalanosRossiUrga2015}.  Our focus is the multivariate innovation
distribution and its static likelihood theory.  Related work
also studies tail
heterogeneity in realized covariance matrices through $F$-Riesz models
\citep{OpschoorLucasRossini:2025}, while \citet{Stollenwerk:2026} finds that
tail-homogeneous $t$-Riesz models can provide a better fit.  CST addresses
the distinct problem of modeling multivariate innovation vectors.

The paper makes four contributions.  First, we derive an exact
mean--covariance standardization, including its admissible parameter region,
and obtain a closed-form joint likelihood.
We also characterize marginal distributions through their characteristic
functions and derive analytical expressions for marginal skewness.  The
standardized form requires finite second moments and is convenient for
interpreting loadings.  A raw representation extends the construction,
identification, and likelihood consistency to all positive tail indices,
although regular joint inference for unrestricted loadings need not survive
when second moments fail.

Second, we characterize the covariance and asymmetric
contributions of each latent subvector in observed space.  We establish
linear irreducibility of the finite-tail-index skew-$t$ subvectors and use
independent-subspace uniqueness \citep{Gutch2012} to determine the full
observational-equivalence class.  The identified contributions yield an
explicit normalization and recovery map for the loading matrix.  This
normalization resolves latent orientation without imposing additional
structure on observed covariance.

Third, we derive analytical scores and the observed
Hessian, accounting for the nonlinear standardization.  Building on
\citet{Arellano-Valle2010a}, we show that Fisher information can be computed
using one-dimensional integrals for subvectors of any fixed finite
dimension.  This reduces the numerical integration burden while retaining
the full multivariate dependence structure.

Fourth, we establish consistency and asymptotic normality of maximum
likelihood in the finite-variance model,
with delta-method inference for loadings, skewness coordinates, and smooth
marginal functionals.  We also analyze the consequences of imposing
symmetry incorrectly: near the CT
subfamily, location can respond at first order to omitted asymmetry, whereas
loading and tail-parameter distortions begin at second order.  These local
orders help interpret the effects of a false symmetry restriction, but do
not imply that CT recovers CST loadings exactly under fixed asymmetry.

Simulations assess the likelihood approximation and compare CT and CST
estimation on the same asymmetric samples.  An application to ten
realized-volatility innovation series illustrates the value of allowing
heterogeneous tails and asymmetry.  The empirical comparison uses common
HAR residuals, with observed-sector restrictions examined separately as a
robustness exercise.  The fitted CST distributions improve on their
symmetric counterparts and capture differences in tail thickness and
positive asymmetry across latent sources.  The analysis illustrates how
these features can be modeled without imposing sector-based equality
restrictions on the baseline loading matrix.

Section \ref{sec:family} presents the standardized family and its marginal
properties.  Section \ref{sec:identification} develops identification and
normalization, and Sections
\ref{sec:inference}--\ref{sec:symmetry-implications} develop likelihood
inference and the consequences of false symmetry.  Section
\ref{sec:raw-cst} treats the raw family and the limits of regular inference.
Sections \ref{sec:simulation} and \ref{sec:application} present the numerical
evidence, and Section \ref{sec:conclusion} concludes.  Proofs and supporting
calculations are collected in the appendices and Supplementary Material.

\section{The standardized CST family}\label{sec:family}

The standardized construction assigns separate roles to location,
covariance, and higher-order shape.  By centering each latent subvector and
fixing its covariance at the identity, we ensure that changes in skewness
or tail thickness do not mechanically change the first two observed
moments.  This interpretation requires $\nu_k>2$, maintained through
Section \ref{sec:symmetry-implications}.  Section \ref{sec:raw-cst} extends
the construction beyond finite variance and explains which identification
and likelihood results survive.

\subsection{A standardized skew-\texorpdfstring{$t$}{t} subvector}\label{subsec:block}

Let $X\in\mathbb{R}^d$ follow the multivariate skew-$t$
distribution $\ST_d(0,\Omega,\eta,\nu)$ of \citet{Azzalini2003a}, where $\Omega\succ0$ is a
scale matrix, $\eta\in\mathbb{R}^d$ is the canonical skewing
coefficient, and $\nu>2$.  Its density is
\begin{equation}\label{eq:raw-density}
 f_{\ST}(x;\Omega,\eta,\nu)
 =2t_d(x;0,\Omega,\nu)
 T_1\!\left[
   \left\{\frac{\nu+d}{\nu+x^{\prime}\Omega^{-1}x}\right\}^{1/2}
   \eta^{\prime}x;\nu+d
 \right],
\end{equation}
where $t_d$ is the $d$-variate Student-$t$ density and $T_1$ is the
univariate Student-$t$ distribution function.  If $\eta=\bm 0_d$,
the distribution reduces to the usual multivariate Student-$t$.  For
$\nu>2$,
\begin{equation}\label{eq:raw-moments}
  \E(X)=a_\nu\omega\delta,
  \qquad
  \Var(X)=\frac{\nu}{\nu-2}\Omega-\E(X)\E(X)^{\prime},
\end{equation}
where
\[
 \omega=\diag(\Omega)^{1/2},\qquad
 a_\nu=\sqrt{\frac{\nu}{\pi}}
 \frac{\Gamma\{(\nu-1)/2\}}{\Gamma(\nu/2)},\qquad
 \delta
  =\frac{\omega^{-1}\Omega\eta}
  {(1+\eta^{\prime}\Omega\eta)^{1/2}}.
\]

When $\nu>3$, let
$\kappa_{3,i}^{X}=\E[\{X_i-\E(X_i)\}^3]$.  The third central moment and
Pearson skewness of $X_i$ have the factorized representation
\begin{subequations}\label{eq:raw-marginal-skewness}
\begin{align}
 \kappa_{3,i}^{X}
 &=\frac{a_\nu\omega_i^3\delta_i}{\nu-3}
 \left\{
   \frac{3\nu}{\nu-2}
   +\left[2a_\nu^2(\nu-3)-\nu\right]\delta_i^2
 \right\},
 \label{eq:raw-marginal-third-moment}\\
 \operatorname{skew}(X_i)
 &=\frac{\kappa_{3,i}^{X}}
 {\omega_i^3
  \left\{\tfrac{\nu}{\nu-2}-a_\nu^2\delta_i^2\right\}^{3/2}},
 \qquad i=1,\ldots,d.
 \label{eq:raw-marginal-skewness-coefficient}
\end{align}
\end{subequations}
For fixed $\nu>3$, the Pearson skewness in
\eqref{eq:raw-marginal-skewness-coefficient} is an odd, strictly increasing
function of $\delta_i$.  Its sign therefore agrees with that of $\delta_i$,
and it vanishes exactly when $\delta_i=0$.  This gives the bounded shape
coordinate a direct interpretation, even though its numerical value is not
the Pearson coefficient.  Supplementary Appendix
\textcolor{blue}{S.1.3} establishes monotonicity.
Proposition \ref{prop:marginal-skewness} below shows how the contributions
of several such sources combine in an observed CST margin.

The following result parameterizes a raw subvector by its mean shift $m$
and degrees of freedom, while fixing its covariance at the identity.  It is
closely related to the centered skew-$t$ parameterization of
\citet{ArellanoValleAzzalini2013}, but is tailored to the convolution: each
latent subvector is constrained to have exactly zero mean and identity covariance
after centering.

\begin{proposition}[Mean-shift standardization]\label{prop:standardization}
Fix $m\in\mathbb{R}^d$ and $\nu>2$, and define
\begin{equation}\label{eq:standardization}
 \Omega(m,\nu)=\frac{\nu-2}{\nu}(I_d+mm^{\prime}),
 \qquad
 \eta(m,\nu)=
 \frac{\Omega(m,\nu)^{-1}m}
 {\{a_\nu^2-m^{\prime}\Omega(m,\nu)^{-1}m\}^{1/2}}.
\end{equation}
If the admissibility condition
\begin{equation}\label{eq:admissibility}
 a_\nu^2-m^{\prime}\Omega(m,\nu)^{-1}m>0
\end{equation}
holds and $X\sim\ST_d\{0,\Omega(m,\nu),\eta(m,\nu),\nu\}$, then
$\E(X)=m$ and $\Var(X)=I_d$.  Therefore $Z=X-m$ satisfies
$\E(Z)=0$ and $\Var(Z)=I_d$.
\end{proposition}

The admissibility condition has a simple geometric form.  With
$c_\nu=(\nu-2)/\nu$, the Sherman-Morrison formula gives
\[
 m^{\prime}\Omega(m,\nu)^{-1}m=\frac{\|m\|^2}{c_\nu(1+\|m\|^2)}.
\]
For a univariate Student-$t_\nu$ variable $T_\nu$,
$a_\nu=\E|T_\nu|$ and $\E(T_\nu^2)=1/c_\nu$; strict Cauchy-Schwarz
therefore gives $0<c_\nu a_\nu^2<1$.  Condition
\eqref{eq:admissibility} is therefore
equivalent to
\begin{equation}\label{eq:admissible-ball}
 \|m\|^2<\bar m_\nu^2,
 \qquad
 \bar m_\nu^2=\frac{c_\nu a_\nu^2}{1-c_\nu a_\nu^2}.
\end{equation}
Thus, for a fixed $\nu$, the mean-shift parameter lies in an open Euclidean
ball.  The smooth map
\begin{equation}\label{eq:unconstrained-m-map}
 m=\bar m_\nu\frac{u}{(1+u^{\prime}u)^{1/2}},
 \qquad u\in\mathbb R^d,
\end{equation}
enforces this joint admissibility restriction during numerical
optimization.  The ball is rotation invariant, which will also allow the
loading orientation and skewness coordinates to be normalized together.

The standardization is also constructive.  If
$\omega=\diag\{\Omega(m,\nu)\}^{1/2}$, the associated bounded skewness
parameter is
\begin{equation}\label{eq:delta-map}
  \delta(m,\nu)=a_\nu^{-1}\omega^{-1}m.
\end{equation}
For fixed $\nu$, the componentwise map between $m$ and $\delta$ is
one-to-one, subject to the joint restriction in
\eqref{eq:admissible-ball}.  Indeed,
$\delta_i=m_i/\{a_\nu\sqrt{c_\nu(1+m_i^2)}\}$, with inverse
$m_i=a_\nu\sqrt{c_\nu}\,\delta_i/
\{1-a_\nu^2c_\nu\delta_i^2\}^{1/2}$ on the admissible set.  We use
$(m,\nu)$ as the primitive subvector
parameters because they enforce the first two moments exactly, while
$\delta$ remains available as a bounded measure of skewness.
Here ``skewness'' denotes directional distributional asymmetry, not a
standardized third moment.  The coordinates $(m,\delta)$ remain well defined
for every $\nu>2$, including $2<\nu\leq3$, where third moments do not exist.

\subsection{Convolution construction and likelihood}\label{subsec:construction}

Let $Z_k=Z_k(m_k,\nu_k)\in\mathbb{R}^{n_k}$, $k=1,\ldots,K$, be
mutually independent standardized subvectors obtained from Proposition
\ref{prop:standardization}.  Put $n=\sum_k n_k$ and
$Z=(Z_1^{\prime},\ldots,Z_K^{\prime})^{\prime}$.  For a nonsingular loading matrix
$\Xi\in\mathbb{R}^{n\times n}$, define
\begin{equation}\label{eq:cst-definition}
  Y=\mu+\Xi Z
  \sim \CST_{\bm n}^{\mathrm{std}}
  \left(\mu,\Xi,\{m_k,\nu_k\}_{k=1}^K\right),
  \qquad \bm n=(n_1,\ldots,n_K)^{\prime}.
\end{equation}
By construction,
\begin{equation}\label{eq:first-two-moments}
  \E(Y)=\mu,
  \qquad
  \Var(Y)=\Xi\Xi^{\prime}
  =\sum_{k=1}^K\Xi_{\bullet k}\Xi_{\bullet k}^{\prime},
\end{equation}
where $E_k\in\mathbb{R}^{n\times n_k}$ selects subvector $k$ and
$\Xi_{\bullet k}=\Xi E_k$.

For an observation $y$, let $z=\Xi^{-1}(y-\mu)$,
$z_{k}=E_k^{\prime}z$, and $x_{k}=z_{k}+m_k$.  The log likelihood
contribution is
\begin{equation}\label{eq:likelihood}
  \ell(\theta)
  =-\frac12\log|\Xi\Xi^{\prime}|
   +\sum_{k=1}^K
   \log f_{\ST}\{x_{k};\Omega(m_k,\nu_k),
                         \eta(m_k,\nu_k),\nu_k\}.
\end{equation}
The invertible loading transformation makes the latent coordinates
recoverable from an observation, so the joint likelihood factors into
subvector densities and a Jacobian term.  The convolution in observed
space therefore retains the computational simplicity of a product
likelihood: evaluating it requires neither numerical convolution nor
integration over latent variables.

\subsection{Marginal characteristic functions and skewness}
\label{subsec:marginals}

A linear projection generally mixes contributions from several latent
subvectors, so its distribution need not belong to the same CST family.
The characteristic function preserves the independent-source structure
under projection.  It therefore provides a common route to marginal
densities, probabilities, and shape measures without imposing a marginal
closure property.

\begin{theorem}[Marginal characteristic function]
\label{thm:marginal-cf}
Let $S\in\mathbb{R}^{r\times n}$ have full row rank and define
$Y_S=SY$, $\mu_S=S\mu$, and $\Xi_{S,k}=S\Xi E_k$.  Then
\begin{equation}\label{eq:marginal-cf}
 \varphi_{Y_S}(u)
 =\exp(iu^{\prime}\mu_S)
  \prod_{k=1}^K\varphi_{Z_k}(\Xi_{S,k}^{\prime}u),
 \qquad u\in\mathbb R^r,
\end{equation}
where, for
$W_k\sim\operatorname{Gamma}(\nu_k/2,\nu_k/2)$ in shape--rate form,
\begin{equation}\label{eq:block-cf}
 \varphi_{Z_k}(t)
 =2\exp(-it^{\prime}m_k)\E_{W_k}\!\left[
   \exp\!\left\{-\frac{t^{\prime}\Omega(m_k,\nu_k)t}{2W_k}\right\}
   \Phi\!\left\{\frac{it^{\prime}\omega_k\delta_k}{\sqrt{W_k}}\right\}
 \right],
 \qquad t\in\mathbb R^{n_k}.
\end{equation}
Here $\Phi$ is evaluated through its analytic continuation.
\end{theorem}

The product in \eqref{eq:marginal-cf} reduces observed marginal analysis to
the subvector characteristic functions.  For the $j$th univariate margin, put
$\beta_{j,k}=E_k^{\prime}\Xi^{\prime}e_j$.  Then
\begin{equation}\label{eq:univariate-cf}
 \varphi_{Y_j}(s)=e^{is\mu_j}
 \prod_{k=1}^K\varphi_{Z_k}(s\beta_{j,k}).
\end{equation}
If $\varphi_{Y_j}\in L^1(\mathbb R)$, Fourier inversion gives
\begin{equation}
 f_{Y_j}(y)
 =\frac1\pi\int_0^\infty
   \operatorname{Re}\{e^{-isy}\varphi_{Y_j}(s)\}\,ds.
   \label{eq:univariate-density}
\end{equation}
Because $Y_j$ has a continuous distribution, the Gil--Pelaez formula gives
\begin{equation}
 F_{Y_j}(y)
 =\frac12-\frac1\pi\int_0^\infty
   \frac{\operatorname{Im}\{e^{-isy}\varphi_{Y_j}(s)\}}{s}\,ds.
   \label{eq:univariate-cdf}
\end{equation}
The last integral is understood in the usual improper sense.  These
expressions allow marginal probabilities to be computed from the joint
model, preserving the same exposure and tail parameters across different
projections.  Multivariate Fourier inversion similarly applies to $Y_S$
whenever $\varphi_{Y_S}\in L^1(\mathbb R^r)$.

The same projection notation gives an analytical Pearson skewness
coefficient.  Define
\begin{equation}\label{eq:marginal-projection-quantities}
 s_{jk}=\beta_{j,k}^{\prime}m_k,
 \qquad
 q_{jk}=\|\beta_{j,k}\|^2,
 \qquad
 \sigma_j^2=\sum_{k=1}^Kq_{jk}=(\Xi\Xi^{\prime})_{jj},
\end{equation}
and let $\mathcal E_j=\{k:q_{jk}>0\}$ collect the subvectors to which the
$j$th margin is exposed.

\begin{proposition}[Univariate marginal skewness]
\label{prop:marginal-skewness}
If $\nu_k>3$ for every $k\in\mathcal E_j$, then the third central moment
and Pearson skewness of $Y_j$ are
\begin{align}
 \kappa_{3,j}
 &=\sum_{k\in\mathcal E_j}
   \frac{s_{jk}}{\nu_k-3}
   \left\{
    3q_{jk}
    +\left(2\nu_k-3-\frac{\nu_k}{a_{\nu_k}^2}\right)s_{jk}^2
   \right\},
   \label{eq:cst-marginal-third-moment}\\
 \operatorname{skew}(Y_j)
 &=\frac{\kappa_{3,j}}{\sigma_j^3}.
   \label{eq:cst-marginal-skewness}
\end{align}
If $\nu_k\leq3$ for any $k\in\mathcal E_j$, then
$\E|Y_j-\mu_j|^3=\infty$ and Pearson skewness is not defined.
\end{proposition}

In the invariant notation developed in Section \ref{subsec:invariants},
$s_{jk}=b_{k,j}$ and $q_{jk}=G_{k,jj}$.  Hence
\eqref{eq:cst-marginal-skewness} depends only on identified observed-space
subvector contributions.  Observed skewness is an aggregate of source-specific contributions:
positive and negative terms in \eqref{eq:cst-marginal-third-moment} can
reinforce or offset one another.  A small marginal skewness coefficient
can consequently conceal substantial asymmetry in the underlying sources.  Likewise,
$\varphi_{Z_k}(s\beta_{j,k})$ depends on the orientation of the loading
exposure relative to $m_k$, whereas under CT it depends only on the exposure
norm.  Supplementary Section \textcolor{blue}{S.1} proves
Theorem \ref{thm:marginal-cf} and Proposition
\ref{prop:marginal-skewness} and gives the directional tail-ratio and
tail-aggregation results.  In particular, an observed margin is governed in
its most remote tails by the smallest $\nu_k$ among its exposed subvectors,
with one-sided constants adding when several subvectors share that minimum.
Thus similar marginal tail indices can coexist with heterogeneous latent
tail sources; recovering that heterogeneity requires the joint distribution.

\subsection{Geometry, interpretation, and special cases}\label{subsec:geometry}

Mean--covariance standardization makes it possible to distinguish exposure
from distributional shape.  The matrix
$G_k=\Xi_{\bullet k}\Xi_{\bullet k}^{\prime}$ measures subvector $k$'s
contribution to covariance, while $\nu_k$ and $m_k$ determine the tail
decay and directional asymmetry of the shocks transmitted through that
subspace.  Holding $\Xi$ fixed therefore allows tail thickness and
asymmetry to change without altering the mean or covariance.  Conversely,
two loading matrices with the same aggregate covariance can allocate
exposure differently across latent tail sources and generate different
joint distributions.

The distinction matters even when asymmetry is absent.  Coordinates within
a standardized subvector are uncorrelated, but they share a random scale;
a large scale realization can produce large movements in several
coordinates simultaneously.  Independence holds across subvectors, so the
partition determines which latent coordinates share these extreme
movements.  A general loading matrix transmits them to overlapping sets of
observed variables.  The resulting dependence can therefore be organized
around latent groups without requiring observed variables to belong to
disjoint clusters.

The familiar special cases isolate these mechanisms.  Setting every
$m_k=0$ gives CT, which retains heterogeneous tails and shared subvector
scales while imposing central symmetry.  With $K=1$, CST is an affine
multivariate skew-$t$ distribution; setting $m_1=0$ also gives the
multivariate Student-$t$.  As all $\nu_k\to\infty$, the subvectors converge
to skew-normal distributions, and imposing $m_k=0$ as well gives the
Gaussian distribution.  These limits hold for fixed mean shifts in the
interior of the limiting admissible region.  The benchmarks distinguish variation in tail
decay from the additional effects of asymmetry and multiple tail sources.

\begin{figure}
	\centering
	\includegraphics[
		width=\textwidth,
		height=0.86\textheight,
		keepaspectratio
	]{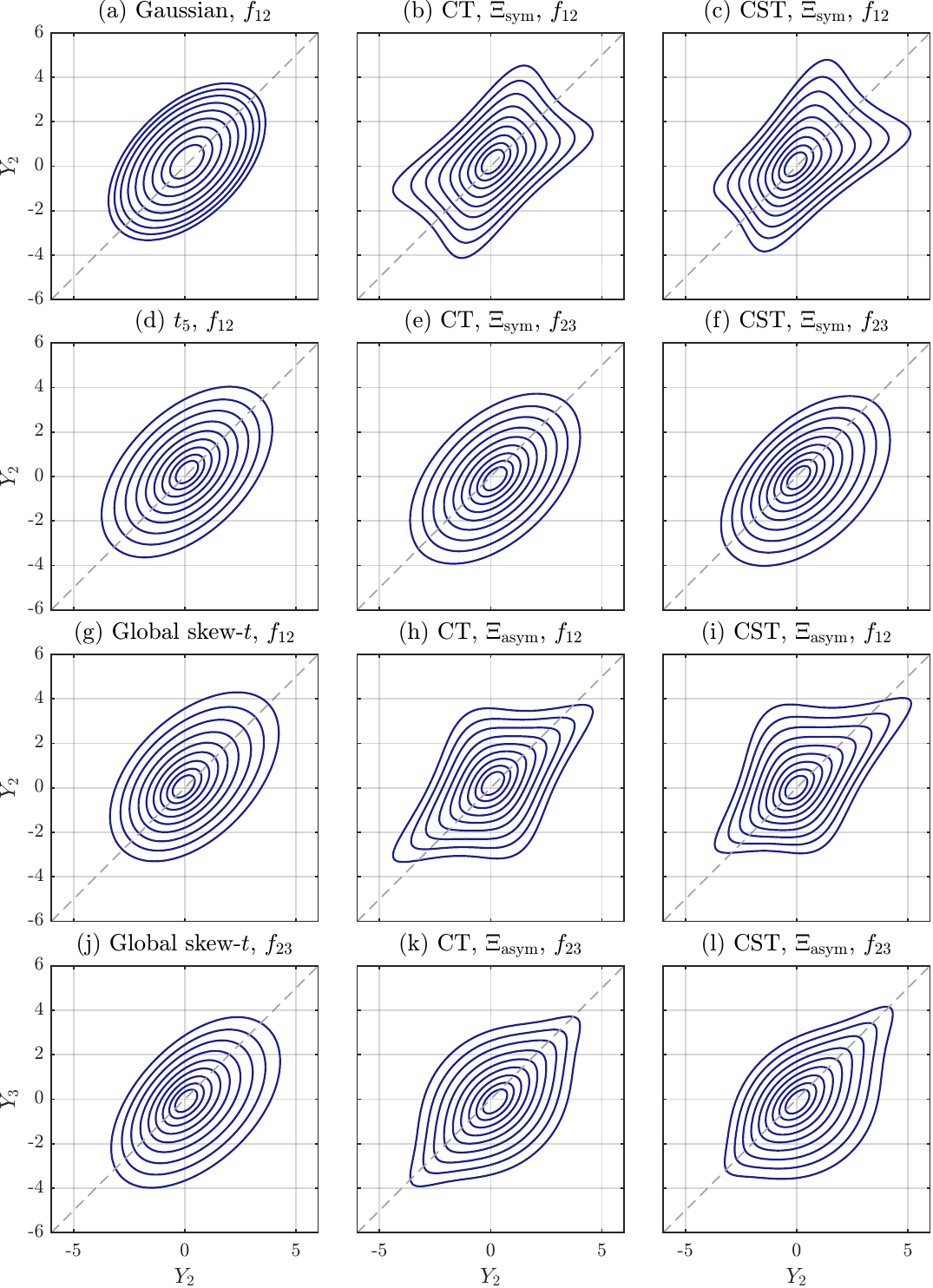}	
\caption{Bivariate marginal density contours under Gaussian, multivariate-$t$,
global skew-$t$, CT, and CST specifications.  The CT and CST panels compare
the symmetric and asymmetric loading matrices defined in the text. Here,
$f_{12}$ and $f_{23}$ denote the bivariate marginals of $(Y_1,Y_2)$ and
$(Y_2,Y_3)$, respectively; both are computed by direct marginalization.  The dashed line is the
45-degree reference line.}
	\label{fig:Contour-plots}
\end{figure}

Figure \ref{fig:Contour-plots} holds
$\mu=(0.10,0.20,-0.10)^{\prime}$ and the equicorrelation covariance matrix
$\Sigma_{1/2}$ fixed to isolate differences beyond the first two moments.
The CT and CST designs have subvector dimensions $(1,2)^{\prime}$ and
degrees of freedom $(4,5)^{\prime}$.  We compare the principal square root
$\Xi_{\mathrm{sym}}=\Sigma_{1/2}^{1/2}$ with the lower Cholesky factor
$\Xi_{\mathrm{asym}}=\operatorname{chol}_{L}(\Sigma_{1/2})$; both satisfy
$\Xi\Xi^{\prime}=\Sigma_{1/2}$.  CST uses $m_1=0.45$ and
$m_2=(0.40,-0.30)^{\prime}$, while CT sets these shifts to zero.  Each
bivariate density is obtained by integrating the trivariate joint density
over the omitted coordinate, so the illustration does not require closure
of CST under marginalization.

The Gaussian and global $t_5$ benchmarks have elliptical contours:
heavier tails change the radial decay while preserving the same basic
geometry.  CT already departs from this pattern.  With the symmetric
loading square root, $f_{12}$ has rounded-diamond contours while $f_{23}$
is much closer to elliptical.  The difference reflects how each observed
pair combines the scalar and bivariate tail sources.  Thus even within one
distribution, similar second-moment dependence need not imply similar
density geometry across pairs.

Comparing CT with CST under the same loading matrix isolates the effect of
asymmetry.  CST preserves the non-elliptical structure but redistributes
density unequally across directions.  The effect differs between $f_{12}$
and $f_{23}$ because their loading exposures have different projections
onto $m_k$, as in Section \ref{subsec:marginals}.  A common latent skewness
vector can consequently produce pronounced asymmetry in one observed
margin and a milder effect in another.  This variation is governed jointly
by skewness and exposure, rather than by a separate skewness parameter for
each observed variable.

Changing the loading square root produces a further change in geometry.
The more angular CT contours under $\Xi_{\mathrm{asym}}$ remain centrally
symmetric: an asymmetric loading matrix does not itself generate skewness
from symmetric sources.  Its role is to change how the independent tail
sources enter the observed variables.  CST then adds directional imbalance
to that geometry.  These comparisons show why choosing a covariance square
root can carry distributional content in a convolution model.  Section
\ref{sec:identification} distinguishes such changes in exposure from the
within-subvector rotations that leave the distribution unchanged.

\section{Identification and normalized parameters}\label{sec:identification}

Identification must distinguish the contribution of a latent subvector
from the coordinates chosen to represent it.  An observed distribution can
identify a subspace and its asymmetric direction while leaving the basis
within that subspace undetermined.  Specifically, if
$Q_k\in O(n_k)$ is orthogonal, then
\begin{equation}\label{eq:rotation}
  (\Xi_{\bullet k},m_k)
  \mapsto
  (\Xi_{\bullet k}Q_k,Q_k^{\prime}m_k)
\end{equation}
leaves the subvector contribution unchanged in distribution.  Equal-dimensional
subvectors may also be permuted.  Skewness therefore does not eliminate the
usual independent-subspace indeterminacy.  Instead, it couples the
orientation of the loading submatrix to the coordinates used to describe its
skewness.

\subsection{Rotation-invariant subvector contributions}\label{subsec:invariants}

For each subvector define
\begin{equation}\label{eq:invariants}
  b_k=\Xi_{\bullet k}m_k,
  \qquad
  G_k=\Xi_{\bullet k}\Xi_{\bullet k}^{\prime}.
\end{equation}
Both objects are invariant under \eqref{eq:rotation}.  They also have direct
observed-space interpretations: $G_k$ is the part of $\Var(Y)$ contributed
by subvector $k$, while $b_k$ records the direction in which that subvector's raw
skewness is transmitted to $Y$.  Moreover,
\[
 \Xi_{\bullet k}\omega_k\delta_k=a_{\nu_k}^{-1}b_k,
 \qquad
 \Xi_{\bullet k}\Omega(m_k,\nu_k)\Xi_{\bullet k}^{\prime}
 =\frac{\nu_k-2}{\nu_k}(G_k+b_kb_k^{\prime}).
\]
The subvector characteristic function shows that the observed contribution of
subvector $k$ depends on its loading and
mean-shift coordinates through $(\nu_k,b_k,G_k)$.  The matrix $G_k$ is
its covariance contribution, while $b_k$ is its observed-space
asymmetric direction.  If $G_k$ has rank $n_k$, then the skewness
magnitude also has the invariant representation
\begin{equation}\label{eq:m-norm-invariant}
  \|m_k\|^2=b_k^{\prime}G_k^+b_k,
\end{equation}
where $G_k^+$ is the Moore-Penrose inverse.
Combining this identity with the admissible ball in
\eqref{eq:admissible-ball} gives an observed-space characterization of
admissibility:
\begin{equation}\label{eq:invariant-admissibility}
 G_k\succeq0,\qquad \rank(G_k)=n_k,\qquad
 b_k\in\operatorname{col}(G_k),\qquad
 b_k^{\prime}G_k^+b_k<\bar m_{\nu_k}^2.
\end{equation}
In addition, $\sum_{k=1}^{K}G_k=\Var(Y)$.  Hence the invariant representation is
not only interpretable but also directly constrained: each asymmetric
direction must lie in the covariance subspace through which that subvector
affects the observations.

\Needspace{6\baselineskip}
The following algebraic result explains why identifying an additional
asymmetric direction does not, by itself, determine more loading
coordinates.  It applies once the subvector-specific invariants have been
recovered from the distribution.

\begin{proposition}[Static loading equivalence in CST and CT]
\label{prop:loading-equivalence}
Fix one subvector, suppress its subscript, and write $d=n_k$ and
$L=\Xi_{\bullet k}\in\mathbb R^{n\times d}$.  Let $G\succeq0$ have rank
$d$ and let $b\in\operatorname{col}(G)$.  Suppose $L$ has full column rank
and
\begin{equation}\label{eq:factor-pair}
 G=LL^{\prime},\qquad b=Lm.
\end{equation}
Every other full-column-rank pair $(\widetilde L,\widetilde m)$ satisfying
the same two equations can be written uniquely as
\begin{equation}\label{eq:factor-pair-orbit}
 \widetilde L=LQ,\qquad \widetilde m=Q^{\prime}m,
 \qquad Q\in O(d).
\end{equation}
Conversely, every $Q\in O(d)$ in \eqref{eq:factor-pair-orbit} preserves
$(b,G)$.  Consequently, the CST invariants $(b,G)$ leave the same
$d(d-1)/2$-dimensional loading orbit as the CT loading invariant $G$ alone.
\end{proposition}

The proof is given in Appendix \ref{app:loading-equivalence-proof}.

Proposition \ref{prop:loading-equivalence} distinguishes additional
distributional identification from additional loading identification.  CST
identifies $b$, which carries information absent under CT, including the
latent skewness magnitude $\|m\|^2=b^{\prime}G^+b$.  It does not, however,
select an orientation of $L$, because the unknown coordinates of $m$ adjust
one-for-one with that orientation.  On the rank-$d$ stratum, $(G,b)$ has
$d$ more free coordinates than $G$, but $(L,m)$ also has $d$ more free
coordinates than $L$.  The unidentified dimension is therefore unchanged.
Using $\delta$ instead does not change this conclusion.  Under a rotation
$Q$, its coordinates become $\delta(Q^{\prime}m,\nu)$ through
\eqref{eq:delta-map}; because the marginal scale factors also change, this
is generally not the linear transformation $Q^{\prime}\delta(m,\nu)$.

\subsection{Identification theorem}\label{subsec:id-theorem}

To make the loading structure explicit, partition the rows and columns of
$\Xi$ according to $\bm n=(n_1,\ldots,n_K)^{\prime}$:
\begin{equation}\label{eq:loading-block-structure}
 \Xi=
 \begin{bmatrix}
  \Xi_{11}&\Xi_{12}&\cdots&\Xi_{1K}\\
  \Xi_{21}&\Xi_{22}&\cdots&\Xi_{2K}\\
  \vdots&\vdots&\ddots&\vdots\\
  \Xi_{K1}&\Xi_{K2}&\cdots&\Xi_{KK}
 \end{bmatrix}
 =\bigl[\Xi_{\bullet1},\ldots,\Xi_{\bullet K}\bigr].
\end{equation}
Here $\Xi_{ij}=E_i^{\prime}\Xi E_j\in\mathbb R^{n_i\times n_j}$,
and the $k$th column group
$\Xi_{\bullet k}=\Xi E_k\in\mathbb R^{n\times n_k}$ contains the
loadings of all observed variables on the latent subvector $Z_k$.
Assumption \ref{ass:identification-normalization} below normalizes each
diagonal loading submatrix $\Xi_{kk}$ to be symmetric positive definite.
The off-diagonal blocks $\Xi_{ij}$, $i\ne j$, remain unrestricted, subject
to nonsingularity of $\Xi$; in particular, $\Xi_{ij}$ need not equal
$\Xi_{ji}^{\prime}$.

To fix subvector labels, define the rotation-invariant scalar
\begin{equation}\label{eq:varsigma}
 \varsigma_k
 =\delta_k^{\prime}\bar\Omega_k^{-1}\delta_k
 =\frac{\|m_k\|^2}
 {a_{\nu_k}^2 c_{\nu_k}(1+\|m_k\|^2)},
 \qquad
 \bar\Omega_k=\omega_k^{-1}\Omega_k\omega_k^{-1},
 \qquad
 c_{\nu_k}=\frac{\nu_k-2}{\nu_k}.
\end{equation}
Write $\zeta_k=(\nu_k,\varsigma_k)^{\prime}$ for the corresponding invariant
label pair.
The following assumption formally records the label and orientation
conventions used to select one representative from each observational
equivalence class.

\Needspace{8\baselineskip}
\begin{assumption}[Identification normalization]
\label{ass:identification-normalization}
The subvector-size vector $\bm n$ is known, and the following conditions hold:
\begin{enumerate}[label=(I\arabic*),leftmargin=*,itemsep=0.25em]
 \item $n_1\leq\cdots\leq n_K$; equal-dimensional subvectors have distinct
 invariant pairs $\zeta_k$ and are ordered lexicographically by those pairs.
 \item For every $k$, $D_k=E_k^{\prime}G_kE_k\succ0$, and the
 within-subvector orientation is chosen so that
 $\Xi_{kk}=E_k^{\prime}\Xi E_k=D_k^{1/2}$, the symmetric positive-definite
 square root of $D_k$.
\end{enumerate}
\end{assumption}

Part (I2) of Assumption \ref{ass:identification-normalization} combines an
invariant transversality requirement with a
just-identifying orientation normalization.  Indeed,
$D_k=\Xi_{kk}\Xi_{kk}^{\prime}$.  Thus $D_k\succ0$ requires the observed
coordinates assigned to subvector $k$ to load nonsingularly on its latent subspace; given
this requirement, the polar decomposition supplies a unique within-subvector
rotation for which $\Xi_{kk}=D_k^{1/2}\succ0$.  Equivalently, requiring
$\Xi_{kk}$ to be symmetric positive definite already implies
$D_k=\Xi_{kk}^2\succ0$.  The normalization imposes
$n_k(n_k-1)/2$ equalities, exactly the dimension of the within-subvector
orthogonal group in Proposition \ref{prop:loading-equivalence}.  This is the
same count as in CT \citep{Hansen2026}: the normalization removes no
distributional feature and no more loading freedom than observational
equivalence requires.

Two properties connect this algebraic normalization to identification from
the distribution.  A single standardized subvector must reveal its tail
index and mean-shift direction up to rotation.  It must also resist a
further decomposition into independent pieces; otherwise the proposed
subvector partition would not describe the underlying independent sources.

\begin{lemma}[Identification of one standardized subvector]
\label{lem:block-identification}
Let $Q\in O(d)$.  If
\[
 Q^{\prime}Z(m,\nu)\overset{d}{=}Z(\widetilde m,\widetilde\nu),
\]
then $\widetilde\nu=\nu$ and $\widetilde m=Q^{\prime}m$.
\end{lemma}

\begin{lemma}[Linear irreducibility]
\label{lem:irreducibility}
Every nondegenerate standardized skew-$t$ subvector with finite degrees of
freedom is linearly irreducible: no nonsingular linear transformation can
partition it into two nonempty independent subvectors.
\end{lemma}

Lemma \ref{lem:block-identification} is proved by projecting the subvector onto
an arbitrary unit direction.  The polynomial tail exponent identifies
$\nu$, while the ratio of the right- and left-tail constants is strictly
monotone in the projected mean shift.  Lemma \ref{lem:irreducibility} uses
the common mixing variable inside a skew-$t$ subvector.  For a radial threshold
$r\to\infty$, the joint exceedance probability of two nonempty transformed
subvectors has order $r^{-\nu}$, whereas independence would imply order
$r^{-2\nu}$.  These arguments are given in Appendix
\ref{app:identification}.

Together with independent-subspace uniqueness, these properties identify
the latent contributions up to permutation and rotation.  The normalization
then turns those contributions into a unique reported parameter, as the
following theorem makes explicit.

\begin{theorem}[Identification and exact recovery]\label{thm:identification}
Fix the subvector-size vector $\bm n$.  Let two admissible CST parameter
values have nonsingular loading matrices and finite degrees of freedom
$\nu_k\in(2,\infty)$.  If they generate the same distribution of $Y$,
then their locations agree and there exist a dimension-preserving
permutation $\pi$ and orthogonal matrices $Q_k\in O(n_k)$ such that
\[
 \widetilde\nu_{\pi(k)}=\nu_k,
 \qquad
 \widetilde\Xi_{\bullet\pi(k)}=\Xi_{\bullet k}Q_k,
 \qquad
 \widetilde m_{\pi(k)}=Q_k^{\prime}m_k.
\]
Consequently, the distribution identifies the collection
$\{(n_k,\nu_k,b_k,G_k):k=1,\ldots,K\}$ up to a dimension-preserving
permutation.  Under Assumption
\ref{ass:identification-normalization}, the normalized parameter is unique.
In particular, each loading submatrix and mean-shift vector are recovered from
the invariants by
\begin{equation}\label{eq:recovery}
 \Xi_{kk}=D_k^{1/2},
 \qquad
 \Xi_{\bullet k}=G_kE_kD_k^{-1/2},
 \qquad
 m_k=D_k^{-1/2}E_k^{\prime}b_k.
\end{equation}
\end{theorem}

The recovery map separates what the distribution determines from how that
information is reported.  Before normalization, the identified object is
the rotational equivalence class in \eqref{eq:rotation}, represented by
$(b_k,G_k)$ together with the tail indices and subvector dimensions.
Under part (I2) of Assumption \ref{ass:identification-normalization},
\eqref{eq:recovery} selects a unique set of loading and skewness coordinates.
The norm of $m_k$ and its observed image $b_k$ are intrinsic, whereas its
individual coordinates, and those of $\delta_k$, depend on the selected
orientation.  This distinction guides interpretation: comparisons of
reported coordinates require a common normalization, while comparisons of
the invariant contributions do not.

\begin{remark}[Ordering and empirical labels]\label{rem:labels}
The ordering in part (I1) of Assumption
\ref{ass:identification-normalization} is an identifying convention, not an
economic ranking.
An empirical implementation may optimize over a convenient provisional
labeling and sort or otherwise relabel the subvectors after estimation.  A
simultaneous permutation of subvectors, loading-column groups, and subvector parameters
does not change the likelihood or any fitted observed-data quantity.  If
asset-aligned labels are retained for presentation, those labels should be
understood as a reporting normalization rather than an intrinsically
identified economic ordering.
\end{remark}

\subsection{The normalized parameter space}\label{subsec:parameters}

For the block structure in \eqref{eq:loading-block-structure}, let
$\gamma$ collect every entry of the off-diagonal blocks and the
lower-triangular entries of the symmetric diagonal blocks.  The number
of free loading parameters is
\begin{equation}\label{eq:p-gamma}
 p_\gamma=n^2-\sum_{k=1}^K\frac{n_k(n_k-1)}{2}.
\end{equation}
Write
\[
 G_\Xi=\frac{\partial\vecop(\Xi)}{\partial\gamma^{\prime}},
 \qquad
 \theta=(\mu^{\prime},\gamma^{\prime},m_1^{\prime},\ldots,m_K^{\prime},\nu_1,\ldots,\nu_K)^{\prime}.
\]
For the linear free-entry parameterization, $G_\Xi$ is a constant
selection/duplication matrix.  The global reporting normalization imposes
\eqref{eq:admissibility}, nonsingularity of $\Xi$, positive definiteness
of its diagonal blocks, and the ordering in part (I1) of Assumption
\ref{ass:identification-normalization}.  Likelihood theory is developed on a
fixed smooth label chart whose equal-dimensional subvectors occupy disjoint
neighborhoods of their invariant pairs.  Thus labels are not repeatedly
resorted as the parameter varies locally.  The global identification result
uses the lexicographic reporting convention, whereas regular local
likelihood theory only requires the full invariant pairs to remain separated.

The parameter count in \eqref{eq:p-gamma} removes exactly the rotational
freedom identified in Proposition \ref{prop:loading-equivalence}.  All
off-diagonal loading entries remain free, so the normalization permits
general observed covariance and overlapping exposures to latent sources.
Economic restrictions such as repeated sector loadings are additional
modeling choices.  Keeping these choices separate allows the empirical
analysis to assess a parsimonious exposure structure without confusing it
with the normalization required for identification.

\section{Likelihood derivatives and information}\label{sec:inference}

The explicit likelihood provides a practical estimation method once its
derivatives respect the standardization and loading normalization.
We derive the score and observed curvature in those coordinates, then use
subvector independence to organize Fisher information.  The main
computational gain comes from reducing the required expectations to
one-dimensional integrals, even for multivariate subvectors.

\subsection{Analytical score}\label{subsec:score}

Throughout this section, $Y$ denotes a generic random observation; lowercase
$x$ is reserved for a deterministic argument of a score function.  Let
\[
 A=\Xi^{-1},\qquad
 Z=A(Y-\mu),\qquad
 Z_k=E_k^{\prime}Z,\qquad
 X_k=Z_k+m_k .
\]
For $x\in\mathbb R^{n_k}$, define
\[
 \begin{split}
 Q_k(x)&=x^{\prime}\Omega_k^{-1}x,\qquad
 v_k(x)=\frac{\nu_k+n_k}{\nu_k+Q_k(x)},\\
 q_k(x)&=v_k(x)^{1/2}\eta_k^{\prime}x,\qquad
 \rho_k(x)=
 \frac{t_1\{q_k(x);\nu_k+n_k\}}
      {T_1\{q_k(x);\nu_k+n_k\}},\\
 \mathcal R_k(x)&=
 1+\frac{q_k(x)\rho_k(x)}{\nu_k+n_k}.
 \end{split}
\]
The subvector argument score combines the radial Student-$t$ contribution
with the directional correction induced by skewness:
\begin{equation}\label{eq:block-argument-score}
 g_k(x)
 =-v_k(x)\mathcal R_k(x)\Omega_k^{-1}x
  +v_k(x)^{1/2}\rho_k(x)\eta_k.
\end{equation}
Put
$H=\{g_1(X_1)^{\prime},\ldots,g_K(X_K)^{\prime}\}^{\prime}$.
For each subvector, let $\mathcal r_k(x)$ denote the raw skew-$t$ score with
respect to
\[
 u_k^{\mathrm{raw}}(x)
 =(x^{\prime},\vech(\Omega_k)^{\prime},\eta_k^{\prime},\nu_k)^{\prime},
 \qquad
 R_k=\left\{
 \frac{\partial u_k^{\mathrm{raw}}(X_k)}
 {\partial(m_k^{\prime},\nu_k)}
 \right\}^{\prime}.
\]
Thus $R_k$ is the chain-rule matrix induced by the standardization.  Explicit
forms of $\mathcal r_k(\cdot)$ and $R_k$ are given in Supplementary Appendix
\textcolor{blue}{S.2.1}.

\begin{proposition}[Analytical score]
\label{prop:score}
Under the normalized parameterization, the exact score contributions of
$Y$ are
 \begin{align}
 s_\mu(Y;\theta)&=-A^{\prime}H,\label{eq:score-mu}\\
 s_\gamma(Y;\theta)&=G_\Xi^{\prime}
   \vecop(-A^{\prime}-A^{\prime}HZ^{\prime}),\label{eq:score-gamma}\\
 s_{(m_k^{\prime},\nu_k)^{\prime}}(Y;\theta)
   &=R_k\mathcal r_k(X_k),
 \qquad k=1,\ldots,K.\label{eq:score-block}
 \end{align}
\end{proposition}

The location and loading identities in \eqref{eq:score-mu}--\eqref{eq:score-gamma}
have the same outer form as in an independent-components likelihood, but $H$ is the score of
the standardized, rather than raw, skew-$t$ subvectors.  The last identity is
where the mean-shift standardization enters explicitly.  In particular,
$R_k$ differentiates $X_k$, $\Omega_k$, and $\eta_k$ jointly with
respect to $(m_k,\nu_k)$; ignoring any of these paths does not produce the
score of the standardized distribution.

\subsection{Observed Hessian}\label{subsec:hessian}

Let
\[
 \vartheta_k=(m_k^{\prime},\nu_k)^{\prime},
 \qquad
 J_k(X_k)
 =\left.\frac{\partial g_k(x)}{\partial x^{\prime}}\right|_{x=X_k},
 \qquad
 \mathcal H_k^{\mathrm{raw}}(X_k)
 =\left.
 \frac{\partial\mathcal r_k(x)}
      {\partial u_k^{\mathrm{raw}}(x)^{\prime}}
 \right|_{x=X_k}
\]
and $J_Z=\sum_kE_kJ_k(X_k)E_k^{\prime}$.  Differentiating Proposition
\ref{prop:score} gives
$\mathcal H(Y;\theta)=\partial^2\ell(Y;\theta)/
\partial\theta\partial\theta^{\prime}$.  Two useful exact blocks are
\begin{align}\label{eq:block-hessian-chain}
 \mathcal H_{\mu\mu}(Y;\theta)=A^{\prime}J_ZA,
 \qquad&
 \mathcal H_{\vartheta_k\vartheta_k}(Y;\theta)
 =R_k\mathcal H_k^{\mathrm{raw}}(X_k)R_k^{\prime}
  +\mathcal K_k(Y),\notag\\
 \mathcal K_k(Y)
 =\sum_j\mathcal r_{k,j}(X_k)
   \frac{\partial^2u_{k,j}^{\mathrm{raw}}(X_k)}
        {\partial\vartheta_k\partial\vartheta_k^{\prime}}.
\end{align}
The term $\mathcal K_k(Y)$ captures the curvature of the standardization
map.  Changing a mean-shift or tail parameter changes the subvector
argument, dispersion, and skewing coefficient together.  The Hessian must
therefore account for the curvature of this joint adjustment as well as
the curvature of the raw log density.

The remaining blocks follow directly by differentiating
\eqref{eq:score-mu}--\eqref{eq:score-block}.  Because the free-entry
parameterization is linear,
$G_\Xi$ is constant and
\begin{equation}\label{eq:hessian-loading-map}
 \mathcal H_{\gamma\gamma}
 =G_\Xi^{\prime}\mathcal H_{\Xi\Xi}G_\Xi,
 \quad
 \mathcal H_{\gamma a}
 =G_\Xi^{\prime}\mathcal H_{\Xi a},
 \quad
 \mathcal H_{a\gamma}
 =\mathcal H_{a\Xi}G_\Xi,
\end{equation}
where $a$ denotes $\mu$ or a subvector-specific parameter.  Thus the linear
loading normalization introduces only selection and duplication of entries;
the additional curvature arises from the nonlinear shape standardization.
Supplementary Appendix \textcolor{blue}{S.2.1} records the raw scores and
standardization differentials used in computation.  Automatic differentiation
may instead be applied directly in normalized coordinates, provided that the
standardization map remains in the computational graph.  The sample-average Hessian
and observed information per observation are
\begin{equation}\label{eq:sample-hessian}
 \mathcal H_T(\theta)=T^{-1}\sum_{t=1}^T\mathcal H(Y_t;\theta),
 \qquad
 \mathcal J_T(\theta)=-\mathcal H_T(\theta).
\end{equation}

\subsection{Fisher information and one-dimensional reduction}
\label{subsec:information}

The independence of the latent subvectors yields a useful decomposition of
Fisher information.  Define the raw within-subvector moments
\begin{equation}\label{eq:primitive-moments}
 H_k=\E\{g_k(X_k)g_k(X_k)^{\prime}\},\qquad
 C_k=\E\{g_k(X_k)\mathcal r_k(X_k)^{\prime}\},\qquad
 \mathcal V_k=\E\{\mathcal r_k(X_k)\mathcal r_k(X_k)^{\prime}\}.
\end{equation}
For the loading score, put
\[
 q_\Xi=\vecop(A^{\prime}HZ^{\prime}),\qquad
 a_\Xi=\vecop(A^{\prime}),\qquad
 q_{\Xi,kk}
 =\vecop\{A^{\prime}E_kg_k(X_k)Z_k^{\prime}E_k^{\prime}\},
\]
and define
\begin{equation}\label{eq:loading-moments}
 Q_\Xi=\E(q_\Xi q_\Xi^{\prime}),
 \qquad
 L_k=\E\{g_k(X_k)q_{\Xi,kk}^{\prime}\},
 \qquad
 M_k=\E\{q_{\Xi,kk}\mathcal r_k(X_k)^{\prime}\}.
\end{equation}
The score identities imply $\E\{g_k(X_k)\}=0$,
$\E\{g_k(X_k)Z_k^{\prime}\}=-I_{n_k}$, and
$\E(q_\Xi)=-a_\Xi$.

\begin{proposition}[Subvector Fisher information]
\label{prop:information}
Suppose the score moments in \eqref{eq:primitive-moments}-
\eqref{eq:loading-moments} exist.  Under the temporary subvector ordering
$\vartheta_k=(m_k^{\prime},\nu_k)^{\prime}$, the potentially nonzero blocks of
$\mathcal I(\theta)=\E\{s(Y;\theta)s(Y;\theta)^{\prime}\}$ are
\begin{align}
 \mathcal I_{\mu\mu}
 &=A^{\prime}\left(\sum_{k=1}^KE_kH_kE_k^{\prime}\right)A,
 &
 \mathcal I_{\mu\gamma}
 &=\left(\sum_{k=1}^KA^{\prime}E_kL_k\right)G_\Xi,
 \label{eq:info-row-one}\\
 \mathcal I_{\mu\vartheta_k}
 &=-A^{\prime}E_kC_kR_k^{\prime},
 &
 \mathcal I_{\gamma\gamma}
 &=G_\Xi^{\prime}(Q_\Xi-a_\Xi a_\Xi^{\prime})G_\Xi,
 \label{eq:info-row-two}\\
 \mathcal I_{\gamma\vartheta_k}
 &=-G_\Xi^{\prime}M_kR_k^{\prime},
 &
 \mathcal I_{\vartheta_k\vartheta_k}
 &=R_k\mathcal V_kR_k^{\prime}.
 \label{eq:info-row-three}
\end{align}
For $k\ne l$, $\mathcal I_{\vartheta_k\vartheta_l}=0$, and the remaining
blocks follow by symmetry.  Every scalar primitive needed to construct
$H_k,C_k,\mathcal V_k,L_k,M_k$, and $Q_\Xi$ is a finite linear
combination of closed-form radial coefficients and one-dimensional
integrals over $(-1,1)$.
\end{proposition}

To see the reduction, suppress the subvector subscript, whiten the raw variable,
and rotate so that $\Omega^{1/2}\eta=\tau e_1$.  For
$V\sim t_d(0,I_d,\nu)$ under the symmetric kernel, the directional
coordinate
\[
 B=\frac{V_1}{\sqrt{\nu+V^{\prime}V}}
\]
has density
\begin{equation}\label{eq:direction-density}
 f_B(b)=
 \frac{(1-b^2)^{(\nu+d-1)/2-1}}
 {\mathrm B\{1/2,(\nu+d-1)/2\}},
 \qquad -1<b<1.
\end{equation}
The skewing argument becomes $\sqrt{\nu+d}\,\tau b$.  Conditional
rotational symmetry integrates the coordinates orthogonal to the skewness
direction analytically.  The required moments through order four, including
the odd moments needed for cross-information, reduce to Kronecker-delta
tensors multiplied by beta-function coefficients and their derivatives.
The remaining numerical terms are finite linear combinations of integrals
of the form
\begin{equation}\label{eq:one-dimensional-integral}
 \int_{-1}^1
 \chi(cb)T_1(cb;\nu+d)b^p
 (1-b^2)^{\lambda}
 \{-\log(1-b^2)\}^{h}f_B(b)\,\mathrm{d}b,
 \quad c=\sqrt{\nu+d}\,\tau.
\end{equation}
Here $p$ and $h$ are nonnegative integers, $\lambda$ is a nonnegative
half-integer, and the scalar function $\chi$ is generated by the raw score
components (including Student-$t$ density, distribution-function, and
degrees-of-freedom derivative terms).
Increasing the subvector dimension changes the radial coefficients and
the number of matrix entries, but the numerical integration still involves
only the scalar direction $b$.  Pure cross-subvector coordinate factors
are evaluated from $\E(Z_k)=0$ and
$\E(Z_kZ_k^{\prime})=I_{n_k}$, while the remaining radial tensors have
explicit beta-function coefficients.  This makes analytical information
feasible without multidimensional quadrature; its matrix assembly retains
the full dependence among parameters.  Supplementary Appendix
\textcolor{blue}{S.3.2} gives the reconstruction details.

The zero cross-subvector entries
$\mathcal I_{\vartheta_k\vartheta_l}=0$ reflect independence of the latent
subvector scores, but they do not imply that estimates of different subvectors are
asymptotically independent.  Location and loading scores couple the subvectors,
so the corresponding off-diagonal entries of $\mathcal I(\theta)^{-1}$ are
generally nonzero.  This distinction matters empirically: subvector-specific
tail and asymmetry estimates can interact through uncertainty about the
common exposure matrix even when their primitive scores are orthogonal.

\section{Maximum likelihood theory}
\label{sec:asymptotics}

Identification determines the population target of maximum likelihood;
likelihood regularity determines how precisely that target can be
estimated.  The results below connect the normalized representation to
consistency, Gaussian limits, and inference for economically interpretable
functions of the parameters.

Let $Y_1,\ldots,Y_T$ be iid and define
\[
 \ell_T(\theta)=T^{-1}\sum_{t=1}^T\ell_t(\theta),
 \qquad
 \widehat\theta_T\in\arg\max_{\theta\in\Theta}\ell_T(\theta).
\]
The subvector-size vector $\bm n$ is treated as fixed.  The theory concerns
finite degrees of freedom and the regular interior of an
orientation-normalized, label-fixed chart.  Gaussian and skew-normal limits
are boundary cases and are not covered by the following local normal
approximation.  The lower bound above two serves both the mean--covariance
standardization and square-integrability of general cross-subvector loading
scores.  Section \ref{sec:raw-cst} later separates these issues: raw-CST
consistency extends to every positive tail index, whereas the present joint
regular normal approximation does not.

\begin{assumption}[Regular normalized parameter space]
\label{ass:regular-space}
The set $\Theta$ is a compact subset of a fixed smooth labeling chart of the
orientation-normalized parameter space, and
$\theta_0\in\operatorname{int}(\Theta)$.  For equal-dimensional subvectors, there
are fixed open sets $\mathcal Z_k\subset\mathbb R^2$ such that
$\zeta_k(\theta)\in\mathcal Z_k$ for every $\theta\in\Theta$ and
$\overline{\mathcal Z}_k\cap\overline{\mathcal Z}_\ell=\varnothing$ whenever
$k\ne\ell$ and $n_k=n_\ell$.
There is an $\epsilon>0$ such
that, uniformly over $\theta\in\Theta$ and $k=1,\ldots,K$,
\begin{equation}\label{eq:regular-bounds}
 \nu_k\geq2+\epsilon,
 \quad
 \sigma_{\min}\{\Xi(\gamma)\}\geq\epsilon,
 \quad
 \lambda_{\min}(\Xi_{kk})\geq\epsilon,
 \quad
 a_{\nu_k}^2-m_k^{\prime}\Omega_k^{-1}m_k\geq\epsilon.
\end{equation}
\end{assumption}

Compactness supplies finite upper bounds for the continuous parameters,
including $\nu_k$.  The other restrictions keep the likelihood away from
singular loadings, degenerate diagonal loading submatrices, the skewness-admissibility
boundary, and label collisions.  The disjoint label neighborhoods fix a
smooth local chart even when two equal-dimensional subvectors have the same tail
index but different invariant asymmetry magnitudes.  This distinction is
important: the lexicographic rule in part (I1) of Assumption
\ref{ass:identification-normalization} selects a global reporting order, but
sorting by that rule at every sample would not be a smooth operation at an
equality of tail indices.  Local likelihood theory instead keeps the
well-separated labels fixed within their respective neighborhoods.

\begin{theorem}[Consistency]\label{thm:consistency}
Suppose $Y_1,\ldots,Y_T$ are iid from the standardized CST distribution with
true parameter $\theta_0$ in the fixed chart.  Under Assumption
\ref{ass:regular-space}, every measurable maximum likelihood estimator
satisfies
\begin{equation}\label{eq:consistency}
 \widehat\theta_T\xrightarrow{p}\theta_0.
\end{equation}
\end{theorem}

The proof uses two features established earlier.  First, Theorem
\ref{thm:identification} makes the population criterion uniquely maximized
at $\theta_0$.  Second, the bounded skewing argument and polynomial tails
give the global envelope
\begin{equation}\label{eq:global-likelihood-envelope}
 \sup_{\theta\in\Theta}|\ell(y;\theta)|
 \leq C_0+C_1\log(1+\|y\|^2),
\end{equation}
which is integrable because the standardized CST distribution has finite
second moments.  Identification and this envelope play different roles:
the former fixes the unique population maximizer, while the latter
controls sample fluctuations of the criterion uniformly over the parameter
space.

\begin{proposition}[Local likelihood regularity]
\label{prop:local-regularity}
Under Assumption \ref{ass:regular-space}, there is an open convex
neighborhood $\mathcal N$ of $\theta_0$, with closure inside the regular
parameter space, on which the likelihood is twice continuously
differentiable almost surely and
\begin{equation}\label{eq:local-envelope-statement}
 \E_{\theta_0}\!\left[
  \sup_{\theta\in\mathcal N}\|s_t(\theta)\|^2\right]<\infty,
 \qquad
 \E_{\theta_0}\!\left[
  \sup_{\theta\in\mathcal N}\|\mathcal H_t(\theta)\|\right]<\infty.
\end{equation}
For some $r>2$,
\begin{equation}\label{eq:uniform-local-moment}
 \sup_{\theta\in\mathcal N}
 \E_\theta(1+\|Y\|)^r<\infty.
\end{equation}
Moreover, $\theta\mapsto f_\theta$ is twice continuously differentiable as
a map from $\mathcal N$ into $L^1(\mathbb R^n)$.  Consequently,
\begin{equation}\label{eq:information-identity}
 \E_{\theta_0}s_t(\theta_0)=0,
 \qquad
 -\E_{\theta_0}\mathcal H_t(\theta_0)
 =\E_{\theta_0}\{s_t(\theta_0)s_t(\theta_0)^{\prime}\}
 =\mathcal I(\theta_0).
\end{equation}
\end{proposition}

Heavy-tailed observations do not automatically produce equally
heavy-tailed scores.  The radial weights damp large within-subvector
arguments: $v_k^{1/2}\|x_k\|$ and $v_k\|x_k\|^2$ are uniformly bounded
on $\mathcal N$, as are the skewing argument and its inverse Mills ratio.
The remaining cross-subvector loading terms grow at most linearly in
another subvector's coordinates, while tail-index derivatives grow only
logarithmically.  This explains both the regularity in Proposition
\ref{prop:local-regularity} and the importance of finite second moments
for unrestricted loading inference.

\begin{assumption}[Information nonsingularity]\label{ass:information}
$\mathcal I(\theta_0)$ is positive definite.
\end{assumption}

Assumption \ref{ass:information} separates regular estimation for a fixed
subvector partition from singular model-selection problems.  It rules out local
score collinearities and, together with Assumption
\ref{ass:regular-space}, excludes Gaussian or skew-normal limits, singular
loading matrices, and collisions of ordered subvector labels.  Theorem
\ref{thm:normality} therefore supports inference conditional on a specified
finite-$\nu$ subvector structure; choosing the number or dimensions of the
subvectors is a distinct, generally nonregular problem.

\begin{theorem}[Asymptotic normality]\label{thm:normality}
Under Assumptions \ref{ass:regular-space} and \ref{ass:information},
\begin{equation}\label{eq:mle-limit}
 \sqrt T(\widehat\theta_T-\theta_0)
 \Rightarrow N\{0,\mathcal I(\theta_0)^{-1}\}.
\end{equation}
Moreover, the outer-product and negative observed-Hessian estimators
\begin{equation}\label{eq:feasible-information}
 \widehat{\mathcal I}_T
 =T^{-1}\sum_{t=1}^Ts_t(\widehat\theta_T)s_t(\widehat\theta_T)^{\prime},
 \qquad
 \widehat{\mathcal J}_T
 =-T^{-1}\sum_{t=1}^T\mathcal H_t(\widehat\theta_T)
\end{equation}
both converge in probability to $\mathcal I(\theta_0)$.
\end{theorem}

The analytical matrix in Proposition \ref{prop:information}, evaluated at
$\widehat\theta_T$, provides a third plug-in estimator of the same limiting
information.  Their agreement links the theoretical calculation to
implementation: persistent discrepancies can reveal problems in derivative
evaluation or optimization, while finite-sample differences need not
indicate a failure of the asymptotic result.

The same limit immediately gives inference for smooth functions of the
normalized parameter.  For a continuously differentiable target
$a(\theta)$, Theorem
\ref{thm:normality} gives
\begin{equation}\label{eq:delta-method}
 \sqrt T\{a(\widehat\theta_T)-a(\theta_0)\}
 \Rightarrow
 N\{0,D_a\mathcal I(\theta_0)^{-1}D_a^{\prime}\},
 \qquad
 D_a=\frac{\partial a(\theta_0)}{\partial\theta^{\prime}}.
\end{equation}
This covers $\vecop(\Xi)$, the bounded skewness coordinates $\delta_k$,
and the invariant targets $(b_k,G_k)$.  It also covers smooth marginal
densities, distribution functions, quantiles away from zero marginal
density, and tail probabilities computed from Section
\ref{subsec:marginals} and Supplementary Section
\textcolor{blue}{S.1.1}.  In particular, define the full-parameter
Jacobian
\[
 D_{\Xi,0}
 =\left.\frac{\partial\vecop(\Xi)}{\partial\theta^{\prime}}
  \right|_{\theta=\theta_0}.
\]
Its loading-coordinate columns contain $G_\Xi$, and its other columns are
zero.  Then
\begin{equation}\label{eq:loading-asymptotic-variance}
 \Avar\{\sqrt T\vecop(\widehat\Xi-\Xi_0)\}
 =V_{\Xi,\mathrm{CST}}
 =D_{\Xi,0}\mathcal I(\theta_0)^{-1}D_{\Xi,0}^{\prime}.
\end{equation}

\section{Consequences of imposing false symmetry}
\label{sec:symmetry-implications}

The common CT--CST loading normalization makes reported loadings comparable,
but it does not make the restricted CT estimator consistent for an
asymmetric CST parameter.  Stack the subvector asymmetry coordinates as
$m=(m_1^{\prime},\ldots,m_K^{\prime})^{\prime}$ and hold
$(\mu_0,\Xi_0,\nu_0)$ fixed.  Let
$\phi_{\mathrm{CT}}^*(m)=(\mu^*(m),\gamma^*(m),\nu^*(m))$ be the pseudo-true
CT parameter under the resulting CST law, with
$\Xi^*(m)=\Xi\{\gamma^*(m)\}$.  Under the local regularity conditions in
Supplementary Proposition \textcolor{blue}{S.1}, reflection
symmetry gives
\begin{equation}\label{eq:ct-local-orders-main}
\begin{aligned}
 \mu^*(m)-\mu_0&=A_\mu m+o(\|m\|^2),
 &\qquad
 \Xi^*(m)-\Xi_0&=O(\|m\|^2),\\
 \nu^*(m)-\nu_0&=O(\|m\|^2).
\end{aligned}
\end{equation}
Here
$A_\mu=\mathcal I_{\mu\mu}^{-1}\mathcal I_{\mu m}$, with the information
blocks evaluated at $(\mu,\gamma,\nu,m)=(\mu_0,\gamma_0,\nu_0,0)$,
where $\Xi(\gamma_0)=\Xi_0$.
The different orders follow from reflection.  Reversing all skewness
directions reverses the pseudo-true location displacement but leaves the
loading and tail parameters unchanged.  Location can therefore respond
linearly to omitted asymmetry, while the first derivatives of the loading
and tail distortions vanish at symmetry.  The resulting local robustness
is a property of the likelihood under reflection; the common normalization
only makes the reported loadings comparable.

Under the additional uniform sampling conditions in Supplementary Assumption
\textcolor{blue}{S.1}, the CT estimator differs from its
triangular pseudo-true value by $O_p(T^{-1/2})$.  Consequently, along a
vanishing-asymmetry sequence, root-$T$ centering at the CST truth is guaranteed
by $\|m_T\|=o(T^{-1/2})$ for location but by the weaker condition
$\|m_T\|=o(T^{-1/4})$ for loadings and tail indices.  There is consequently
a range of local departures for which loading inference remains centered
at the CST truth even though location bias need not be negligible relative
to sampling uncertainty.

For fixed nonzero asymmetry, the pseudo-true distortions need not vanish.
Restricted CT inference must then be centered at
$\phi_{\mathrm{CT}}^*(m)$ and use the usual sandwich covariance, rather than
the inverse information matrix \citep{White1982}.  Supplementary Section
\textcolor{blue}{S.4} gives the complete result and its proof.
The paired experiment in Section \ref{subsec:simulation-misspecified}
illustrates the different effects on location, loadings, and tail parameters
when CT and CST are fitted to the same asymmetric samples.

\section{Raw CST with arbitrary positive tail indices}
\label{sec:raw-cst}

The finite-variance restriction supports the interpretation of loadings
as covariance contributions.  The convolution construction itself requires
only independent subvector distributions and an invertible loading map.
A unit-dispersion representation therefore extends CST to tail indices
for which means or variances may not exist.  For each subvector let
\begin{equation}\label{eq:raw-cst-block}
 U_k\sim\ST_{n_k}(0,I_{n_k},\alpha_k,\nu_k),
 \qquad \alpha_k\in\mathbb R^{n_k},\qquad \nu_k>0.
\end{equation}
Write
\begin{equation}\label{eq:raw-bounded-shape}
 \delta_k^{\mathrm r}
 =\frac{\alpha_k}{(1+\alpha_k^{\prime}\alpha_k)^{1/2}},
 \qquad \|\delta_k^{\mathrm r}\|<1,
\end{equation}
and take the $U_k$ to be mutually independent.

\begin{definition}[Unit-dispersion raw CST]\label{def:raw-cst}
Writing $U=(U_1^{\prime},\ldots,U_K^{\prime})^{\prime}$, define
\begin{equation}\label{eq:raw-cst-definition}
 Y=\mu^{\mathrm r}+\Xi^{\mathrm r}U
 \sim\CST_{\bm n}^{\mathrm{raw}}
 \left(\mu^{\mathrm r},\Xi^{\mathrm r},
       \{\alpha_k,\nu_k\}_{k=1}^K\right),
 \qquad |\Xi^{\mathrm r}|\ne0.
\end{equation}
\end{definition}

Setting each subvector's dispersion to the identity entails no loss of
generality because any positive-definite dispersion matrix for that subvector
can be absorbed into the corresponding
loading columns while co-transforming the shape coefficient.  If
$u=(\Xi^{\mathrm r})^{-1}(y-\mu^{\mathrm r})$, the log likelihood is
\begin{equation}\label{eq:raw-cst-loglikelihood}
 \ell^{\mathrm r}(y;\theta^{\mathrm r})
 =-\frac12\log|\Xi^{\mathrm r}\Xi^{\mathrm r\prime}|
  +\sum_{k=1}^K\log f_{\ST}(u_k;I_{n_k},\alpha_k,\nu_k),
\end{equation}
and is explicit for every positive tail index.  The interpretation of
its parameters changes: $\mu^{\mathrm r}$ is a location parameter and
$\Xi^{\mathrm r}\Xi^{\mathrm r\prime}$ is an aggregate dispersion matrix.
Neither is generally the corresponding moment, even when that moment
exists; the transformation below supplies the mean and covariance when
they are finite.  The raw model is normalized by the same positive-definite
diagonal loading-submatrix orientation used for standardized CST.  Its complete
rotation-invariant identification and recovery result, which does not use
moments, is Supplementary Theorem \textcolor{blue}{S.1}.

For the exact correspondence, define the raw observed-space quantities
$S_k=\Xi^{\mathrm r}_{\bullet k}\Xi^{\mathrm r\prime}_{\bullet k}$
and $d_k=\Xi^{\mathrm r}_{\bullet k}\delta_k^{\mathrm r}$.  All square
roots below are symmetric positive definite.

\begin{theorem}[Exact raw--standardized correspondence]
\label{thm:raw-standardized-map}
Suppose $\nu_k>2$ for every $k$, put $c_\nu=(\nu-2)/\nu$, and start
from a normalized raw parameter.  For each subvector define
\begin{equation}\label{eq:raw-block-moments}
 r_k=a_{\nu_k}\delta_k^{\mathrm r},\qquad
 V_k=c_{\nu_k}^{-1}I_{n_k}-r_kr_k^{\prime}.
\end{equation}
Then $r_k=\E(U_k)$ and $V_k=\Var(U_k)\succ0$.  Put
\begin{equation}\label{eq:raw-map-polar}
 A_k=E_k^{\prime}\Xi^{\mathrm r}_{\bullet k}V_k^{1/2},\qquad
 Q_k=A_k^{\prime}(A_kA_k^{\prime})^{-1/2}.
\end{equation}
The normalized mean--covariance parameters generating the same distribution are
\begin{align}
 \mu
 &=\mu^{\mathrm r}+\sum_{k=1}^K
       \Xi^{\mathrm r}_{\bullet k}r_k,
 \label{eq:raw-map-location}\\
 \Xi_{\bullet k}
   &=\Xi^{\mathrm r}_{\bullet k}V_k^{1/2}Q_k,
 &m_k&=Q_k^{\prime}V_k^{-1/2}r_k.
 \label{eq:raw-map-loading-shape}
\end{align}
Their invariant contributions satisfy
\begin{align}
 b_k&=a_{\nu_k}d_k,
 &G_k&=c_{\nu_k}^{-1}S_k-a_{\nu_k}^2d_kd_k^{\prime},
 \label{eq:raw-standardized-invariant-map}\\
 d_k&=a_{\nu_k}^{-1}b_k,
 &S_k&=c_{\nu_k}(G_k+b_kb_k^{\prime}).
 \label{eq:standardized-raw-invariant-map}
\end{align}
With $\rho_k^{\mathrm r}=d_k^{\prime}S_k^+d_k$, the standardized and raw
label invariants coincide:
\begin{equation}\label{eq:raw-standardized-label-map}
 \varsigma_k=\rho_k^{\mathrm r}.
\end{equation}

Conversely, from normalized standardized parameters let
$\Omega_k=\Omega(m_k,\nu_k)$, $\eta_k=\eta(m_k,\nu_k)$, and
\begin{equation}\label{eq:standardized-map-raw-pre}
 L_k=\Xi_{\bullet k}\Omega_k^{1/2},\qquad
 \alpha_k^{\circ}=\Omega_k^{1/2}\eta_k,\qquad
 B_k=E_k^{\prime}L_k,\qquad
 H_k=B_k^{\prime}(B_kB_k^{\prime})^{-1/2}.
\end{equation}
The inverse normalized raw parameters are
\begin{equation}\label{eq:standardized-map-raw}
 \mu^{\mathrm r}=\mu-\sum_{k=1}^K\Xi_{\bullet k}m_k,\qquad
 \Xi^{\mathrm r}_{\bullet k}=L_kH_k,\qquad
 \alpha_k=H_k^{\prime}\alpha_k^{\circ}.
\end{equation}
The two maps are smooth inverses on every fixed normalized labeling chart.
\end{theorem}

The polar factors $Q_k$ and $H_k$ restore the diagonal loading-submatrix orientation
after rescaling; omitting them would generally produce the same distribution
but not its normalized representative.  The theorem shows that, when all
$\nu_k>2$, raw and standardized CST are the same family in different
coordinates.

The correspondence separates a change of coordinates from a change in
inferential scope.  For every finite $\nu_k>0$, the raw maximum likelihood
estimator is consistent on a compact, fixed normalized chart with uniformly
nondegenerate loadings and tail indices bounded away from zero.  A
logarithmic likelihood envelope suffices, so this result does not require
finite moments.  When all $\nu_k>2$, the smooth correspondence also
transfers the joint $\sqrt T$-normality result at an interior true parameter
with nonsingular information.

The obstacle below the finite-variance boundary is more specific.  A
cross-subvector loading derivative combines a score from one subvector
with a coordinate from another.  If $K\geq2$ and that second subvector
has $\nu_k\leq2$, the resulting score can have infinite second moment.
The joint finite-information Gaussian argument then fails in those
directions, even though the density and consistency remain available.
Single-subvector or suitably restricted models may avoid this obstruction.
Supplementary Theorem \textcolor{blue}{S.2} states the precise
conditions and scope.

\section{Simulation study}\label{sec:simulation}

\subsection{Design and reported estimands}\label{subsec:simulation-design}

The simulations examine two implications of the theory.  Under correct
specification, we assess how well the analytical information describes
sampling dispersion and how quickly the Gaussian approximation captures
the two tails of the estimator's distribution.  Under false symmetry, we
examine how a symmetric likelihood reallocates the omitted asymmetry across
location, loading, and tail parameters.  Using the same CST design for both
experiments connects the inference results in Section \ref{sec:asymptotics}
with the misspecification analysis in Section \ref{sec:symmetry-implications}.

The design combines a scalar subvector with a bivariate subvector,
$\bm n=(1,2)^{\prime}$, in an observed dimension of three.  It therefore
retains both shared scale variation within a subvector and mixing across
independent tail sources:
\[
 Y_t=\mu_0+\Xi_0 Z_t,
 \qquad Z_t=(Z_{1t}^{\prime},Z_{2t}^{\prime})^{\prime}.
\]
The true parameters are
\[
 \mu_0=(0.10,0.20,0.30)^{\prime},
 \qquad
 \Xi_0=
 \begin{pmatrix}
 0.60&0.30&0.10\\
 0.50&0.70&0.20\\
 0.40&0.20&0.80
 \end{pmatrix},
\]
$\nu_0=(4,5)^{\prime}$, $m_{1,0}=0.45$, and $m_{2,0}=(0.40,-0.30)^{\prime}$.  The diagonal loading
submatrices are symmetric positive definite, implementing the common
just-identifying normalization for CT and CST.  The implied bounded skewness
parameters are
\[
 \delta_{1,0}=0.5803,
 \qquad
 \delta_{2,0}=(0.5052,-0.3909)^{\prime},
\]
so both subvectors have moderate admissible asymmetry.  Raw skew-$t$ draws are
generated using the standardization in Proposition
\ref{prop:standardization}, and $m_{k,0}$ is then subtracted to obtain a
zero-mean identity-covariance subvector.

Estimation uses the normalized coordinates $\mu$, $\gamma$, $(m_1,m_2)$,
and $(\nu_1,\nu_2)$.  The reported loadings are reconstructed from
$\gamma$, while $\delta_k$ is obtained from $(m_k,\nu_k)$ through the
standardization map.  Reporting these transformations checks whether the
likelihood theory remains informative for the quantities used to interpret
the fitted model.  Tail thickness is reported as $1/\nu_k$, with zero
corresponding to the skew-normal boundary, or the Gaussian boundary under
symmetry.  We use 50,000 replications at
$T\in\{1{,}000,2{,}000,4{,}000,8{,}000\}$; the paired CT--CST comparison
uses the same number of replications at $T=8{,}000$.

For a scalar estimand $a_j(\theta)$, the analytical standard deviation is
\begin{equation}\label{eq:simulation-astd}
 \operatorname{aSD}_j
 =\left\{T^{-1}D_{j,0}\mathcal I(\theta_0)^{-1}D_{j,0}^{\prime}\right\}^{1/2},
 \qquad
 D_{j,0}=\frac{\partial a_j(\theta_0)}{\partial\theta^{\prime}}.
\end{equation}
Evaluating $\mathcal I(\theta_0)$ and $D_{j,0}$ at the truth isolates the
accuracy of the first-order approximation, including the delta method for
transformed parameters.  The comparison is subject to Monte Carlo and
numerical integration error, but does not include the uncertainty from
estimating the information matrix.  Except for the transformed skewness
coordinates, the left and right columns report the frequencies with which
$(\widehat a_j-a_{j,0})/\operatorname{aSD}_j$ falls below $-1.96$ or above
$1.96$.  For $\delta_k$, critical values are obtained by mapping draws from
the joint asymptotic distribution of $(\widehat m_k,\widehat\nu_k)$ through
the bounded nonlinear transformation.  This produces asymmetric mapped
intervals when the local transformation is noticeably curved.

\subsection{Correctly specified likelihood inference}
\label{subsec:simulation-correct}

Table \ref{tab:Finite-sample-properties-of} supports the information
formula across parameters with different roles in the model.  Location is
already well centered at the smallest sample size, while the modest biases
in loadings and shape parameters generally diminish as the sample grows.
The Monte Carlo standard deviations decline at approximately the
$T^{-1/2}$ rate and approach their analytical counterparts.  Agreement for
the reported loadings and bounded skewness coordinates also supports the
Jacobian transformations needed to translate primitive parameter uncertainty
into uncertainty about model interpretation.

\begin{table}
\caption{Finite-sample properties of maximum likelihood estimators.}\label{tab:Finite-sample-properties-of}

\input{tables/spd_n12_N50000_T1000_8000.tex}
\end{table}

The two tail frequencies reveal a more gradual convergence than the
standard deviations alone suggest.  For example, at $T=1{,}000$ the
$\Xi_{11}$ sampling distribution has a nearly correct dispersion but
unequal left and right rejection frequencies, 0.0175 and 0.0420, against a
nominal 0.025 in each tail.  Both move toward their nominal values as $T$
increases.  Several skewness coordinates show a similar imbalance even
when their analytical standard deviations are accurate.  Matching the
scale of the estimator's distribution therefore does not by itself ensure
accurate one-sided probability statements.

Some tail imbalance remains at $T=8{,}000$, notably for $\Xi_{11}$ and
$\delta_{2,1}$.  Agreement in dispersion supports the analytical
information as a measure of precision, while one-sided inference remains
more sensitive to the parameter and sample size.

\subsection{Fitting CT to skewed CST data}
\label{subsec:simulation-misspecified}

The second experiment fits CT and CST to the same asymmetric samples.
Bias and RMSE are measured against the CST truth, so CT's error includes
both sampling variation and the distortion caused by restricting the
population likelihood.  Pairing the fits makes the comparison less
sensitive to differences between simulated samples.

Table \ref{tab:Loading-matrix-estimation-under} shows that the cost of
false symmetry is concentrated in location and tail shape.  The RMSEs for
the first two location coordinates are about six times those under CST.
The CT estimates also increase $1/\nu_k$, using heavier symmetric tails to
accommodate observations generated by asymmetric sources.  A symmetric
likelihood can therefore partly fit skewness by moving its center and
thickening both tails, even though neither adjustment reproduces the
directional imbalance.  The resulting location distortion illustrates why
the center of a misspecified likelihood need not recover the data-generating
mean.

\begin{table}[!htbp]
\caption{CT and CST estimation under a skewed CST data-generating process}\label{tab:Loading-matrix-estimation-under}
\input{tables/ct_vs_cst_T8000_sameDGP_table.tex}
\end{table}

\Needspace{3\baselineskip}
Loading estimates are much less sensitive in this design: their CT/CST
RMSE ratios remain close to one, although the small biases are not all
zero.  This contrast is consistent with Supplementary Proposition
\textcolor{blue}{S.1}, which eliminates the linear response
of loadings to omitted asymmetry near the CT subfamily.  The simulation
uses fixed nonzero asymmetry, however, so it illustrates the pattern of
distortions rather than verifying the proposition's local rates.

The practical implication is that apparent stability of estimated
loadings can coexist with substantial misspecification of location and
tail shape.  Comparing covariance or exposure estimates alone may therefore
understate the cost of imposing symmetry.  The common normalization makes
the CT and CST loadings comparable, but the broader distributional
comparison remains essential.

\FloatBarrier
\section{Realized-volatility application}\label{sec:application}

The application examines how heterogeneous tails and asymmetry contribute
to the joint distribution of volatility innovations after predictable
persistence has been removed.

\subsection{Data and common first-stage residuals}
\label{subsec:application-data}

We use daily realized volatilities for the S\&P 500 index (SPX) and nine
U.S. stocks from January 2001 through December 2020.  The stocks form three
groups: Health Care (LLY, MRK, JNJ), Industrials (MMM, GE, HON), and
Information Technology (AAPL, IBM, INTC).  Realized variances $RV_{i,t}$
are constructed using realized kernels following
\citet{Barndorff-Nielsen2008,Barndorff-Nielsen2009a} and converted to
annualized volatility in percentage points,
$Y_{i,t}=100\sqrt{252\,RV_{i,t}}$.  The effective sample contains
$T=4{,}950$ observations after aligning the series and forming lags.

Following the HAR approach of \citet{Corsi2009} and the realized-volatility
literature \citep{Andersen2007,Bollerslev2016}, we estimate the same
conditional-mean specification for every competing innovation distribution:
\begin{equation}\label{eq:har-model}
 Y_t=\beta_0+\beta_dY_t^{(d)}+\beta_wY_t^{(w)}
     +\beta_mY_t^{(m)}+\varepsilon_t.
\end{equation}
Here $Y_t^{(d)}=Y_{t-1}$, while $Y_t^{(w)}$ and $Y_t^{(m)}$ average lags
2--5 and 6--22, respectively.  The coefficient matrices are diagonal and
estimated by asset-by-asset OLS.  Every second-stage distribution is then
fitted to the same centered residual vectors, so comparisons reflect the
innovation law rather than differences in the treatment of persistence.

\begin{table}[!htbp]
\caption{HAR coefficients and key statistics}\label{tab:HAR-coefficients-and}
\input{tables/empirical_RK_HAR_coefficients_table.tex}
\end{table}

Table \ref{tab:HAR-coefficients-and} shows small first-order residual
autocorrelations together with pronounced right skewness and excess
kurtosis, even in the winsorized descriptive statistics.  This motivates
heavy-tailed asymmetric innovation laws, although small first-order
autocorrelations do not establish serial independence.  The distributional
problem is therefore to capture large,
asymmetric departures from predictable volatility.  These innovations also
have a clear cross-sectional organization: SPX is positively correlated
with every stock, and within-sector correlations are generally stronger
than cross-sector correlations (Supplementary Table
\textcolor{blue}{S.1}).  This pattern motivates
examining shared market and sector tail sources.

\subsection{Competing innovation distributions}
\label{subsec:application-models}

We compare Gaussian, multivariate-$t$, global Azzalini--Capitanio skew-$t$,
CT, and CST distributions.  The first three establish what can be explained
by a common tail index and global asymmetry.  For CT and CST, the Cluster
specification uses $\bm n=(1,3,3,3)^{\prime}$, with one market component
and three sector subvectors.  The Hetero specification uses ten scalar
subvectors.  Cluster allows coordinates within a sector subvector to share
a scale-mixture shock, whereas Hetero assigns each latent coordinate its
own independent scale shock.  Their comparison therefore examines the value
of grouping tail sources, beyond allowing heterogeneous univariate shapes.

All baseline convolution specifications estimate a general loading matrix,
so any observed series can be exposed to several latent subvectors.
The reported matrix $\bar\Xi$ has unit row norms, with positive marginal
residual scales collected in the separately estimated diagonal matrix
$\Lambda$.  The diagonal loading submatrices of $\bar\Xi$ are normalized
to be symmetric positive definite.  Thus the orientation convention is
applied in marginally standardized observation coordinates; the effective
loading matrix in residual units is $\Lambda\bar\Xi$.  Market
and sector names provide reporting labels, as discussed in Remark
\ref{rem:labels}.  We compare joint log likelihood and BIC.  Table
\ref{tab:Second-stage-maximum-likelihood} also separates the sum of fitted
marginal log likelihoods, $\ell_m$, from the residual dependence
contribution, $\ell_c=\ell-\ell_m$.  Because the fitted margins differ,
this decomposition describes where fit gains arise.  It does not compare
copulas evaluated at a common set of marginal transforms.  The likelihood
and BIC comparisons measure in-sample fit on common estimated residuals;
the iid theory does not by itself justify two-stage time-series inference.

\subsection{Empirical results}\label{subsec:application-results}

Table \ref{tab:Second-stage-maximum-likelihood} shows that heavy tails,
asymmetry, and the grouping of tail sources each matter for fit.  Replacing
Gaussian innovations with a multivariate $t$ yields the largest improvement,
but a common tail index leaves substantial gains available.  Cluster-CST
has the highest joint likelihood and the lowest BIC among the baseline
specifications.  Its advantage over Cluster-CT isolates the contribution
of asymmetry within the same latent partition: allowing skewness lowers BIC
by 1,421 points after accounting for the additional parameters.

\begin{table}[!tp]
\caption{Baseline second-stage estimates}\label{tab:Second-stage-maximum-likelihood}
\input{tables/table_empirical_second_stage_likelihood_v1.tex}
\end{table}

\begin{table}[!tp]
\caption{Estimated standardized loading matrices for the convolution specifications}\label{tab:Estimated-loading-matrices}
\input{tables/table_empirical_all_Xi_estimates_v1.tex}
\end{table}

The tail estimates clarify why one common tail index can be restrictive.  Under
Cluster-CST, the Health Care subvector has the heaviest tails and
Information Technology the lightest, with fitted indices of about 2.33 and
3.04.  The same ranking appears under Cluster-CT.  Since the latent
subvectors have identity covariance, this difference concerns the decay of
extreme shocks rather than their variance scale.  It allows distinct
sources of volatility innovations to generate extremes at different rates,
even when their contributions overlap in observed variables.  Several
fitted indices lie below three, so their Pearson skewness coefficients do
not exist.  The reported $\delta$ coordinates remain well-defined measures
of asymmetry and should not be equated with the sample skewness statistics.

Asymmetry addresses a separate feature: unusually large upward volatility
innovations are poorly represented by symmetric shocks.  The positive
skewness of the HAR residuals persists after filtering, and the marginal
diagnostics in Supplementary Table \textcolor{blue}{S.4} show that CST
substantially reduces the distortion left by CT.  The PIT means move closer
to one half, the average marginal Kolmogorov--Smirnov statistic falls by
more than half, and the normal-score transforms become much less skewed.
This links the likelihood improvement to a better description of the
imbalance between upward and downward volatility innovations.  The PIT
variances remain below the uniform benchmark $1/12$, however, so the
relative improvement coexists with residual marginal miscalibration.

\Needspace{6\baselineskip}
The comparison between Cluster-CST and Hetero-CST is particularly
informative about dependence.  Their summed marginal log likelihoods are
almost identical, yet Cluster-CST achieves a higher joint likelihood with
fewer parameters.  The gain is therefore concentrated in the dependence
contribution, although similar marginal likelihoods do not imply identical
fitted margins.  This is consistent with the value of
shared scale shocks within multivariate tail subvectors: allowing every
scalar source its own tail and skewness parameters does not reproduce the
same joint behavior.  In this application, grouping latent coordinates
captures useful dependence as well as reducing the parameter count.

The loading matrices in Table \ref{tab:Estimated-loading-matrices} show how
these sources reach the observed series.  Broad positive exposure to the
market component coexists with more concentrated sector exposures.
This organization is similar under CT and CST, suggesting that asymmetry
refines the distribution of shocks transmitted through an economically
recognizable exposure pattern.  Similar loading patterns can nevertheless
produce different joint tail behavior when the latent sources differ in
tail thickness and asymmetry.  The sector interpretation concerns these
exposure patterns; individual latent skewness coordinates depend on the
chosen orientation, while $(b_k,G_k)$ describe the invariant contributions.

The findings also survive a substantial simplification of the loading
matrix.  Supplementary Tables \textcolor{blue}{S.2}
and \textcolor{blue}{S.3} impose repeated coefficients
within observed sector blocks, using the structure described in
Supplementary Section \textcolor{blue}{S.6} and motivated by
\citet{Archakov2024}.  The restricted Cluster-CST fit has a lower BIC than
its flexible counterpart and retains the tail ranking, positive asymmetry,
and broad market-sector exposure pattern.  Thus the main distributional
findings do not depend on estimating every exposure separately; a compact
loading structure can retain the features that distinguish CST from the
alternatives.

\FloatBarrier

\section{Conclusion}\label{sec:conclusion}

CST provides a way to model heterogeneous tail sources and directional
asymmetry while keeping means, covariances, and higher-order shape
separately interpretable.  Independent standardized skew-$t$ subvectors
generate shared tail shocks within latent groups, and a general loading
matrix allows their effects to overlap across observed variables.  This
combination explains why similar covariance patterns can coexist with
different marginal asymmetries and joint density contours.

The identification analysis locates the distributional information in the
observed-space contributions $(b_k,G_k)$.  Asymmetry adds an identified
direction but leaves the within-subvector loading orbit intact because
the latent shape coordinates co-rotate.  Normalizing that orbit gives a
unique reported parameter without imposing additional covariance
restrictions.  Analytical derivatives and the one-dimensional information
reduction then make likelihood estimation and regular inference feasible
on a fixed normalized chart.

The theoretical and numerical results also clarify what stable loading
estimates can conceal.  Near symmetry, omitted skewness can shift location
at first order while affecting loadings only at second order.  The paired
simulation illustrates this imbalance: exposure estimates change relatively
little even as location and tail estimates absorb substantial distortion.
In the realized-volatility application, asymmetry improves marginal
calibration, and grouped tail sources improve joint fit relative to scalar
sources with similar marginal fit.  Both findings emphasize the information
in higher-order shape beyond covariance alone.

The raw formulation extends the family, identification, and likelihood
consistency to every positive tail index under the stated conditions.
The finite-variance boundary remains consequential for joint Gaussian
inference because unrestricted loading scores can lose square
integrability.  Developing inference beyond that boundary, selecting
subvector structures from the data, accounting jointly for serial
dependence and first-stage estimation, and developing dynamic multivariate
risk models based on CST are directions for further work.

\section*{Use of AI tools}

During the preparation of this work, the authors used generative AI tools
to assist with manuscript review, copy-editing, and numerical checks of
mathematical derivations.  The authors reviewed and revised the AI-assisted
material and take full responsibility for the final manuscript.

\bibliographystyle{apalike}
\bibliography{CST_ref}

\appendix

\section{Proofs for identification}\label{app:identification}

\subsection{Proof of Proposition \ref{prop:loading-equivalence}}
\label{app:loading-equivalence-proof}

If $\widetilde L\widetilde L^{\prime}=LL^{\prime}=G$, then
$\operatorname{col}(\widetilde L)=\operatorname{col}(L)$, so the full-column-rank
factorizations satisfy $\widetilde L=LQ$ for a unique nonsingular $Q$.
Substitution into the equality of the two Gram matrices and multiplication
by a left inverse of $L$ and its transpose give $QQ^{\prime}=I_d$; hence
$Q\in O(d)$.  The equality $L m=LQ\widetilde m$ then gives
$\widetilde m=Q^{\prime}m$.  The converse follows by direct substitution.
Because the action is free when $L$ has full column rank, its orbit has
dimension $\dim O(d)=d(d-1)/2$. \hfill$\square$

\subsection{Proof of Theorem \ref{thm:identification}}

The proof is organized in five steps.  The first two establish
equivariance and identification of a single subvector.  The third shows that
a finite-$\nu$ skew-$t$ subvector cannot be decomposed linearly into smaller
independent subvectors.  The final two steps invoke independent-subspace
uniqueness and apply the normalization.

\subsection*{Step 1: rotational equivariance}

Fix an orthogonal $Q\in O(d)$ and put $\widetilde m=Q^{\prime}m$.  Directly from
\eqref{eq:standardization},
\[
 \Omega(\widetilde m,\nu)=Q^{\prime}\Omega(m,\nu)Q,
 \qquad
 \eta(\widetilde m,\nu)=Q^{\prime}\eta(m,\nu).
\]
The multivariate skew-$t$ density is equivariant under orthogonal
transformations, so
\[
 Q^{\prime}Z(m,\nu)\overset{d}{=}Z(Q^{\prime}m,\nu).
\]
It follows that
$(\Xi_{\bullet k}Q)Z(Q^{\prime}m_k,\nu_k)$ and
$\Xi_{\bullet k}Z(m_k,\nu_k)$ have the same distribution. Both $b_k$ and
$G_k$ are unchanged.

\subsection*{Step 2: identification of one standardized subvector}

We show that
$Z(m,\nu)\overset{d}{=}Z(\widetilde m,\widetilde\nu)$ implies
$(m,\nu)=(\widetilde m,\widetilde\nu)$.  Fix a unit vector
$u\in\mathbb{R}^d$, set $s=u^{\prime}m$, and consider $u^{\prime}Z(m,\nu)$.  This is a
univariate skew-$t$ variable with location $-s$, dispersion
\[
 \sigma_\nu(s)^2=c_\nu(1+s^2),
 \qquad c_\nu=(\nu-2)/\nu,
\]
and bounded skewness parameter
\[
 \delta_\nu(s)=\frac{s}{a_\nu\sigma_\nu(s)}.
\]
Let $\alpha_\nu(s)=\delta_\nu(s)/\{1-\delta_\nu(s)^2\}^{1/2}$.  Its
density $f_{s,\nu}$ has right- and left-tail constants
\begin{align*}
 \lim_{y\to\infty}y^{\nu+1}f_{s,\nu}(y)
 & =2\kappa_\nu\sigma_\nu(s)^\nu
 T_1\{\alpha_\nu(s)\sqrt{\nu+1};\nu+1\},\\
 \lim_{y\to\infty}y^{\nu+1}f_{s,\nu}(-y)
 & =2\kappa_\nu\sigma_\nu(s)^\nu
 T_1\{-\alpha_\nu(s)\sqrt{\nu+1};\nu+1\},
\end{align*}
where $\kappa_\nu$ is defined in Supplementary Equation
\textup{(\textcolor{blue}{S.3})}.
Both constants are strictly positive under admissibility.  The tail
exponent therefore identifies $\nu$.  For fixed $\nu$, the ratio of the
two constants is
\[
 \mathcal R_\nu(s)=
 \frac{T_1\{\alpha_\nu(s)\sqrt{\nu+1};\nu+1\}}
 {T_1\{-\alpha_\nu(s)\sqrt{\nu+1};\nu+1\}}.
\]
Because
\[
 \frac{\partial\delta_\nu(s)}{\partial s}
 =\{a_\nu\sqrt{c_\nu}(1+s^2)^{3/2}\}^{-1}>0,
\]
and both subsequent maps from $\delta$ to $\alpha$ and from $\alpha$ to
$\mathcal R_\nu$ are strictly increasing, $\mathcal R_\nu(s)$ identifies
$s=u^{\prime}m$.
Equality for every unit $u$ identifies $m$. Combining this argument with
Step 1 shows that
\[
 Q^{\prime}Z(m,\nu)\overset{d}{=}Z(\widetilde m,\widetilde\nu)
 \quad\Longrightarrow\quad
 \widetilde\nu=\nu,\quad \widetilde m=Q^{\prime}m.
\]

\subsection*{Step 3: linear irreducibility}

Centering does not affect independence, so consider the raw
representation $X=W^{-1/2}U$, where $U$ is nondegenerate skew normal
and $W\sim\operatorname{Gamma}(\nu/2,\nu/2)$ is independent of $U$.
Suppose a nonsingular transformation partitions $X$ into two nonempty
independent subvectors, $BX=(X_1^{\prime},X_2^{\prime})^{\prime}$, and write
$BU=(U_1^{\prime},U_2^{\prime})^{\prime}$.  The density of $BU$ is positive,
so for some $c>0$,
\[
 p_c=\Pr(\|U_1\|>c,\ \|U_2\|>c)>0.
\]
Since $\Pr(W\leq x)\asymp x^{\nu/2}$ as $x\downarrow0$, there is
$C_1>0$ such that, for all sufficiently large $r$,
\[
 \Pr(\|X_1\|>r,\ \|X_2\|>r)
 \geq p_c\Pr(W<c^2r^{-2})\geq C_1r^{-\nu}.
\]
For $j=1,2$, skew normality gives $\E\|U_j\|^\nu<\infty$ and hence
\[
 \Pr(\|X_j\|>r)
 =\E\Pr\{W<\|U_j\|^2r^{-2}\mid U_j\}=O(r^{-\nu}).
\]
Independence of $X_1$ and $X_2$ would make their joint exceedance
$O(r^{-2\nu})$, a contradiction. Each finite-$\nu$ subvector is therefore
linearly irreducible.

\subsection*{Step 4: uniqueness of the invariant decomposition}

Let $\theta$ and $\widetilde\theta$ generate the same distribution.
The standardized subvectors have zero mean, so $\mu=\widetilde\mu$ and
\[
 \Xi Z\overset{d}{=}\widetilde\Xi\widetilde Z.
\]
Both sides are decompositions into independent subspaces with existing
covariance.  Every subspace is non-Gaussian (because $\nu_k<\infty$) and
linearly irreducible by Step 3.  These are precisely the conditions of the
independent-subspace uniqueness theorem of \citet{Gutch2012}, which implies a
dimension-preserving permutation $\pi$ and
nonsingular matrices $B_k$ such that
\[
 \widetilde Z_{\pi(k)}\overset{d}{=}B_kZ_k,
 \qquad
 \widetilde\Xi_{\bullet\pi(k)}=\Xi_{\bullet k}B_k^{-1}.
\]
Since both subvector representations have identity covariance,
$B_kB_k^{\prime}=I_{n_k}$.  Thus $B_k$ is orthogonal.  With
$Q_k=B_k^{\prime}$, Step 2
gives
\[
 \widetilde\nu_{\pi(k)}=\nu_k,
 \quad
 \widetilde m_{\pi(k)}=Q_k^{\prime}m_k,
 \quad
 \widetilde\Xi_{\bullet\pi(k)}=\Xi_{\bullet k}Q_k.
\]
This proves identification of $(\nu_k,b_k,G_k)$ up to permutation.
Proposition \ref{prop:loading-equivalence} then shows that $(b_k,G_k)$
determines the pair $(\Xi_{\bullet k},m_k)$ only up to the same joint
orthogonal action displayed in \eqref{eq:rotation}.

\subsection*{Step 5: normalization and recovery}

Part (I1) of Assumption \ref{ass:identification-normalization} fixes the
remaining permutation.  Proposition
\ref{prop:loading-equivalence} shows that it remains only to choose one
representative of each orthogonal orbit.  For one subvector, let
$A_k=E_k^{\prime}\Xi_{\bullet k}$, so that under part (I2) of Assumption
\ref{ass:identification-normalization}
$D_k=A_kA_k^{\prime}\succ0$.  The left polar decomposition is
$A_k=D_k^{1/2}O_k$ for a unique orthogonal matrix $O_k$.  Applying the
admissible within-subvector rotation $Q_k=O_k^{\prime}$ therefore produces a
representative whose diagonal loading submatrix is $D_k^{1/2}\succ0$.
For any normalized representative,
\[
 D_k=\Xi_{kk}\Xi_{kk}^{\prime}=\Xi_{kk}^2.
\]
Uniqueness of the symmetric positive-definite square root gives
$\Xi_{kk}=D_k^{1/2}$. Furthermore,
\[
 G_kE_k=\Xi_{\bullet k}\Xi_{kk},
 \qquad
 E_k^{\prime}b_k=\Xi_{kk}m_k.
\]
Solving these equations gives \eqref{eq:recovery}. Hence the normalized
pair is unique, and Proposition \ref{prop:standardization} uniquely
determines $(\Omega_k,\eta_k,\delta_k)$ from $(m_k,\nu_k)$.  This
completes the proof. \hfill$\square$

\section{Proofs for maximum likelihood theory}\label{app:mle}

\subsection{Proof of Theorem \ref{thm:consistency}}\label{app:consistency}

Let $L(\theta)=\E_{\theta_0}\ell(Y;\theta)$.  The parameter-space bounds
imply, uniformly over $\Theta$, bounded norms of $\Xi^{-1}$, $\mu$,
$m_k$, $\Omega_k$, $\Omega_k^{-1}$, and $\eta_k$.  Thus
\[
 \|z(\theta,y)\|+\|x_k(\theta,y)\|
 \leq C(1+\|y\|),
 \qquad
 Q_k(\theta,y)\leq C(1+\|y\|^2).
\]
The skewing argument is uniformly bounded.  Indeed, Cauchy-Schwarz
gives
\[
 q_k(\theta,y)^2
 \leq(\nu_k+n_k)\eta_k^{\prime}\Omega_k\eta_k,
\]
and
\[
 \eta_k^{\prime}\Omega_k\eta_k
 =\frac{m_k^{\prime}\Omega_k^{-1}m_k}
 {a_{\nu_k}^2-m_k^{\prime}\Omega_k^{-1}m_k}.
\]
The uniform admissibility gap therefore bounds the right-hand side.
Because $\Theta$ is compact,
$T_1\{q_k(\theta,y);\nu_k+n_k\}$ is uniformly bounded
away from zero.  The determinant and normalizing terms are also bounded,
so there are finite constants $C_0,C_1$ such that
\begin{equation}\label{eq:log-envelope}
 \sup_{\theta\in\Theta}|\ell(y;\theta)|
 \leq C_0+C_1\log(1+\|y\|^2).
\end{equation}
The true CST distribution has finite second moments, making this
envelope integrable.  Continuity and a uniform law of large numbers yield
$\sup_{\theta\in\Theta}|\ell_T(\theta)-L(\theta)|\xrightarrow{p}0$.

Finally,
\[
 L(\theta)-L(\theta_0)
 =-D_{\mathrm{KL}}(f_{\theta_0}\|f_\theta)\leq0.
\]
Equality requires equality of the two densities almost everywhere,
which implies $\theta=\theta_0$ by Theorem
\ref{thm:identification} and the fixed orientation and labeling chart.  The
argmax theorem establishes consistency.
\hfill$\square$

\subsection{Proofs of Proposition \ref{prop:local-regularity} and
Theorem \ref{thm:normality}}\label{app:normality}

Choose an open convex neighborhood $\mathcal N$ of $\theta_0$ whose
closure lies inside the regular parameter space.  The standardization
maps and $\Xi^{-1}$ have bounded first and second derivatives on
$\overline{\mathcal N}$.  Write $x_k=x_k(\theta,Y)$.  Moreover,
\[
 v_k(x_k)^{1/2}\|x_k\|\leq C,
 \qquad
 v_k(x_k)\|x_k\|^2\leq C,
 \qquad
 |q_k(x_k)|\leq C.
\]
The inverse Mills ratio $\rho_k(x_k)$ and the first two derivatives of
$\log T_1\{q_k(x_k);\nu_k+n_k\}$ are consequently bounded.  Combining
\eqref{eq:block-argument-score} with the raw-score formulas and
standardization differentials in Supplementary Appendix
\textcolor{blue}{S.2.1} gives
\begin{align}
 \sup_{\theta\in\mathcal N}\|s(Y;\theta)\|^2
 &\leq C\{1+\|Y\|^2+\log^2(1+\|Y\|^2)\},\label{eq:score-envelope}\\
 \sup_{\theta\in\mathcal N}\|\mathcal H(Y;\theta)\|
 &\leq C\{1+\|Y\|^2+\log^2(1+\|Y\|^2)\},\label{eq:hessian-envelope}
\end{align}
where $\mathcal H$ is the observed Hessian.  Both right-hand sides are
integrable under $P_{\theta_0}$.  We next verify the stronger uniform property
needed for differentiation under the density integral.  Let
\[
 \underline\nu=\inf_{\theta\in\mathcal N,\,k}\nu_k>2
\]
and choose $r\in(2,\underline\nu)$.  A raw skew-$t$ density is bounded by
twice its symmetric Student-$t$ kernel.  The compact parameter bounds,
followed by the bounded affine standardization and loading maps, therefore
give
\[
 \sup_{\theta\in\mathcal N}\E_\theta(1+\|Y\|)^r<\infty,
\]
which is \eqref{eq:uniform-local-moment}.

Pointwise in $y$, the first two density derivatives satisfy, entrywise,
\begin{equation}\label{eq:density-derivative-identities}
 \dot f_\theta(y)=f_\theta(y)s(y;\theta),
 \qquad
 \ddot f_\theta(y)=f_\theta(y)
 \{\mathcal H(y;\theta)+s(y;\theta)s(y;\theta)^{\prime}\}.
\end{equation}
The bounds \eqref{eq:score-envelope}--\eqref{eq:hessian-envelope}, now
integrated under $P_\theta$, and the uniform $r$th moment make the $L^1$
tails of both derivatives uniformly negligible over $\mathcal N$.  On each
compact set of $y$ values, the density and the derivatives in
\eqref{eq:density-derivative-identities} are jointly continuous and uniformly
bounded.  Truncation followed by dominated convergence thus makes the two
derivative maps $L^1$-continuous.  Applying the fundamental theorem of
calculus along line segments in the convex set $\mathcal N$ proves that
$\theta\mapsto f_\theta$ is twice continuously differentiable into $L^1$.
Differentiating $\int f_\theta(y)\,dy=1$ once and twice now gives
\[
 \E_{\theta_0}s(Y;\theta_0)=0,
 \qquad
 -\E_{\theta_0}\mathcal H(Y;\theta_0)
 =\E_{\theta_0}\{s(Y;\theta_0)s(Y;\theta_0)^{\prime}\}
 =\mathcal I(\theta_0).
\]

Consistency places $\widehat\theta_T$ in $\mathcal N$ with probability
tending to one. Expanding the first-order condition around $\theta_0$
gives
\[
 0=\frac1{\sqrt T}\sum_{t=1}^Ts_t(\theta_0)
   +\overline{\mathcal H}_T
    \sqrt T(\widehat\theta_T-\theta_0),
\]
where $\overline{\mathcal H}_T$ is the Hessian averaged along the line
segment joining $\theta_0$ and $\widehat\theta_T$. By
\eqref{eq:score-envelope}, the multivariate central limit theorem gives
\[
 T^{-1/2}\sum_{t=1}^Ts_t(\theta_0)
 \Rightarrow N\{0,\mathcal I(\theta_0)\}.
\]
The uniform law implied by \eqref{eq:hessian-envelope} gives
$\overline{\mathcal H}_T\xrightarrow{p}-\mathcal I(\theta_0)$.  Positive
definiteness of the information and Slutsky's theorem yield
\eqref{eq:mle-limit}.

Finally, \eqref{eq:score-envelope} and \eqref{eq:hessian-envelope}
also imply uniform laws for $s_ts_t^{\prime}$ and $\mathcal H_t$ on
$\mathcal N$. Evaluating them at the consistent estimator proves
consistency of the outer-product and observed-Hessian information
estimators. \hfill$\square$

\end{document}


\title{\textbf{Supplementary Material for\\
Convolution Skew-$t$ Distributions}}
\author{\parbox{0.95\textwidth}{\centering
  Xinxian Chen$^{\ddagger}$ \quad Peter Reinhard Hansen$^{\mathsection}$\thanks{Corresponding author: Peter Reinhard Hansen. Email: \texttt{hansen@unc.edu}.} \quad Chen Tong$^{\ddagger}$\\[0.70cm]
  \small $^{\ddagger}$Department of Finance, School of Economics,\\[0.15cm]
  Xiamen University, China\\[0.3cm]
  $^{\mathsection}$Department of Economics,\\[0.15cm]
  University of North Carolina at Chapel Hill, United States}}
\date{October 25, 2026}
\maketitle

\noindent
This supplement develops the calculations and proofs underlying the
main paper.  The marginal results connect projected skewness to directional
tail behavior; the derivative and information calculations support likelihood
implementation.  The false-symmetry and raw-CST results state the conditions
under which the main conclusions extend to misspecification and infinite
variance.  Additional empirical evidence examines loading restrictions and
marginal calibration.  Both simulation experiments are in the main paper.
Definitions and equation references without an S prefix refer to
the main paper.

\section{Standardization and marginal calculations}
\label{app:supp-distribution}

\subsection{Additional marginal-tail results}
\label{app:marginal-results}

The marginal formulas in Section \textcolor{blue}{2.3} describe how
independent latent contributions combine under projection.  The results
here explain which contributions dominate the far tails and how their
asymmetries determine the balance between the two sides.  Proofs of
Theorem \textcolor{blue}{1} and Proposition
\textcolor{blue}{2} follow below.  We retain the notation
$\beta_{j,k}=E_k^{\prime}\Xi^{\prime}e_j$.

The projected subvector contribution reveals the central distinction between
the two convolution distributions.  Under CT, spherical symmetry implies that
$\varphi_{Z_k}(s\beta_{j,k})$ depends on $\beta_{j,k}$ only through
$\|\beta_{j,k}\|$.  In CST, it generally depends on
\begin{equation}\label{eq:projected-shape}
 \beta_{j,k}^{\prime}m_k,
 \qquad
 \beta_{j,k}^{\prime}\Omega(m_k,\nu_k)\beta_{j,k},
 \qquad
 \beta_{j,k}^{\prime}\omega_k\delta_k.
\end{equation}
The observed marginal shape therefore depends not only on the magnitude of
the loading exposure but also on its orientation relative to the subvector
skewness direction.  When $m_k=0$, dependence on that orientation
disappears and the projected contribution reduces to its symmetric CT
counterpart.

This directional interpretation can be made exact from the tail signature
of a standardized subvector.  Let $u$ be a unit vector, put $s=u^{\prime}m$, and write
$U=u^{\prime}Z(m,\nu)$.  Define
\[
 \sigma_\nu(s)^2=c_\nu(1+s^2),\qquad
 \delta_\nu(s)=\frac{s}{a_\nu\sigma_\nu(s)},\qquad
 \alpha_\nu(s)=\frac{\delta_\nu(s)}
 {\{1-\delta_\nu(s)^2\}^{1/2}}.
\]
The two density tails have the common polynomial exponent $\nu+1$, while
their limiting ratio is
\begin{equation}\label{eq:directional-tail-ratio}
 \lim_{x\to\infty}\frac{f_U(x)}{f_U(-x)}
 =\mathcal R_\nu(s)
 :=\frac{T_1\{\alpha_\nu(s)\sqrt{\nu+1};\nu+1\}}
 {T_1\{-\alpha_\nu(s)\sqrt{\nu+1};\nu+1\}}.
\end{equation}
The function $\mathcal R_\nu(s)$ is strictly increasing and equals one only
at $s=0$.  Thus $\nu$ controls the rate of tail decay, whereas $u^{\prime}m$
controls the right-left imbalance in direction $u$.  This separates two
aspects of extreme behavior: how quickly tail probabilities decay and how
that probability is divided between opposing directions.  CST allows both
aspects to vary across latent subvectors.

For an observed direction $a\ne0$, put
\[
 V_k(a)=a^{\prime}\Xi_{\bullet k}Z_k,
 \qquad
 \mathcal E(a)=\{k:a^{\prime}\Xi_{\bullet k}\ne0\},
 \qquad
 \nu_*(a)=\min_{k\in\mathcal E(a)}\nu_k.
\]
Nonsingularity of $\Xi$ makes $\mathcal E(a)$ nonempty.
For $k\in\mathcal E(a)$, let
\[
 d_{k,+}(a)=\lim_{x\to\infty}x^{\nu_k+1}f_{V_k(a)}(x),
 \qquad
 d_{k,-}(a)=\lim_{x\to\infty}x^{\nu_k+1}f_{V_k(a)}(-x).
\]
They are explicit.  With
$\beta_k=\Xi_{\bullet k}^{\prime}a$, $q_k=\|\beta_k\|$,
$u_k=\beta_k/q_k$, $s_k=u_k^{\prime}m_k$, and
\begin{equation}\label{eq:tail-kappa}
 \kappa_\nu
 =\frac{\nu^{\nu/2}\Gamma\{(\nu+1)/2\}}
       {\sqrt\pi\,\Gamma(\nu/2)},
\end{equation}
the directional-tail calculation leading to
\eqref{eq:directional-tail-ratio} gives
\begin{equation}\label{eq:block-density-tail-constants}
 d_{k,\pm}(a)
 =2\kappa_{\nu_k}
  \{q_k\sigma_{\nu_k}(s_k)\}^{\nu_k}
  T_1\{\pm\alpha_{\nu_k}(s_k)\sqrt{\nu_k+1};\nu_k+1\}>0.
\end{equation}

\begin{lemma}[Observed-direction tail aggregation]
\label{lem:tail-aggregation}
Let $\mathcal M(a)=\{k\in\mathcal E(a):\nu_k=\nu_*(a)\}$.  Then
\begin{align}
 \Pr\{a^{\prime}(Y-\mu)>x\}
 &\sim \frac{x^{-\nu_*(a)}}{\nu_*(a)}
       \sum_{k\in\mathcal M(a)}d_{k,+}(a),
 \label{eq:projection-right-tail}\\
 \Pr\{a^{\prime}(Y-\mu)<-x\}
 &\sim \frac{x^{-\nu_*(a)}}{\nu_*(a)}
       \sum_{k\in\mathcal M(a)}d_{k,-}(a).
 \label{eq:projection-left-tail}
\end{align}
\end{lemma}

Thus a projection is governed by the smallest tail index among its exposed
subvectors; when several subvectors attain that index, their one-sided constants
add.  The proof in Subsection
\ref{app:tail-aggregation} supplies the one-big-jump bounds for these
two-sided variables.  Consequently, a univariate tail fit may reveal only
the heaviest exposed subvector and conceal lighter-tailed sources.  The
multivariate decomposition is what permits the full collection of
subvector-specific tail indices and asymmetric directions to be recovered.

Equation \textup{(\textcolor{blue}{16})} requires only one-dimensional Fourier
inversion.  For the trivariate illustration in Figure
\textcolor{blue}{1}, the explicit joint density offers a direct
alternative: a bivariate density is obtained by integrating over the omitted
coordinate.  For example,
\begin{equation}\label{eq:direct-bivariate-marginal}
 f_{Y_1,Y_2}(y_1,y_2)=\int_{-\infty}^{\infty}
 f_Y(y_1,y_2,y_3)\,dy_3,
\end{equation}
and analogously for other pairs.

\subsection{Proof of Proposition \textcolor{blue}{1}}

Write $\Omega=\Omega(m,\nu)$ and $\eta=\eta(m,\nu)$. The conventional
shape parameter is $\alpha=\omega\eta$, and
$\bar\Omega=\omega^{-1}\Omega\omega^{-1}$. Substitution of
\textup{(\textcolor{blue}{5})} into the shape transformation gives
\[
 \frac{\bar\Omega\alpha}
 {(1+\alpha^{\prime}\bar\Omega\alpha)^{1/2}}
 =a_\nu^{-1}\omega^{-1}m=\delta.
\]
The first identity in \textup{(\textcolor{blue}{3})} therefore yields
$\E(X)=a_\nu\omega\delta=m$. In addition,
\[
 \frac{\nu}{\nu-2}\Omega(m,\nu)=I_d+mm^{\prime}.
\]
The second identity in \textup{(\textcolor{blue}{3})} gives
$\Var(X)=I_d+mm^{\prime}-mm^{\prime}=I_d$.  Centering by $m$
proves the result.
\hfill$\square$

\subsection{Strict monotonicity of raw marginal skewness}
\label{app:raw-skewness-monotonicity}

Fix $\nu>3$ and suppress the positive factor
$a_\nu/(\nu-3)$ in \textup{(\textcolor{blue}{4b})}.  Put
\[
 D_\nu=\frac{\nu}{\nu-2},\qquad
 A_\nu=3D_\nu,\qquad
 B_\nu=2a_\nu^2(\nu-3)-\nu.
\]
Apart from that positive factor, raw Pearson skewness is
\[
 h_\nu(\delta)
 =\frac{\delta(A_\nu+B_\nu\delta^2)}
 {(D_\nu-a_\nu^2\delta^2)^{3/2}},
 \qquad -1<\delta<1.
\]
Direct differentiation gives
\begin{equation}\label{eq:raw-skewness-derivative-proof}
 h_\nu'(\delta)
 =\frac{A_\nu D_\nu+
       (3B_\nu D_\nu+2A_\nu a_\nu^2)\delta^2}
 {(D_\nu-a_\nu^2\delta^2)^{5/2}}.
\end{equation}
The denominator is positive, and the numerator is affine in $\delta^2$.
Its value at $\delta^2=0$ is $A_\nu D_\nu>0$.  At $\delta^2=1$ it equals
\begin{equation}\label{eq:raw-skewness-endpoint-proof}
 3D_\nu\{2a_\nu^2(\nu-2)-D_\nu(\nu-3)\}.
\end{equation}
Strict log-convexity of the gamma function implies
\[
 \Gamma(\nu/2)^2
 <\Gamma\{(\nu-1)/2\}\Gamma\{(\nu+1)/2\}
 =\frac{\nu-1}{2}\Gamma\{(\nu-1)/2\}^2,
\]
and hence
\[
 a_\nu^2>\frac{2\nu}{\pi(\nu-1)}>\frac2\pi.
\]
An elementary differentiation shows that
\[
 \sup_{\nu>3}\frac{\nu(\nu-3)}{2(\nu-2)^2}
 =\frac9{16}<\frac2\pi.
\]
Therefore $2a_\nu^2(\nu-2)^2>\nu(\nu-3)$, which makes
\eqref{eq:raw-skewness-endpoint-proof} strictly positive.  An affine
function positive at both endpoints of $[0,1]$ is positive throughout that
interval.  Equation \eqref{eq:raw-skewness-derivative-proof} thus proves
that raw Pearson skewness is strictly increasing in $\delta$ on $(-1,1)$.
\hfill$\square$

\subsection{Derivation of the marginal characteristic function}
\label{app:marginal-cf-proof}

For $Y_S=S\mu+\sum_k\Xi_{S,k}Z_k$, independence immediately implies
\[
 \E\exp(iu^{\prime}Y_S)
 =\exp(iu^{\prime}\mu_S)
  \prod_k\E\exp\{i(\Xi_{S,k}^{\prime}u)^{\prime}Z_k\},
\]
which is \textup{(\textcolor{blue}{13})}.  To obtain \textup{(\textcolor{blue}{14})}, use the
scale-mixture representation
\[
 X_k\overset{d}{=}W_k^{-1/2}U_k,
 \qquad
 W_k\sim\operatorname{Gamma}(\nu_k/2,\nu_k/2),
\]
where $U_k$ is the associated skew-normal vector, independent of $W_k$.
Conditional on $W_k$,
\[
 \E\{\exp(it^{\prime}X_k)\mid W_k\}
 =2\exp\left\{-\frac{t^{\prime}\Omega_kt}{2W_k}\right\}
 \Phi\left(\frac{it^{\prime}\omega_k\delta_k}{\sqrt{W_k}}\right).
\]
Taking expectations over $W_k$ and using $Z_k=X_k-m_k$ gives
\textup{(\textcolor{blue}{14})}. Standard Fourier inversion applies whenever the
resulting characteristic function is integrable. In particular, for
$S=e_j^{\prime}$,
\[
 f_{Y_j}(y)=\frac{1}{\pi}\int_0^\infty
 \operatorname{Re}\{\exp(-isy)\varphi_{Y_j}(s)\}\,ds.
\]

\subsection{Proof of Proposition \textcolor{blue}{2}}
\label{app:marginal-skewness-proof}

For subvector $k$, put
$U_{jk}=\beta_{j,k}^{\prime}Z_k$, $s_{jk}=\beta_{j,k}^{\prime}m_k$, and
$q_{jk}=\|\beta_{j,k}\|^2$.  The corresponding raw projection
$\beta_{j,k}^{\prime}X_k$ is a univariate skew-$t$ variable with dispersion
and bounded skewness parameter
\[
 \sigma_{jk}^2
 =c_{\nu_k}(q_{jk}+s_{jk}^2),
 \qquad
 \delta_{jk}=\frac{s_{jk}}{a_{\nu_k}\sigma_{jk}}.
\]
Substituting these expressions into the factorized univariate third central
moment in \textup{(\textcolor{blue}{4a})} gives
\begin{align*}
 \E(U_{jk}^3)
 &=a_{\nu_k}\delta_{jk}\sigma_{jk}^3
 \left[
  \frac{\nu_k}{\nu_k-3}(3-\delta_{jk}^2)
  -\frac{3\nu_k}{\nu_k-2}
  +2a_{\nu_k}^2\delta_{jk}^2
 \right]\\
 &=\frac{s_{jk}}{\nu_k-3}
 \left\{
  3q_{jk}
  +\left(2\nu_k-3-\frac{\nu_k}{a_{\nu_k}^2}\right)s_{jk}^2
 \right\}.
\end{align*}
The variables $U_{j1},\ldots,U_{jK}$ are independent and centered.
Consequently, all mixed terms in the third central moment of
$Y_j-\mu_j=\sum_kU_{jk}$ vanish, so their third moments add.  Their variances
also add to $\sum_kq_{jk}=(\Xi\Xi^{\prime})_{jj}$.  Division by this variance
to the power $3/2$ proves \textup{(\textcolor{blue}{19})}--
\textup{(\textcolor{blue}{20})}.  If an exposed subvector has $\nu_k\leq3$,
then $\nu_*(e_j)\leq3$.  Lemma \ref{lem:tail-aggregation} gives positive
two-sided tail constants of order $x^{-\nu_*(e_j)}$ for $Y_j-\mu_j$; hence
the tail-integral identity
$\E|Y_j-\mu_j|^3=3\int_0^\infty x^2\Pr(|Y_j-\mu_j|>x)\,dx$ diverges.
Pearson skewness is therefore undefined.
\hfill$\square$

\subsection{Proof of Lemma \ref{lem:tail-aggregation}}
\label{app:tail-aggregation}

Fix $a\ne0$ and abbreviate $V_k=V_k(a)$.  Integrating the density-tail
limits in \eqref{eq:block-density-tail-constants} gives, for every exposed
subvector,
\begin{equation}\label{eq:block-survival-tail-proof}
 \Pr(V_k>x)\sim\frac{d_{k,+}(a)}{\nu_k}x^{-\nu_k},
 \qquad
 \Pr(V_k<-x)\sim\frac{d_{k,-}(a)}{\nu_k}x^{-\nu_k}.
\end{equation}
These relations are unaffected by the finite centering used to pass from a
raw subvector to $Z_k$.  The variables $\{V_k:k\in\mathcal E(a)\}$ are
independent.

Let $J=|\mathcal E(a)|$ and choose $0<\varepsilon<1/J$.  For two distinct
exposed subvectors $k$ and $l$, independence and
\eqref{eq:block-survival-tail-proof} give
\begin{equation}\label{eq:two-large-jumps-proof}
 \Pr\{|V_k|>\varepsilon x,\ |V_l|>\varepsilon x\}
 =O(x^{-\nu_k-\nu_l})=o(x^{-\nu_*(a)}).
\end{equation}
For the lower bound, take the union, over $k\in\mathcal M(a)$, of the
events
\[
 V_k>\{1+(J-1)\varepsilon\}x,
 \qquad |V_l|\leq\varepsilon x\quad(l\ne k).
\]
Each such event implies $\sum_lV_l>x$, and their pairwise intersections are
$o(x^{-\nu_*(a)})$ by \eqref{eq:two-large-jumps-proof}.  Regular variation
therefore yields
\begin{equation}\label{eq:tail-liminf-proof}
 \liminf_{x\to\infty}x^{\nu_*(a)}
 \Pr\!\left(\sum_lV_l>x\right)
 \geq\{1+(J-1)\varepsilon\}^{-\nu_*(a)}
 \sum_{k\in\mathcal M(a)}\frac{d_{k,+}(a)}{\nu_*(a)}.
\end{equation}

For the upper bound, if $\sum_lV_l>x$ and at most one $|V_k|$ exceeds
$\varepsilon x$, that component must exceed
$\{1-(J-1)\varepsilon\}x$.  The event that two or more components exceed
$\varepsilon x$ in absolute value is negligible by
\eqref{eq:two-large-jumps-proof}; subvectors with $\nu_k>\nu_*(a)$ also have
$o(x^{-\nu_*(a)})$ tails.  Hence
\begin{equation}\label{eq:tail-limsup-proof}
 \limsup_{x\to\infty}x^{\nu_*(a)}
 \Pr\!\left(\sum_lV_l>x\right)
 \leq\{1-(J-1)\varepsilon\}^{-\nu_*(a)}
 \sum_{k\in\mathcal M(a)}\frac{d_{k,+}(a)}{\nu_*(a)}.
\end{equation}
Letting $\varepsilon\downarrow0$ in
\eqref{eq:tail-liminf-proof}--\eqref{eq:tail-limsup-proof} proves
\eqref{eq:projection-right-tail}.  Applying the same argument to $-V_k$
proves \eqref{eq:projection-left-tail}.
\hfill$\square$

\section{Raw derivatives, score, and Hessian details}
\subsection{Raw subvector derivatives}\label{app:raw-score}

For a raw subvector, define $Q=x^{\prime}\Omega^{-1}x$,
$v=(\nu+d)/(\nu+Q)$, $q=v^{1/2}\eta^{\prime}x$,
$\rho=t_1(q;\nu+d)/T_1(q;\nu+d)$, and
$\mathcal R=1+q\rho/(\nu+d)$. In addition to the argument score in
\textup{(\textcolor{blue}{31})}, the raw shape and symmetric matrix
scores are
\begin{align}
 s_\eta^{\mathrm{raw}}
 &=v^{1/2}\rho x,\label{eq:raw-eta-score}\\
 S_\Omega^{\mathrm{raw}}
 &=\frac12\{v\mathcal R\Omega^{-1}xx^{\prime}\Omega^{-1}
             -\Omega^{-1}\},\qquad
 s_\Omega^{\mathrm{raw}}
 =D_d^{\prime}\vecop(S_\Omega^{\mathrm{raw}}),
 \label{eq:raw-omega-score}
\end{align}
where $D_d$ is the duplication matrix. Holding $(\Omega,\eta)$ fixed,
the raw degrees-of-freedom score is
\begin{equation}\label{eq:raw-nu-score}
 \begin{split}
 s_\nu^{\mathrm{raw}}
 ={}&\frac12\psi\left(\frac{\nu+d}{2}\right)
 -\frac12\psi\left(\frac\nu2\right)-\frac{d}{2\nu}
 -\frac12\log\left(1+\frac Q\nu\right)
 +\frac{vQ}{2\nu}\\
 &+\frac{\rho q}{2}
 \left(\frac{1}{\nu+d}-\frac{1}{\nu+Q}\right)
 +\left.\frac{\partial}{\partial a}
 \log T_1(q;a)\right|_{a=\nu+d}.
 \end{split}
\end{equation}
Equations \textup{(\textcolor{blue}{31})} and
\eqref{eq:raw-eta-score}--\eqref{eq:raw-nu-score} form the raw score
$\mathcal r=(g^{\prime},s_\Omega^{\mathrm{raw}\prime},
s_\eta^{\mathrm{raw}\prime},s_\nu^{\mathrm{raw}})^{\prime}$.

The chain-rule matrix $R_k$ is also explicit. Suppress $k$, put
$c=(\nu-2)/\nu$, and write
\[
 \Omega=c(I+mm^{\prime}),\quad w=\Omega^{-1}m,\quad
 M=m^{\prime}w,\quad \Delta=a_\nu^2-M,\quad \eta=w\Delta^{-1/2}.
\]
For arbitrary perturbations $(dm,d\nu)$,
\begin{align*}
 d\Omega
 &=c(dm\,m^{\prime}+m\,dm^{\prime})+\frac{2}{\nu^2}(I+mm^{\prime})d\nu,\\
 dw&=\Omega^{-1}dm-\Omega^{-1}(d\Omega)w,\\
 dM&=2w^{\prime}dm-w^{\prime}(d\Omega)w,\\
 d\Delta&=2a_\nu\dot a_\nu d\nu-dM,\\
 d\eta&=\Delta^{-1/2}dw-\frac12w\Delta^{-3/2}d\Delta,
\end{align*}
where
\[
 \dot a_\nu=a_\nu\left\{
 \frac{1}{2\nu}
 +\frac12\psi\left(\frac{\nu-1}{2}\right)
 -\frac12\psi\left(\frac\nu2\right)
 \right\}.
\]
Together with $dx=dm$ and the derivative of the final raw coordinate $\nu$,
these differentials give the rows of
\[
 R_k=
 \begin{pmatrix}
  J_{u,m,k}^{\prime}\\ j_{u,\nu,k}^{\prime}
 \end{pmatrix},
 \quad
 J_{u,m,k}=
 \frac{\partial(X_{kt}^{\prime},\vech(\Omega_k)^{\prime},\eta_k^{\prime},\nu_k)^{\prime}}
 {\partial m_k^{\prime}},
 \quad
 j_{u,\nu,k}=
 \frac{\partial(X_{kt}^{\prime},\vech(\Omega_k)^{\prime},\eta_k^{\prime},\nu_k)^{\prime}}
 {\partial\nu_k}.
\]

\subsection{Proof of the score identities}\label{app:score-proof}

For this proof, reintroduce the observation index and write
$Z_t=A(Y_t-\mu)$, $Z_{kt}=E_k^{\prime}Z_t$,
$X_{kt}=Z_{kt}+m_k$, and
$H_t=\{g_1(X_{1t})^{\prime},\ldots,g_K(X_{Kt})^{\prime}\}^{\prime}$.
Since $Z_t=A(Y_t-\mu)$, a location perturbation gives
$dZ_t=-A\,d\mu$. The determinant in \textup{(\textcolor{blue}{12})} does not
depend on $\mu$, so
\[
 d\ell_t=H_t^{\prime}dZ_t=-H_t^{\prime}A\,d\mu,
\]
which yields $s_\mu(Y_t;\theta)=-A^{\prime}H_t$.  For a loading perturbation,
$dA=-A(d\Xi)A$, $dZ_t=-A(d\Xi)Z_t$, and
\[
 d\{-\tfrac12\log|\Xi\Xi^{\prime}|\}=-\operatorname{tr}(A\,d\Xi).
\]
Therefore
\[
 d\ell_t=-\operatorname{tr}(A\,d\Xi)-H_t^{\prime}A(d\Xi)Z_t,
\]
and
\[
 s_\Xi(Y_t;\theta)
 =\vecop(-A^{\prime}-A^{\prime}H_tZ_t^{\prime}),
 \qquad
 s_\gamma(Y_t;\theta)=G_\Xi^{\prime}s_\Xi(Y_t;\theta).
\]
Finally, $(m_k,\nu_k)$ enters subvector $k$ through
$(X_{kt},\Omega_k,\eta_k,\nu_k)$. The chain rule gives
\[
 s_{(m_k^{\prime},\nu_k)^{\prime}}(Y_t;\theta)
 =R_k\mathcal r_k(X_{kt}).
\]
These are exactly the score identities in
\textup{(\textcolor{blue}{32})}--\textup{(\textcolor{blue}{34})}, proving Proposition
\textcolor{blue}{4}. \hfill$\square$

The formulas above provide the ingredients for either direct Hessian
assembly or automatic differentiation in the normalized coordinates.  In
the latter implementation, the maps
\(m_k\mapsto\{\Omega_k,\eta_k,X_{kt}\}\) must remain inside the
computational graph so that the curvature term in equation
\textup{(\textcolor{blue}{35})} is retained.

\section{Fisher-information assembly and reduction}
We first assemble the information blocks from the score identities and then
reduce their primitive expectations to one-dimensional integrals.

\subsection{Assembly of the information blocks}
\label{app:information-assembly}

Let
\[
 a_\Xi=\vecop(A^{\prime}),\qquad
 q_{\Xi,t}=\vecop(A^{\prime}H_tZ_t^{\prime}),\qquad
 Q_\Xi=\E(q_{\Xi,t}q_{\Xi,t}^{\prime}),
\]
and set $s_\Xi(Y_t;\theta)=-a_\Xi-q_{\Xi,t}$, so
$s_\gamma(Y_t;\theta)=G_\Xi^{\prime}s_\Xi(Y_t;\theta)$.  The regular score identity for one
standardized subvector gives
\[
 \E\{g_k(X_{kt})\}=0,
 \qquad
 \E\{g_k(X_{kt})Z_{kt}^{\prime}\}=-I_{n_k}.
\]
Independence and zero means then imply
$\E(q_{\Xi,t})=-a_\Xi$.  Consequently,
\[
 \E\{s_\Xi(Y_t;\theta)s_\Xi(Y_t;\theta)^{\prime}\}
 =\E\{(a_\Xi+q_{\Xi,t})
       (a_\Xi+q_{\Xi,t})^{\prime}\}
 =Q_\Xi-a_\Xi a_\Xi^{\prime}.
\]
Premultiplication and postmultiplication by $G_\Xi^{\prime}$ and $G_\Xi$ yield
\textup{(\textcolor{blue}{41})}'s loading block.

For location,
$s_\mu(Y_t;\theta)=-A^{\prime}\sum_kE_kg_k(X_{kt})$.
Cross-subvector products vanish because the
subvectors and their scores are independent and centered.  Hence
\[
 \E\{s_\mu(Y_t;\theta)s_\mu(Y_t;\theta)^{\prime}\}
 =A^{\prime}\left(\sum_kE_kH_kE_k^{\prime}\right)A.
\]
Only the within-subvector component
\[
 q_{\Xi,kk,t}
 =\vecop\{A^{\prime}E_kg_k(X_{kt})Z_{kt}^{\prime}E_k^{\prime}\}
\]
contributes to the
cross moment between $g_k(X_{kt})$ and the loading score; the other subvector terms
factor through a zero score mean.  This gives
\[
 \E\{s_\mu(Y_t;\theta)s_\gamma(Y_t;\theta)^{\prime}\}
 =\left(\sum_kA^{\prime}E_kL_k\right)G_\Xi.
\]
Likewise,
$s_{\vartheta_k}(Y_t;\theta)=R_k\mathcal r_k(X_{kt})$ implies
\[
 \E\{s_\mu(Y_t;\theta)s_{\vartheta_k}(Y_t;\theta)^{\prime}\}
 =-A^{\prime}E_kC_kR_k^{\prime},
\qquad
 \E\{s_\gamma(Y_t;\theta)s_{\vartheta_k}(Y_t;\theta)^{\prime}\}
 =-G_\Xi^{\prime}M_kR_k^{\prime},
\]
and
\[
 \E\{s_{\vartheta_k}(Y_t;\theta)
       s_{\vartheta_k}(Y_t;\theta)^{\prime}\}
 =R_k\mathcal V_kR_k^{\prime}.
\]
For $k\ne l$, independence and the zero score means give
\[
 \E\{s_{\vartheta_k}(Y_t;\theta)
       s_{\vartheta_l}(Y_t;\theta)^{\prime}\}=0.
\]
These calculations establish
the information-block formulas in Proposition \textcolor{blue}{5}.

\subsection{Reduction of information moments}\label{app:information}

Each entry of the Fisher information is an expectation of products of
the raw subvector scores, standardized-subvector scores, and terms of the form
$g_k(X_k)Z_k^{\prime}$. Cross-subvector expectations factor by independence. It is
therefore enough to treat an expectation within one raw subvector.

Let $X^\circ$ denote the raw skew-$t$ vector after whitening and an
orthogonal rotation chosen so that the skewness direction is $\tau e_1$.
For any integrable scalar $F$, the perturbation-of-symmetry representation
of \textup{(\textcolor{blue}{2})} gives
\begin{equation}\label{eq:skew-expectation}
 \E_{\ST}\{F(X^\circ)\}
 =2\E_t\!\left[
   T_1\{q(V);\nu+d\}F(V)
 \right],
 \qquad V\sim t_d(0,I_d,\nu),
\end{equation}
where
\[
 q(V)=cB,\qquad
 c=\sqrt{\nu+d}\,\tau,\qquad
 B=\frac{V_1}{(\nu+V^{\prime}V)^{1/2}}.
\]

The conditional reduction can be stated exactly.  Define
\[
 U=\frac{V}{(\nu+V^{\prime}V)^{1/2}}.
\]
The transformation
$V=\sqrt\nu\,U/(1-U^{\prime}U)^{1/2}$ gives the density
\begin{equation}\label{eq:unit-ball-density}
 f_U(u)=
 \frac{\Gamma\{(\nu+d)/2\}}
      {\pi^{d/2}\Gamma(\nu/2)}
 (1-u^{\prime}u)^{\nu/2-1},
 \qquad u^{\prime}u<1.
\end{equation}
In particular, $B=U_1$ has density \textup{(\textcolor{blue}{43})}.  For
$d\ge2$, conditional on $B=b$,
\begin{equation}\label{eq:conditional-ball-decomposition}
 U_{-1}=\sqrt{1-b^2}\,R_\perp,
 \qquad
 f_{R_\perp}(r)=
 \frac{\Gamma\{(\nu+d-1)/2\}}
      {\pi^{(d-1)/2}\Gamma(\nu/2)}
 (1-r^{\prime}r)^{\nu/2-1},
 \quad r^{\prime}r<1,
\end{equation}
and the distribution of $R_\perp$ does not depend on $b$.  Thus $B$ and
$R_\perp$ are independent under this representation.  Moreover,
$\|R_\perp\|^2\sim
\operatorname{Beta}\{(d-1)/2,\nu/2\}$, and its direction is independent
and uniform on the unit sphere in $\mathbb R^{d-1}$.  The case $d=1$
requires no orthogonal component.

These identities make the reduction constructive.  With
$v=(\nu+d)/(\nu+V^{\prime}V)$,
\[
 v^{1/2}V=\sqrt{\nu+d}\,U,
 \qquad
 vVV^{\prime}=(\nu+d)UU^{\prime},
 \qquad
 \log\left(1+\frac{V^{\prime}V}{\nu}\right)
 =-\log(1-U^{\prime}U).
\]
In products generated by $Q_\Xi$, any coordinate from a subvector that carries
no score factor is integrated first.  Independence, $\E(Z_k)=0$, and
$\E(Z_kZ_k^{\prime})=I_{n_k}$ then either annihilate the term or replace the
unscaled coordinates by a known identity tensor.  Every remaining
within-subvector factor contains the damping displayed above.  Consequently,
every primitive information moment is a finite sum of a scalar function of
$cB$, a monomial in $U$, a nonnegative half-integer power of
$1-U^{\prime}U$, and at most a squared logarithm of
$1-U^{\prime}U$.

The conditional coefficients are explicit.  Put $p=d-1$ and
$S=R_\perp^{\prime}R_\perp$.  For $p\geq1$, $r,h\in\{0,1,2,\ldots\}$,
$\lambda>-\nu/2$, and coordinate indices
$i_1,\ldots,i_{2r}\in\{1,\ldots,p\}$,
\begin{align}
 &\E\!\left[
  R_{\perp,i_1}\cdots R_{\perp,i_{2r}}
  (1-S)^\lambda\{-\log(1-S)\}^h
 \right]
 \nonumber\\
 &\quad={}
 (-\partial_\lambda)^h
 \left[
  \frac{\mathrm B(p/2+r,\nu/2+\lambda)}
       {\mathrm B(p/2,\nu/2)}
  \frac{
   \displaystyle\sum_{\pi\in\mathcal P_{2r}}
    \prod_{\{a,b\}\in\pi}\delta_{i_ai_b}}
       {p(p+2)\cdots(p+2r-2)}
 \right],
 \label{eq:conditional-beta-tensor}
\end{align}
where $\mathcal P_{2r}$ is the set of pairings of
$\{1,\ldots,2r\}$; for $r=0$, the pairing sum and denominator are both
one.  Every corresponding odd moment is zero.  Formula
\eqref{eq:conditional-beta-tensor} follows by writing
$R_\perp=S^{1/2}D$, with $D$ uniform on the unit sphere and independent of
$S\sim\operatorname{Beta}(p/2,\nu/2)$.

Under \eqref{eq:conditional-ball-decomposition},
\[
 1-U^{\prime}U=(1-B^2)(1-R_\perp^{\prime}R_\perp).
\]
Thus a monomial with $j$ orthogonal coordinates contributes the factor
$(1-B^2)^{j/2}$, and expanding
\[
 -\log(1-U^{\prime}U)
 =-\log(1-B^2)-\log(1-S)
\]
reduces its conditional expectation to the beta coefficients in
\eqref{eq:conditional-beta-tensor}.  The only remaining expectation is over
$B$ and is precisely of the form \textup{(\textcolor{blue}{44})}.

For completeness, the degrees-of-freedom score introduces the scalar
\[
 d_\nu^{T_1}(q)=
 \left.\frac{\partial}{\partial a}\log T_1(q;a)
 \right|_{a=\nu+d}.
\]
Because $q=cB$ ranges over the compact interval $[-|c|,|c|]$ and
$T_1(q;\nu+d)$ is strictly positive, $d_\nu^{T_1}(q)$ and the other scalar score
functions are continuous and bounded there.  The logarithmic beta
coefficients are finite for $\nu>2$.  Hence all interchanges of conditional
expectation, finite summation, and differentiation used above are justified.
Cross-subvector terms factor by independence, and all remaining vector, matrix,
and four-index arrays are assembled algebraically.  This proves the
information-reduction part of Proposition \textcolor{blue}{5}.
\hfill$\square$

\section{Misspecification under false symmetry}
\label{sec:misspecification}

Imposing CT symmetry sets $m_k=0$ for every subvector and changes the
population likelihood whenever the data-generating CST law is asymmetric.
The common normalization in Proposition \textcolor{blue}{3}
makes the reported CT and CST loadings comparable, but their population
maximizers can still differ.  The result below characterizes how those
differences first emerge near symmetry and states the uniform sampling
conditions needed to relate them to estimation error.

To study those distortions jointly, stack the subvector mean shifts as
$m=(m_1^{\prime},\ldots,m_K^{\prime})^{\prime}$ and hold
$(\mu_0,\Xi_0,\nu_0)$ fixed.  Let $Y_m=\mu_0+\Xi_0Z(m)$ denote the resulting
CST observation, where $\nu_0=(\nu_{10},\ldots,\nu_{K0})^{\prime}$, and
write $P_m$ for its law.  A
normalized CT parameter $\phi$ collects $(\mu,\gamma,\nu)$, with
$\Xi=\Xi(\gamma)$; let $\gamma_0$ be the coordinate satisfying
$\Xi(\gamma_0)=\Xi_0$.  Define
\begin{equation}\label{eq:pseudotrue}
 Q_m(\phi)=\E_m\{\ell_{\mathrm{CT}}(Y_m;\phi)\},
 \qquad
 \phi_{\mathrm{CT}}^*(m)
 =\arg\max_{\phi\in\Theta_{\mathrm{CT}}}Q_m(\phi)
 =(\mu^*(m),\gamma^*(m),\nu^*(m)).
\end{equation}
Write $\Xi^*(m)=\Xi\{\gamma^*(m)\}$.  At $m=0$, CT is correctly
specified and
$\phi_{\mathrm{CT}}^*(0)=(\mu_0,\gamma_0,\nu_0)$.
For an estimator
$\widehat\phi_{\mathrm{CT}}=(\widehat\mu_{\mathrm{CT}},
\widehat\gamma_{\mathrm{CT}},\widehat\nu_{\mathrm{CT}})$, write
$\widehat\Xi_{\mathrm{CT}}=\Xi(\widehat\gamma_{\mathrm{CT}})$.

For the information calculations, order the full CST parameter as
$\theta=(\phi^{\prime},m^{\prime})^{\prime}$ and let
$\theta_0=(\phi_{\mathrm{CT}}^*(0)^{\prime},0^{\prime})^{\prime}$.
Partition the per-observation information at symmetry as
\begin{equation}\label{eq:ct-information-partition}
 \mathcal I_0=
 \begin{pmatrix}
  \mathcal I_{\phi\phi}&\mathcal I_{\phi m}\\
  \mathcal I_{m\phi}&\mathcal I_{mm}
 \end{pmatrix}.
\end{equation}
All information blocks below are evaluated at $\theta_0$.
The following assumption makes the uniform sampling requirements explicit.

\begin{assumption}[Uniform local CT M-estimation]
\label{ass:uniform-ct-mestimation}
There are neighborhoods $\mathcal M$ of zero and $\Phi_0$ of
$\phi_{\mathrm{CT}}^*(0)$, with compact closures contained in fixed regular
charts and $\overline\Phi_0\subset\operatorname{int}(\Theta_{\mathrm{CT}})$,
such that the following conditions hold.  For an iid sample from $P_m$, write
\[
 \widehat Q_{T,m}(\phi)=T^{-1}\sum_{t=1}^T
   \ell_{\mathrm{CT}}(Y_t;\phi),\qquad
 \widehat{\mathcal H}_{T,m}(\phi)=T^{-1}\sum_{t=1}^T
   \mathcal H_\phi^{\mathrm{CT}}(Y_t;\phi).
\]
\begin{enumerate}
 \item[(M1)] For every $m\in\mathcal M$, the criterion $Q_m$ has the unique
 interior maximizer $\phi_{\mathrm{CT}}^*(m)\in\Phi_0$, and, for every
 $\varepsilon>0$,
 \begin{equation}\label{eq:uniform-ct-separation}
  \inf_{m\in\mathcal M}\ 
  \inf_{\substack{\phi\in\Theta_{\mathrm{CT}}:\\
       \|\phi-\phi_{\mathrm{CT}}^*(m)\|\geq\varepsilon}}
  \{Q_m(\phi_{\mathrm{CT}}^*(m))-Q_m(\phi)\}>0.
 \end{equation}
 \item[(M2)] For every $\eta>0$,
 \begin{equation}\label{eq:uniform-ct-criterion-ulln}
  \sup_{m\in\mathcal M}P_m^{\otimes T}\!\left\{
   \sup_{\phi\in\Theta_{\mathrm{CT}}}
   |\widehat Q_{T,m}(\phi)-Q_m(\phi)|>\eta\right\}\longrightarrow0.
 \end{equation}
 Differentiation through $Q_m(\phi)=\E_m\ell_{\mathrm{CT}}(Y;\phi)$ is
 valid twice on $\Phi_0$.  The CT score satisfies
 \begin{equation}\label{eq:uniform-ct-score-moment}
  \sup_{m\in\mathcal M}\sup_{\phi\in\Phi_0}
  \E_m\|s_\phi^{\mathrm{CT}}(Y;\phi)\|^2<\infty.
 \end{equation}
 For some $\rho>0$, the population Hessian is jointly continuous on
 $\{(m,\phi):m\in\mathcal M,
 \|\phi-\phi_{\mathrm{CT}}^*(m)\|\leq\rho\}\subset
 \mathcal M\times\Phi_0$, and, for every $\eta>0$,
 \begin{equation}\label{eq:uniform-ct-hessian-ulln}
  \sup_{m\in\mathcal M}P_m^{\otimes T}\!\left[
   \sup_{\|\phi-\phi_{\mathrm{CT}}^*(m)\|\leq\rho}
   \left\|\widehat{\mathcal H}_{T,m}(\phi)
   -\E_m\mathcal H_\phi^{\mathrm{CT}}(Y;\phi)\right\|>\eta
  \right]\longrightarrow0.
 \end{equation}
 \item[(M3)] The matrix
 \begin{equation}\label{eq:uniform-ct-hessian}
  H_{\mathrm{CT}}(m)
  =-\E_m\mathcal H_{\phi}^{\mathrm{CT}}
       \{Y;\phi_{\mathrm{CT}}^*(m)\}
 \end{equation}
 is continuous on $\mathcal M$ and uniformly positive definite.
\end{enumerate}
\end{assumption}

\begin{proposition}[Local effect of false symmetry on the full CT parameter vector]
\label{prop:ct-local-robustness}
Suppose $\theta_0$ lies in the regular interior of Assumption
\textcolor{blue}{2} and that, in a neighborhood of
$\{\phi_{\mathrm{CT}}^*(0),0\}$, $Q_m(\phi)$ is three times continuously
differentiable, its maximizer in \eqref{eq:pseudotrue} is unique and
interior, and $\partial^2Q_0\{\phi_{\mathrm{CT}}^*(0)\}/
\partial\phi\partial\phi^{\prime}$ is nonsingular.  Then the pseudo-true map
is twice continuously differentiable and satisfies
\begin{align}
 \mu^*(-m)-\mu_0
 &=-\{\mu^*(m)-\mu_0\},
 &
 \gamma^*(-m)&=\gamma^*(m),
 \nonumber\\
 \Xi^*(-m)&=\Xi^*(m),
 &
 \nu^*(-m)&=\nu^*(m).
 \label{eq:ct-pseudotrue-parity}
\end{align}
Using the information blocks in \eqref{eq:ct-information-partition}, the
derivative of the full pseudo-true map is
\begin{equation}\label{eq:ct-pseudotrue-derivative}
 \left.\frac{\partial\phi_{\mathrm{CT}}^*(m)}{\partial m^{\prime}}
 \right|_{m=0}
 =\mathcal I_{\phi\phi}^{-1}\mathcal I_{\phi m}
 =\begin{pmatrix}
   \mathcal I_{\mu\mu}^{-1}\mathcal I_{\mu m}\\[2pt]
   0\\[2pt]
   0
  \end{pmatrix}.
\end{equation}
Let $A_\mu=\mathcal I_{\mu\mu}^{-1}\mathcal I_{\mu m}$.  As $m\to0$,
\begin{align}
 \mu^*(m)-\mu_0
 &=A_\mu m+o(\|m\|^2),
 &
 \gamma^*(m)-\gamma_0
 &=O(\|m\|^2),
 \nonumber\\
 \Xi^*(m)-\Xi_0
 &=O(\|m\|^2),
 &
 \nu^*(m)-\nu_0
 &=O(\|m\|^2).
 \label{eq:ct-local-orders}
\end{align}
If, in addition, Assumption \ref{ass:uniform-ct-mestimation} holds, row $T$
consists of iid observations from $P_{m_T}$, and
$\widehat\phi_{\mathrm{CT}}$ is a measurable maximizer of
$\widehat Q_{T,m_T}$ over $\Theta_{\mathrm{CT}}$, then every triangular
sequence $m_T\to0$ satisfies
\begin{align}
 \widehat\mu_{\mathrm{CT}}-\mu_0
 &=A_\mu m_T+O_p(T^{-1/2})+o(\|m_T\|^2),
 \nonumber\\
 \widehat\Xi_{\mathrm{CT}}-\Xi_0
 &=O_p(T^{-1/2})+O(\|m_T\|^2),
 &
 \widehat\nu_{\mathrm{CT}}-\nu_0
 &=O_p(T^{-1/2})+O(\|m_T\|^2).
 \label{eq:ct-local-estimation-rates}
\end{align}
Consequently, all three CT estimators are consistent for their CST
counterparts along any vanishing-asymmetry sequence.  Root-$T$ centering at
the CST truth is guaranteed by $\|m_T\|=o(T^{-1/2})$ for location, but by
the weaker condition $\|m_T\|=o(T^{-1/4})$ for loadings and degrees of
freedom.
\end{proposition}

The different orders follow from reflection, not from loading
identification.  The standardization gives
$Z_k(-m_k,\nu_k)\overset{d}{=}-Z_k(m_k,\nu_k)$, whereas the CT density is
centrally symmetric.  Reversing all skewness directions therefore reverses
the pseudo-true location displacement but leaves the pseudo-true loading and
tail parameters unchanged.  Location can therefore respond linearly to
omitted asymmetry, while the linear terms for $\Xi^*(m)$ and $\nu^*(m)$
must vanish.  The common CT--CST normalization makes these loading matrices
directly comparable; co-rotation itself is not the source of the local
robustness.

For fixed $m\in\mathcal M$, Assumption
\ref{ass:uniform-ct-mestimation} and standard misspecified-likelihood theory
\citep{White1982} give
\begin{equation}\label{eq:ct-limit}
 \widehat\phi_{\mathrm{CT}}\xrightarrow{p}
 \phi_{\mathrm{CT}}^*(m),
 \qquad
 \sqrt T\{\widehat\phi_{\mathrm{CT}}-\phi_{\mathrm{CT}}^*(m)\}
 \Rightarrow
 N\{0,H_{\mathrm{CT}}(m)^{-1}J_{\mathrm{CT}}(m)
 H_{\mathrm{CT}}(m)^{-1}\},
\end{equation}
where $H_{\mathrm{CT}}(m)$ is the negative expected Hessian and
$J_{\mathrm{CT}}(m)$ is the score outer-product matrix, both evaluated at
$\phi_{\mathrm{CT}}^*(m)$.  The sandwich covariance measures sampling
uncertainty around this pseudo-true parameter.  For fixed nonzero asymmetry,
any component of
$(\mu^*(m),\Xi^*(m),\nu^*(m))-(\mu_0,\Xi_0,\nu_0)$ may persist as
$T\to\infty$.  The first-order approximation to mean squared error then
combines persistent squared approximation bias with a sampling-variance
term of order $T^{-1}$.
Increasing sample size can sharpen estimation of the restricted target
without bringing that target closer to the CST truth.  This explains the
local scope of Proposition \ref{prop:ct-local-robustness}: it controls the
response to vanishing asymmetry, not exact recovery under a fixed false
restriction.  Its proof is given in Section \ref{app:false-symmetry}.
\subsection{Proof of Proposition \ref{prop:ct-local-robustness}}
\label{app:false-symmetry}

The standardization maps satisfy
\[
 \Omega(-m_k,\nu_k)=\Omega(m_k,\nu_k),
 \qquad
 \eta(-m_k,\nu_k)=-\eta(m_k,\nu_k).
\]
Reflection equivariance of the skew-$t$ density therefore gives
\[
 Z_k(-m_k,\nu_k)\overset{d}{=}-Z_k(m_k,\nu_k).
\]
Writing $Y_m=\mu_0+\Xi_0Z(m)$, simultaneous reversal of all subvector skewness
directions consequently yields
$Y_{-m}\overset{d}{=}2\mu_0-Y_m$.

Every CT density is centrally symmetric.  Hence, for its population
criterion $Q_m(\mu,\gamma,\nu)=\E_m\ell_{\mathrm{CT}}(Y;\mu,\gamma,\nu)$,
\[
 Q_{-m}(\mu,\gamma,\nu)
 =Q_m(2\mu_0-\mu,\gamma,\nu).
\]
At $m=0$, CT is correctly specified, so
$\phi_{\mathrm{CT}}^*(0)=(\mu_0,\gamma_0,\nu_0)$.  The implicit-function
theorem, the assumed third-order smoothness of $Q_m(\phi)$, and the
nonsingular Hessian make
$\phi_{\mathrm{CT}}^*(m)$ twice continuously differentiable near zero.
Shrink this neighborhood so that both the maximizer and its reflection
in the location coordinate remain in the interior of the same parameter
chart.  The reflected maximizer is a stationary point of $Q_{-m}$;
uniqueness of the local stationary branch then gives
\eqref{eq:ct-pseudotrue-parity}.  This local argument does not require the
entire compact optimization set to be reflection invariant.
To identify its first derivative, define
$\Psi(\phi,m)=\partial Q_m(\phi)/\partial\phi$ and write
$\phi_0=\phi_{\mathrm{CT}}^*(0)$.  At the CT null,
\[
 \frac{\partial\Psi(\phi_0,0)}{\partial\phi^{\prime}}
 =-\mathcal I_{\phi\phi},
 \qquad
 \frac{\partial\Psi(\phi_0,0)}{\partial m^{\prime}}
 =\E_0\{s_\phi(Y;\theta_0)s_m(Y;\theta_0)^{\prime}\}
 =\mathcal I_{\phi m}.
\]
The second equality is the usual derivative-of-expectation identity, applied
to the CST law generating $Y_m$.  Differentiating
$\Psi\{\phi_{\mathrm{CT}}^*(m),m\}=0$ therefore gives
\[
 \left.\frac{\partial\phi_{\mathrm{CT}}^*(m)}{\partial m^{\prime}}
 \right|_{m=0}
 =\mathcal I_{\phi\phi}^{-1}\mathcal I_{\phi m}.
\]
At symmetry the location and subvector-asymmetry scores are odd under reflection,
while the loading and tail scores are even.  Hence
$\mathcal I_{m\psi}=\mathcal I_{\mu\psi}=0$ for
$\psi=(\gamma^{\prime},\nu^{\prime})^{\prime}$, which yields
\eqref{eq:ct-pseudotrue-derivative} and
$A_\mu=\mathcal I_{\mu\mu}^{-1}\mathcal I_{\mu m}$.
The first derivatives at zero of the even maps $\gamma^*(m)$ and $\nu^*(m)$
are therefore zero.  The smooth loading-chart map
$\gamma\mapsto\Xi(\gamma)$ transfers the same parity and second-order rate
to $\Xi^*(m)$.  Similarly, the second derivative at zero of the odd map
$\mu^*(m)-\mu_0$ is zero.  The second-order Taylor expansions give
\eqref{eq:ct-local-orders}.

We finally verify the triangular-array assertion rather than invoking it as
an unstated M-estimation condition.  By (M1) and
\eqref{eq:uniform-ct-criterion-ulln}, the usual separation argument is
uniform in $m$: for every $\varepsilon>0$,
\[
 \sup_{m\in\mathcal M}P_m^{\otimes T}
 \{\|\widehat\phi_{\mathrm{CT}}
       -\phi_{\mathrm{CT}}^*(m)\|>\varepsilon\}\longrightarrow0.
\]
Thus, along $m_T\to0$, the estimator is interior with probability tending
to one and its score vanishes.  Taylor's formula with integral remainder
around $\phi_T^*=\phi_{\mathrm{CT}}^*(m_T)$ gives
\begin{equation}\label{eq:triangular-ct-score-expansion}
 0=T^{-1/2}\sum_{t=1}^T
   s_\phi^{\mathrm{CT}}(Y_{Tt};\phi_T^*)
   +\overline{\mathcal H}_{\phi,T}
    \sqrt T(\widehat\phi_{\mathrm{CT}}-\phi_T^*),
\end{equation}
where
\[
 \overline{\mathcal H}_{\phi,T}
 =\int_0^1\widehat{\mathcal H}_{T,m_T}
   \{\phi_T^*+r(\widehat\phi_{\mathrm{CT}}-\phi_T^*)\}\,dr.
\]
The first term in \eqref{eq:triangular-ct-score-expansion} is $O_p(1)$:
its summands have mean zero because $\phi_T^*$ is an interior population
maximizer, and their second moments are uniformly bounded by
\eqref{eq:uniform-ct-score-moment}.  Uniform consistency, continuity of the
population Hessian, and \eqref{eq:uniform-ct-hessian-ulln} give
\[
 \overline{\mathcal H}_{\phi,T}
 =-H_{\mathrm{CT}}(m_T)+o_p(1).
\]
Condition (M3) makes the inverse of this matrix $O_p(1)$ uniformly along the
sequence.  Solving \eqref{eq:triangular-ct-score-expansion} therefore yields
\[
 \widehat\phi_{\mathrm{CT}}-
 \phi_{\mathrm{CT}}^*(m_T)=O_p(T^{-1/2}).
\]
The smooth map $\Xi(\gamma)$ also transfers this stochastic rate from
$\widehat\gamma_{\mathrm{CT}}$ to $\widehat\Xi_{\mathrm{CT}}$.  Combining
these expansions with \eqref{eq:ct-local-orders} yields
\eqref{eq:ct-local-estimation-rates}, and every component converges to zero
when $m_T\to0$.  Sufficient rates for root-$T$ centering are
$\|m_T\|=o(T^{-1/2})$ for location and $\|m_T\|=o(T^{-1/4})$ for loadings
and degrees of freedom.
\hfill$\square$

\section{Raw-CST theory and proofs}
\label{app:raw-cst}

\subsection{Raw invariants and identification}
\label{app:raw-identification-results}

An orthogonal change of coordinates within subvector $k$ acts as
\begin{equation}\label{eq:raw-rotation}
 (\Xi^{\mathrm r}_{\bullet k},\alpha_k)
 \mapsto
 (\Xi^{\mathrm r}_{\bullet k}Q_k,Q_k^{\prime}\alpha_k),
 \qquad Q_k\in O(n_k).
\end{equation}
The corresponding raw observed-space invariants are
\begin{equation}\label{eq:raw-invariants}
 S_k=\Xi^{\mathrm r}_{\bullet k}
       \Xi^{\mathrm r\prime}_{\bullet k},
 \qquad
 d_k=\Xi^{\mathrm r}_{\bullet k}\delta_k^{\mathrm r}.
\end{equation}
Here $S_k$ is a dispersion contribution, not a covariance contribution.
For any observed direction $a$ exposed to subvector $k$, put
$q_k(a)^2=a^{\prime}S_ka$ and
$\tau_k(a)=a^{\prime}d_k/q_k(a)$.  A nondegenerate projection of subvector $k$
has tail exponent $\nu_k$, scale $q_k(a)$, and right--left tail ratio
\begin{equation}\label{eq:raw-directional-tail-ratio}
 \frac{T_1\!\left[
   \{\tau_k(a)/\sqrt{1-\tau_k(a)^2}\}\sqrt{\nu_k+1};\nu_k+1
 \right]}
 {T_1\!\left[
  -\{\tau_k(a)/\sqrt{1-\tau_k(a)^2}\}\sqrt{\nu_k+1};\nu_k+1
 \right]}.
\end{equation}
This ratio is strictly increasing in $\tau_k(a)$.  Consequently, the tail
signature identifies dispersion and directional asymmetry without using a
mean or covariance.

For equal-dimensional subvectors, define the raw invariant label
\begin{equation}\label{eq:raw-label}
 \zeta_k^{\mathrm r}
 =\left(\nu_k,\rho_k^{\mathrm r}\right)^{\prime},
 \qquad
 \rho_k^{\mathrm r}=\|\delta_k^{\mathrm r}\|^2
 =d_k^{\prime}S_k^+d_k.
\end{equation}
The raw labeling and orientation convention orders distinct
$\zeta_k^{\mathrm r}$ lexicographically within each subvector dimension and
selects the representative satisfying
\begin{equation}\label{eq:raw-orientation-normalization}
 D_k^{\mathrm r}=E_k^{\prime}S_kE_k\succ0,
 \qquad
 \Xi_{kk}^{\mathrm r}
 =E_k^{\prime}\Xi^{\mathrm r}_{\bullet k}
 =(D_k^{\mathrm r})^{1/2}\succ0.
\end{equation}

\begin{theorem}[Raw-CST identification and recovery]
\label{thm:raw-identification}
Fix $\bm n$.  Let two raw-CST parameter values have nonsingular loading
matrices and finite tail indices $\nu_k\in(0,\infty)$.  If they generate the
same distribution of $Y$, then their location
parameters agree and there exist a dimension-preserving permutation $\pi$
and $Q_k\in O(n_k)$ such that
\begin{equation}\label{eq:raw-identification-orbit}
 \widetilde\nu_{\pi(k)}=\nu_k,
 \qquad
 \widetilde\Xi^{\mathrm r}_{\bullet\pi(k)}
   =\Xi^{\mathrm r}_{\bullet k}Q_k,
 \qquad
 \widetilde\alpha_{\pi(k)}=Q_k^{\prime}\alpha_k.
\end{equation}
Hence the distribution identifies
$\{(n_k,\nu_k,d_k,S_k):k=1,\ldots,K\}$ up to permutation.  Under
the convention that equal-dimensional subvectors have distinct labels
\eqref{eq:raw-label}, are ordered by those labels, and satisfy
\eqref{eq:raw-orientation-normalization}, the normalized raw parameter is
unique and is recovered by
\begin{align}
 \Xi_{kk}^{\mathrm r}
  &=(D_k^{\mathrm r})^{1/2},
 &
 \Xi^{\mathrm r}_{\bullet k}
  &=S_kE_k(D_k^{\mathrm r})^{-1/2},
 \label{eq:raw-loading-recovery}\\
 \delta_k^{\mathrm r}
  &=(D_k^{\mathrm r})^{-1/2}E_k^{\prime}d_k,
 &
 \alpha_k
  &=\frac{\delta_k^{\mathrm r}}
  {(1-\|\delta_k^{\mathrm r}\|^2)^{1/2}}.
 \label{eq:raw-shape-recovery}
\end{align}
\end{theorem}

Theorem \ref{thm:raw-identification} does not infer orthogonality from a
covariance matrix.  Its proof instead uses multivariate regular variation.
At each tail index, the tail measure is supported on the union of the
corresponding latent subspaces; nonsingularity of $\Xi^{\mathrm r}$ makes
those subspaces a direct sum, so the maximal linear pieces of that support
separate the subvectors.  The two one-sided tail constants then recover
$(S_k,d_k)$.  Since every skew-$t$ subvector characteristic function is
nonzero, the recovered factors can be cancelled and the argument iterated
over the remaining tail indices.  The proof follows below.

\subsection{Likelihood consistency and the boundary of regular inference}
\label{app:raw-likelihood-scope}

Write the normalized raw parameter as
$\theta^{\mathrm r}=(\mu^{\mathrm r\prime},
\gamma^{\mathrm r\prime},\alpha_1^{\prime},\ldots,
\alpha_K^{\prime},\nu^{\prime})^{\prime}$, where
$\Xi^{\mathrm r}=\Xi^{\mathrm r}(\gamma^{\mathrm r})$.  Let
$\Theta^{\mathrm r}$ be a compact subset of a fixed labeling chart satisfying
\eqref{eq:raw-label}--\eqref{eq:raw-orientation-normalization}.  Assume that,
for some $\epsilon>0$, uniformly over $\Theta^{\mathrm r}$,
\begin{equation}\label{eq:raw-parameter-bounds}
 \nu_k\geq\epsilon,
 \qquad
 \sigma_{\min}(\Xi^{\mathrm r})\geq\epsilon,
 \qquad
 \lambda_{\min}(\Xi_{kk}^{\mathrm r})\geq\epsilon,
\end{equation}
and use fixed disjoint label neighborhoods for equal-dimensional subvectors.
Compactness supplies finite upper bounds for $\nu_k$, $\alpha_k$, and the
remaining parameters.

\begin{theorem}[Raw-CST likelihood scope]
\label{thm:raw-consistency}
Let $Y_1,\ldots,Y_T$ be iid raw CST with true normalized parameter
$\theta_0^{\mathrm r}\in\Theta^{\mathrm r}$, and let
\begin{equation}\label{eq:raw-mle}
 \widehat\theta_T^{\mathrm r}
 \in\arg\max_{\theta^{\mathrm r}\in\Theta^{\mathrm r}}
 T^{-1}\sum_{t=1}^T\ell^{\mathrm r}(Y_t;\theta^{\mathrm r})
\end{equation}
be any measurable maximizer.
\begin{enumerate}[label=(\roman*),leftmargin=*,itemsep=0.3em]
 \item Under \eqref{eq:raw-parameter-bounds}, for arbitrary true
 $\nu_{k,0}>0$,
 \begin{equation}\label{eq:raw-consistency}
  \widehat\theta_T^{\mathrm r}\xrightarrow{p}\theta_0^{\mathrm r}.
 \end{equation}
 \item If $\theta_0^{\mathrm r}\in\operatorname{int}(\Theta^{\mathrm r})$,
 $\min_k\nu_{k,0}>2$, and the standardized model satisfies Assumptions
 \textcolor{blue}{2} and \textcolor{blue}{3} on a compact neighborhood of
 $h(\theta_0^{\mathrm r})$, where $h$ is the map in Theorem
 \textcolor{blue}{5}, then
 \begin{equation}\label{eq:raw-normality}
  \sqrt T(\widehat\theta_T^{\mathrm r}-\theta_0^{\mathrm r})
  \Rightarrow N\{0,(\mathcal I_0^{\mathrm r})^{-1}\},
  \qquad
  \mathcal I_0^{\mathrm r}=J_0^{\prime}
  \mathcal I\{h(\theta_0^{\mathrm r})\}J_0,
 \end{equation}
 where $J_0$ is the Jacobian of $h$ at
 $\theta_0^{\mathrm r}$.
 \item If $K\geq2$ and some $\nu_{l,0}\leq2$, the score of the
 unrestricted loading model is not square integrable in general.  More
 precisely, for $k\ne l$ one can choose an admissible loading tangent whose
 score contains
 \begin{equation}\label{eq:raw-cross-loading-score}
  g_k(U_k)^{\prime}C U_l,
  \qquad
  g_k(u)=\frac{\partial}{\partial u}
        \log f_{\ST}(u;I_{n_k},\alpha_k,\nu_k),
  \qquad k\ne l,\quad C\ne0,
 \end{equation}
 and this term has infinite second moment.  Hence the
 finite-information Gaussian theory in Theorem
 \textcolor{blue}{4} is unavailable in
 those directions.  This assertion does not preclude regular inference in
 single-subvector or suitably restricted submodels.
\end{enumerate}
\end{theorem}

Part (i) follows because the raw log likelihood has a uniform envelope of
order $1+\log(1+\|y\|^2)$, which is integrable under every positive
skew-$t$ tail index.  Part (ii) is invariance of maximum likelihood under the
smooth bijection in Theorem \textcolor{blue}{5}.  Part (iii)
isolates the obstruction: independence of $U_k$ and $U_l$ makes the second
moment in \eqref{eq:raw-cross-loading-score} contain a nonzero second moment
of a projection of $U_l$, which is infinite when $\nu_l\leq2$.  The density,
characteristic-function, tail-aggregation, identification, and consistency
results therefore remain available without finite variance, while covariance
interpretations, the current Fisher-information formulas, and joint Gaussian
inference retain their stated
$\nu_k>2$ scope.  In particular, the theorem makes no claim that every
component has a non-Gaussian limit when a tail index is at most two; it
identifies why the joint regular finite-information result cannot be carried
over unchanged.  The proof follows below.

\subsection{Proof of Theorem \ref{thm:raw-identification}}

We give a moment-free argument based on tail measures.  Fix one raw subvector
and write
\[
 U\sim\ST_d(0,I_d,\alpha,\nu),
 \qquad
 \delta=\frac{\alpha}{(1+\alpha^{\prime}\alpha)^{1/2}}.
\]
From \textup{(\textcolor{blue}{2})}, uniformly over the unit sphere,
\begin{equation}\label{eq:raw-radial-density-limit}
 r^{\nu+d}f_U(rs)
 \longrightarrow
 C_{\nu,d}
 T_1\{\sqrt{\nu+d}\,\alpha^{\prime}s;\nu+d\}>0,
 \qquad r\to\infty,
\end{equation}
where $C_{\nu,d}>0$ does not depend on $s$.  Thus $U$ is multivariate
regularly varying with index $\nu$, and its tail measure has a strictly
positive angular density on the whole sphere.  Throughout this proof the
normalization is
\[
 \Lambda(A)=\lim_{r\to\infty}r^\nu\Pr(r^{-1}U\in A)
\]
on Borel continuity sets $A$ bounded away from zero for which the tail
measure is finite, with the corresponding pushforward definition after a
linear transformation.

Let $L\in\mathbb R^{n\times d}$ have full column rank and set $X=LU$.
The tail measure $\Lambda$ of $X$ is supported on
$\mathcal V=\operatorname{col}(L)$ and has full support on
$\mathcal V\setminus\{0\}$.  For $a$ not orthogonal to $\mathcal V$, set
\[
 q(a)^2=a^{\prime}LL^{\prime}a,
 \qquad
 \tau(a)=\frac{a^{\prime}L\delta}{q(a)},
 \qquad
 \alpha_{\tau(a)}=\frac{\tau(a)}{\{1-\tau(a)^2\}^{1/2}}.
\]
The projection $a^{\prime}X$ is a univariate skew-$t$ variable.  Therefore
the two half-space masses of its tail measure are
\begin{align}
 c_+(a)&:=\Lambda\{x:a^{\prime}x>1\}
 =\frac{2\kappa_\nu}{\nu}q(a)^\nu
 T_1\{\alpha_{\tau(a)}\sqrt{\nu+1};\nu+1\},
 \label{eq:raw-tail-measure-plus}\\
 c_-(a)&:=\Lambda\{x:a^{\prime}x<-1\}
 =\frac{2\kappa_\nu}{\nu}q(a)^\nu
 T_1\{-\alpha_{\tau(a)}\sqrt{\nu+1};\nu+1\}.
 \label{eq:raw-tail-measure-minus}
\end{align}
It follows that
\begin{equation}\label{eq:raw-tail-measure-recovery}
 q(a)^2=
 \left\{\frac{\nu\{c_+(a)+c_-(a)\}}{2\kappa_\nu}\right\}^{2/\nu},
 \qquad
 \frac{c_+(a)}{c_-(a)}
 =\frac{T_1\{\alpha_{\tau(a)}\sqrt{\nu+1};\nu+1\}}
 {T_1\{-\alpha_{\tau(a)}\sqrt{\nu+1};\nu+1\}}.
\end{equation}
The ratio on the right is strictly increasing in $\tau(a)$.  Hence
$\Lambda$ identifies the quadratic form $a^{\prime}Sa$ and the linear
form $a^{\prime}d$ for every $a$, where $S=LL^{\prime}$ and $d=L\delta$.
It therefore identifies $(S,d)$.

Now consider a raw-CST representation and write
$X_k=\Xi^{\mathrm r}_{\bullet k}U_k$ and
$\mathcal V_k=\operatorname{col}(\Xi^{\mathrm r}_{\bullet k})$.
Nonsingularity of $\Xi^{\mathrm r}$ implies the direct-sum decomposition
\begin{equation}\label{eq:raw-direct-sum-proof}
 \mathbb R^n=\mathcal V_1\oplus\cdots\oplus\mathcal V_K.
\end{equation}
Let $\nu_*=\min_k\nu_k$.  Independence, regular variation, and the
one-large-jump argument give the tail measure of the sum as
\begin{equation}\label{eq:raw-minimum-tail-measure}
 \Lambda_*=\sum_{k:\nu_k=\nu_*}\Lambda_k.
\end{equation}
Indeed, the probability that two independent subvectors are simultaneously of
order $r$ is $O(r^{-2\nu_*})$, subvectors with larger tail indices are
$o(r^{-\nu_*})$, and a fixed translation is negligible on the $r$ scale.
Equation \eqref{eq:raw-radial-density-limit} also makes this convergence
valid on all relatively compact continuity sets bounded away from zero.

The support of $\Lambda_*$ is
$\bigcup_{k:\nu_k=\nu_*}(\mathcal V_k\setminus\{0\})$.  By
\eqref{eq:raw-direct-sum-proof}, its maximal linear subspaces are exactly
the $\mathcal V_k$ with $\nu_k=\nu_*$.  Thus the distribution identifies
those subspaces and their dimensions.  Restricting $\Lambda_*$ to one such
subspace isolates $\Lambda_k$, and
\eqref{eq:raw-tail-measure-recovery} identifies its $(S_k,d_k)$.
The factorization argument below shows that $(\nu_k,S_k,d_k)$ determines
the distribution of $X_k$.

It remains to show that the identified factors may be removed before
repeating the argument at the next tail index.  Every raw subvector
characteristic function is nonzero.  To see this, for fixed $t$ the
projection $t^{\prime}X_k$ has a univariate skew-$t$ density $f$ whose even
part satisfies
\[
 \{f(x)+f(-x)\}/2=t_1(x;0,q^2,\nu_k).
\]
Consequently the real part of its characteristic function is the
characteristic function of a symmetric Student-$t$.  Its normal-mixture
representation gives, at scalar argument $s$,
\[
 \operatorname{Re}\varphi_{t^{\prime}X_k}(s)
 =\E\exp\{-s^2q^2/(2W)\}>0,
 \qquad W\sim\operatorname{Gamma}(\nu_k/2,\nu_k/2).
\]
The degenerate case $q=0$ gives the value one.  The factor is therefore
nowhere zero, which justifies cancellation without a moment assumption.

For two observationally equivalent raw-CST representations, the preceding
tail-measure argument matches all subvectors having the smallest tail index and
identifies their distributions.  Cancel their common, nonzero
characteristic-function factors.  The quotient is the characteristic
function of a location shift plus the sum of the remaining independent
subvectors in each representation.  Iterating over the finitely many distinct
tail indices identifies all $(n_k,\nu_k,S_k,d_k)$ up to permutation.  After
every subvector factor has
been cancelled, equality of the remaining characteristic functions gives
$\exp(it^{\prime}\mu^{\mathrm r})=
\exp(it^{\prime}\widetilde\mu^{\mathrm r})$ for every $t$, and hence
$\mu^{\mathrm r}=\widetilde\mu^{\mathrm r}$.

Finally, if two full-column-rank factors satisfy
$LL^{\prime}=\widetilde L\widetilde L^{\prime}=S$, then
$\widetilde L=LQ$ for some $Q\in O(d)$.  Equality
$L\delta=\widetilde L\widetilde\delta=d$ gives
$\widetilde\delta=Q^{\prime}\delta$ and therefore
$\widetilde\alpha=Q^{\prime}\alpha$.  This proves
\eqref{eq:raw-identification-orbit}.

It remains to verify recovery.  From $S_k=\Xi^{\mathrm r}_{\bullet k}
\Xi^{\mathrm r\prime}_{\bullet k}$ and the normalization,
\[
 S_kE_k
 =\Xi^{\mathrm r}_{\bullet k}\Xi_{kk}^{\mathrm r},
 \qquad
 E_k^{\prime}d_k
 =\Xi_{kk}^{\mathrm r}\delta_k^{\mathrm r}.
\]
Because $\Xi_{kk}^{\mathrm r}=(D_k^{\mathrm r})^{1/2}$, these two equations
give \eqref{eq:raw-loading-recovery} and the first identity in
\eqref{eq:raw-shape-recovery}.  The second is the inverse of
\textup{(\textcolor{blue}{57})}.  This completes the proof.
\hfill$\square$

\subsection{Proof of Theorem \textcolor{blue}{5}}

Fix a subvector and suppress its subscript.  For $\nu>2$, let
$c=(\nu-2)/\nu$, $r=a_\nu\delta^{\mathrm r}$, and
$V=c^{-1}I-rr^{\prime}$.  Since
$c\|r\|^2\leq ca_\nu^2<1$, $V\succ0$.  Directly from the raw moment
formula,
\[
 \E(U)=r,
 \qquad
 \Var(U)=V.
\]
Define the provisionally standardized subvector
\[
 Z^\circ=V^{-1/2}(U-r),
 \qquad
 m^\circ=V^{-1/2}r.
\]
Then $\E(Z^\circ)=0$ and $\Var(Z^\circ)=I$.  The Sherman--Morrison formula
gives
\begin{equation}\label{eq:raw-map-sherman-morrison}
 V^{-1}=c(I+m^\circ m^{\circ\prime}).
\end{equation}
Under the linear transformation $V^{-1/2}U$, the raw skew-$t$ dispersion and
canonical shape become
\[
 \Omega^\circ=V^{-1},
 \qquad
 \eta^\circ=V^{1/2}\alpha.
\]
Furthermore,
\[
 a_\nu^2-m^{\circ\prime}V m^\circ
 =a_\nu^2-r^{\prime}r
 =a_\nu^2(1-\|\delta^{\mathrm r}\|^2),
\]
so
\[
 \frac{V m^\circ}
 {\{a_\nu^2-m^{\circ\prime}V m^\circ\}^{1/2}}
 =V^{1/2}\alpha=\eta^\circ.
\]
Together with \eqref{eq:raw-map-sherman-morrison}, this proves that the
provisional subvector is exactly the mean--covariance standardization in
Proposition \textcolor{blue}{1}.

At the observed level,
\[
 \mu^{\mathrm r}+\Xi^{\mathrm r}_{\bullet k}U
 =\mu^{\mathrm r}+\Xi^{\mathrm r}_{\bullet k}r
  +\Xi^{\mathrm r}_{\bullet k}V^{1/2}Z^\circ.
\]
The matrix $Q=A^{\prime}(AA^{\prime})^{-1/2}$ in
\textup{(\textcolor{blue}{61})} is orthogonal and satisfies
$AQ=(AA^{\prime})^{1/2}\succ0$.  Rotating
$Z^\circ$ to $Q^{\prime}Z^\circ$ therefore yields
\textup{(\textcolor{blue}{62})}--\textup{(\textcolor{blue}{63})} and restores
the required diagonal loading-submatrix orientation.

The invariant identities follow without reference to $Q$:
\[
 b=\Xi^{\mathrm r}r=a_\nu d,
 \qquad
 G=\Xi^{\mathrm r}V\Xi^{\mathrm r\prime}
   =c^{-1}S-a_\nu^2dd^{\prime}.
\]
Solving gives \textup{(\textcolor{blue}{65})}.
Moreover,
\[
 \|m^\circ\|^2
 =\frac{ca_\nu^2\|\delta^{\mathrm r}\|^2}
 {1-ca_\nu^2\|\delta^{\mathrm r}\|^2}.
\]
Substitution into \textup{(\textcolor{blue}{28})} gives
$\varsigma=\|\delta^{\mathrm r}\|^2=\rho^{\mathrm r}$, proving
\textup{(\textcolor{blue}{66})}.

Conversely, if $X\sim\ST(0,\Omega,\eta,\nu)$ is the uncentered subvector in
the standardized construction, then
\[
 U=\Omega^{-1/2}X
 \sim\ST(0,I,\Omega^{1/2}\eta,\nu).
\]
Substitution into $Y=\mu+\Xi(X-m)$ gives the provisional raw location and
loading in \textup{(\textcolor{blue}{67})}--
\textup{(\textcolor{blue}{68})}.  The orthogonal factor $H$ satisfies
$BH=(BB^{\prime})^{1/2}\succ0$, so it selects the normalized raw
representative.  Both constructions preserve the distribution and are
inverse on observational-equivalence classes.  Uniqueness of the normalized
representative in Theorems \textcolor{blue}{2} and
\ref{thm:raw-identification} makes them inverse on the reported charts.
Matrix inversion, the positive-definite square root, and polar normalization
are smooth on the stated interiors, proving the final assertion.
\hfill$\square$

\subsection{Proof of Theorem \ref{thm:raw-consistency}}

Let $A^{\mathrm r}=(\Xi^{\mathrm r})^{-1}$ and
$u=A^{\mathrm r}(y-\mu^{\mathrm r})$.  Compactness and
\eqref{eq:raw-parameter-bounds} give, uniformly over
$\Theta^{\mathrm r}$,
\[
 \|u\|\leq C(1+\|y\|),
 \qquad
 \|\alpha_k\|\leq C,
 \qquad
 \epsilon\leq\nu_k\leq C.
\]
The skewing argument in subvector $k$ satisfies
\[
 \left|
 \left\{\frac{\nu_k+n_k}{\nu_k+u_k^{\prime}u_k}\right\}^{1/2}
 \alpha_k^{\prime}u_k
 \right|^2
 \leq(\nu_k+n_k)\|\alpha_k\|^2.
\]
It therefore ranges over a fixed compact set.  Continuity and positivity of
the Student-$t$ distribution function imply that the skewing factor is
uniformly bounded away from zero.  The determinant and gamma-function terms
are also uniformly bounded, and hence
\begin{equation}\label{eq:raw-log-envelope-proof}
 \sup_{\theta^{\mathrm r}\in\Theta^{\mathrm r}}
 |\ell^{\mathrm r}(y;\theta^{\mathrm r})|
 \leq C_0+C_1\log(1+\|y\|^2).
\end{equation}

This envelope is integrable even when moments do not exist.  Indeed, the raw
skew-$t$ density is at most twice its symmetric Student-$t$ kernel, so for
every $\nu>0$,
\[
 \E\log(1+\|U_k\|^2)<\infty.
\]
The elementary bound
$1+\|\mu^{\mathrm r}+\Xi^{\mathrm r}U\|^2
\leq C\prod_k(1+\|U_k\|^2)$ then gives
$\E\log(1+\|Y\|^2)<\infty$.  Continuity and the uniform law of large
numbers yield uniform convergence of the raw sample criterion to
$L^{\mathrm r}(\theta^{\mathrm r})=
\E_{\theta_0^{\mathrm r}}\ell^{\mathrm r}
(Y;\theta^{\mathrm r})$.  The Kullback--Leibler inequality and Theorem
\ref{thm:raw-identification} make $\theta_0^{\mathrm r}$ its unique maximizer
on the fixed normalized chart.  The argmax theorem proves
\eqref{eq:raw-consistency}.

Under part (ii), consistency confines the MLE to a finite-variance
neighborhood with probability tending to one.  Theorem
\textcolor{blue}{5} gives a smooth local diffeomorphism
$h:\theta^{\mathrm r}\mapsto\theta$ with Jacobian
$J_0=\partial h(\theta_0^{\mathrm r})/\partial\theta^{\mathrm r\prime}$.
Since the two likelihoods agree,
\[
 s^{\mathrm r}(Y;\theta_0^{\mathrm r})
 =J_0^{\prime}s\{Y;h(\theta_0^{\mathrm r})\},
 \qquad
 \mathcal I_0^{\mathrm r}=J_0^{\prime}
 \mathcal I\{h(\theta_0^{\mathrm r})\}J_0.
\]
Theorem \textcolor{blue}{4} and the delta method prove
\eqref{eq:raw-normality}.

Finally, differentiate \textup{(\textcolor{blue}{59})} in a loading
direction $H$.  With $g(U)=(g_1(U_1)^{\prime},\ldots,
g_K(U_K)^{\prime})^{\prime}$,
\begin{equation}\label{eq:raw-loading-directional-score-proof}
 D_H\ell^{\mathrm r}
 =-\operatorname{tr}(A^{\mathrm r}H)
  -g(U)^{\prime}A^{\mathrm r}HU.
\end{equation}
Assume $K\geq2$ and fix $k\ne l$.  Choose a nonzero
$C\in\mathbb R^{n_k\times n_l}$ such that
$\Xi^{\mathrm r}_{lk}C$ is symmetric: take
$C=\Xi_{lk}^{\mathrm r\prime}$ when $\Xi^{\mathrm r}_{lk}\ne0$, and
otherwise take any nonzero $C$.  Let $M$ have $(k,l)$ block $C$ and every
other block zero, and set $H=\Xi^{\mathrm r}M$.  Then the only possibly
nonzero diagonal block of $H$ is
$H_{ll}=\Xi^{\mathrm r}_{lk}C$, which is symmetric.  Thus $H$ is an
admissible tangent to the normalized loading chart, while
$A^{\mathrm r}H=M$ and $\operatorname{tr}(M)=0$.  The directional score is
exactly $-g_k(U_k)^{\prime}CU_l$.

Conditional on $U_k$, every nonzero projection of $U_l$ has infinite
second moment when $\nu_l\leq2$.
Moreover, $C^{\prime}g_k(U_k)$ is nonzero with positive probability for
every $C\ne0$.  Otherwise, positivity and continuity of the skew-$t$
density would imply $C^{\prime}g_k(u)=0$ for every $u$.  Choosing $z$ with
$Cz\ne0$ would then make the density constant along every line parallel to
$Cz$, which is incompatible with integrability.
Therefore
\[
 \E\{g_k(U_k)^{\prime}CU_l\}^2=\infty.
\]
This proves part (iii).  The construction uses a cross-subvector tangent and therefore does
not apply to $K=1$ or to a model that fixes the relevant loading direction.
\hfill$\square$

\section{Additional empirical evidence}
\label{app:empirical-supplement}

\subsection{Residual dependence and block-restricted loadings}

Table \ref{tab:Empirical-correlation-matrix} shows positive SPX--stock
correlations and generally stronger within-sector dependence, motivating
the market--sector grouping.

\begin{table}[H]
\caption{Empirical correlation matrix of HAR residuals}\label{tab:Empirical-correlation-matrix}

\input{tables/empirical_RK_HAR_residual_correlation_table.tex}

\end{table}

Tables \ref{tab:Second-stage-estimates-under} and
\ref{tab:Estimated-loading-matrices-1} impose repeated coefficients within
observed groups $\bm r=(1,3,3,3)^{\prime}$.  Following
\citet{Archakov2024}, the restricted matrix $B$ satisfies
\begin{equation}\label{eq:empirical-block-restriction}
 B_{gg}=(a_g-b_g)I_{r_g}+b_g\bm 1_{r_g}\bm 1_{r_g}^{\prime},
 \qquad
 B_{gh}=c_{gh}\bm 1_{r_g}\bm 1_{r_h}^{\prime},\quad g\ne h.
\end{equation}
For a singleton, $B_{gg}=a_g$.  Apply the restriction to covariance
(Gaussian and global $t$), raw dispersion (global skew-$t$), or $\bar\Xi$
(CT and CST).  Only covariance and dispersion require $c_{gh}=c_{hg}$.
The observed grouping $\bm r$ equals the latent partition $\bm n$ for
Cluster but not Hetero.

In $\widehat\varepsilon_t=\Lambda\bar\Xi Z_t$, positive diagonal
$\Lambda$ collects scales and $\diag(\bar\Xi\bar\Xi^{\prime})=\bm 1_n$.
The $1+3\times2+12=19$ loading coefficients satisfy four distinct
row-normalization constraints.  The remaining 15, plus ten scales, ten
asymmetry coordinates, and four or ten tail indices, give 39 Cluster-CST
or 45 Hetero-CST parameters.

\clearpage

\begin{table}
\caption{Second-stage estimates under block-restricted specifications}\label{tab:Second-stage-estimates-under}
\input{tables/table_empirical_second_stage_blockXi_likelihood_v1.tex}
\end{table}

\begin{table}
\caption{Estimated standardized loading matrices under block-$\bar{\Xi}$ restrictions}\label{tab:Estimated-loading-matrices-1}

\input{tables/table_empirical_blockXi_Xi_estimates_v1.tex}
\end{table}

\clearpage
\subsection{Marginal calibration}

Table \ref{tab:supp-pit} examines whether the gains in joint likelihood
also improve the fitted marginal distributions.  Under both flexible and
block-restricted loadings, CST moves the PIT means toward one half,
reduces the marginal Kolmogorov--Smirnov statistics, and substantially
reduces skewness in the normal-score transforms.  The similarity of these
improvements across loading specifications links the marginal gains to
allowing asymmetric innovations; they do not depend on fitting every
loading coefficient separately.  The PIT variances remain below $1/12$,
however, and the normal-score transforms retain some skewness and excess
kurtosis.  These are descriptive in-sample comparisons; the reported KS
statistics are not tests with critical values adjusted for parameter
estimation and serial dependence.

\begin{table}[H]
\centering
\caption{Probability-integral-transform diagnostics for CT and CST distributions}
\label{tab:supp-pit}
\begin{threeparttable}
\footnotesize
\input{tables/table_empirical_PIT_CT_CST_diagnostics.tex}
\end{threeparttable}
\end{table}

\bibliographystyle{apalike}
\bibliography{CST_ref}

%% file: tables/spd_n12_N50000_T1000_8000.tex
{\footnotesize
\renewcommand{\arraystretch}{1.00}
\setlength{\tabcolsep}{1.5pt}

\setlength{\aboverulesep}{0.30ex}
\setlength{\belowrulesep}{0.30ex}
\setlength{\cmidrulesep}{0.20ex}

\resizebox{\textwidth}{!}{%
\begin{tabularx}{\textwidth}{@{}p{1.2cm}c*{10}{Y}@{}}
	\toprule
	& &
	\multicolumn{5}{c}{$T=1{,}000$}
	& \multicolumn{5}{c}{$T=2{,}000$} \\
	\cmidrule(lr){3-7}
	\cmidrule(lr){8-12}
	
			$\theta$ & True
				& Mean & Std & aSD & $\alpha_L^{0.025}$ & $\alpha_R^{0.025}$
				& Mean & Std & aSD & $\alpha_L^{0.025}$ & $\alpha_R^{0.025}$ \\
	\midrule
	
	$\mu_1$
	& 0.1000
	& 0.1001 & 0.0210 & 0.0210 & 0.0249 & 0.0262
	& 0.0999 & 0.0150 & 0.0149 & 0.0264 & 0.0256 \\
	
	$\mu_2$
	& 0.2000
	& 0.2001 & 0.0275 & 0.0274 & 0.0238 & 0.0265
	& 0.1999 & 0.0195 & 0.0194 & 0.0260 & 0.0253 \\
	
	$\mu_3$
	& 0.3000
	& 0.3002 & 0.0285 & 0.0285 & 0.0246 & 0.0252
	& 0.2999 & 0.0202 & 0.0201 & 0.0252 & 0.0254 \\
	
	\midrule
	
	$\Xi_{11}$
	& 0.6000
	& 0.6011 & 0.0354 & 0.0337 & 0.0175 & 0.0420
	& 0.6005 & 0.0244 & 0.0239 & 0.0193 & 0.0351 \\
	
	$\Xi_{21}$
	& 0.5000
	& 0.5011 & 0.0403 & 0.0386 & 0.0272 & 0.0358
	& 0.5005 & 0.0279 & 0.0273 & 0.0273 & 0.0295 \\
	
	$\Xi_{31}$
	& 0.4000
	& 0.4007 & 0.0404 & 0.0386 & 0.0306 & 0.0313
	& 0.4004 & 0.0279 & 0.0273 & 0.0283 & 0.0293 \\
	
	$\Xi_{12}$
	& 0.3000
	& 0.2999 & 0.0241 & 0.0233 & 0.0302 & 0.0282
	& 0.3001 & 0.0167 & 0.0165 & 0.0269 & 0.0267 \\
	
	$\Xi_{22}$
	& 0.7000
	& 0.6995 & 0.0303 & 0.0297 & 0.0268 & 0.0292
	& 0.6999 & 0.0211 & 0.0210 & 0.0243 & 0.0266 \\
	
	$\Xi_{32}$
	& 0.2000
	& 0.2000 & 0.0194 & 0.0189 & 0.0282 & 0.0278
	& 0.1999 & 0.0135 & 0.0134 & 0.0257 & 0.0259 \\
	
	$\Xi_{13}$
	& 0.1000
	& 0.1001 & 0.0204 & 0.0196 & 0.0301 & 0.0297
	& 0.1000 & 0.0142 & 0.0138 & 0.0279 & 0.0274 \\
	
	$\Xi_{33}$
	& 0.8000
	& 0.7994 & 0.0312 & 0.0306 & 0.0259 & 0.0294
	& 0.7997 & 0.0218 & 0.0216 & 0.0251 & 0.0272 \\
	
	\midrule
	
	$m_1$
	& 0.4500
	& 0.4490 & 0.1036 & 0.1013 & 0.0315 & 0.0242
	& 0.4497 & 0.0722 & 0.0716 & 0.0278 & 0.0235 \\
	
	$\delta_1$
	& 0.5803
	& 0.5731 & 0.1063 & 0.1025 & 0.0331 & 0.0181
	& 0.5769 & 0.0735 & 0.0725 & 0.0294 & 0.0195 \\
	
	$\nu_1^{-1}$
	& 0.2500
	& 0.2485 & 0.0334 & 0.0329 & 0.0309 & 0.0232
	& 0.2493 & 0.0235 & 0.0233 & 0.0285 & 0.0227 \\
	
	\midrule
	
	$m_{2,1}$
	& 0.4000
	& 0.3959 & 0.1132 & 0.1109 & 0.0363 & 0.0175
	& 0.3986 & 0.0797 & 0.0784 & 0.0329 & 0.0205 \\
	
	$m_{2,2}$
	& -0.3000
	& -0.2972 & 0.1194 & 0.1168 & 0.0206 & 0.0351
	& -0.2982 & 0.0837 & 0.0826 & 0.0204 & 0.0322 \\
	
	$\delta_{2,1}$
	& 0.5052
	& 0.4935 & 0.1237 & 0.1194 & 0.0371 & 0.0154
	& 0.5000 & 0.0863 & 0.0844 & 0.0334 & 0.0191 \\
	
	$\delta_{2,2}$
	& -0.3909
	& -0.3807 & 0.1420 & 0.1388 & 0.0186 & 0.0349
	& -0.3853 & 0.0997 & 0.0981 & 0.0198 & 0.0327 \\
	
	$\nu_2^{-1}$
	& 0.2000
	& 0.1986 & 0.0226 & 0.0222 & 0.0313 & 0.0235
	& 0.1993 & 0.0158 & 0.0157 & 0.0281 & 0.0232 \\
	
	\midrule
	
	& &
	\multicolumn{5}{c}{$T=4{,}000$}
	& \multicolumn{5}{c}{$T=8{,}000$} \\
	\cmidrule(lr){3-7}
	\cmidrule(lr){8-12}
	
			$\theta$ & True
				& Mean & Std & aSD & $\alpha_L^{0.025}$ & $\alpha_R^{0.025}$
				& Mean & Std & aSD & $\alpha_L^{0.025}$ & $\alpha_R^{0.025}$ \\
	\midrule
	
	$\mu_1$
	& 0.1000
	& 0.1000 & 0.0105 & 0.0105 & 0.0248 & 0.0256
	& 0.1000 & 0.0075 & 0.0074 & 0.0259 & 0.0261 \\
	
	$\mu_2$
	& 0.2000
	& 0.1999 & 0.0138 & 0.0137 & 0.0254 & 0.0256
	& 0.2000 & 0.0097 & 0.0097 & 0.0252 & 0.0254 \\
	
	$\mu_3$
	& 0.3000
	& 0.2999 & 0.0143 & 0.0142 & 0.0258 & 0.0254
	& 0.3000 & 0.0101 & 0.0101 & 0.0249 & 0.0248 \\
	
	\midrule
	
	$\Xi_{11}$
	& 0.6000
	& 0.6004 & 0.0170 & 0.0169 & 0.0200 & 0.0321
	& 0.6000 & 0.0120 & 0.0119 & 0.0218 & 0.0299 \\
	
	$\Xi_{21}$
	& 0.5000
	& 0.5003 & 0.0195 & 0.0193 & 0.0256 & 0.0280
	& 0.5000 & 0.0137 & 0.0136 & 0.0246 & 0.0266 \\
	
	$\Xi_{31}$
	& 0.4000
	& 0.4001 & 0.0194 & 0.0193 & 0.0266 & 0.0257
	& 0.3999 & 0.0137 & 0.0136 & 0.0266 & 0.0249 \\
	
	$\Xi_{12}$
	& 0.3000
	& 0.3000 & 0.0118 & 0.0117 & 0.0276 & 0.0270
	& 0.3000 & 0.0083 & 0.0082 & 0.0259 & 0.0255 \\
	
	$\Xi_{22}$
	& 0.7000
	& 0.7000 & 0.0150 & 0.0148 & 0.0243 & 0.0273
	& 0.7000 & 0.0105 & 0.0105 & 0.0241 & 0.0255 \\
	
	$\Xi_{32}$
	& 0.2000
	& 0.2000 & 0.0095 & 0.0095 & 0.0252 & 0.0253
	& 0.2000 & 0.0067 & 0.0067 & 0.0256 & 0.0248 \\
	
	$\Xi_{13}$
	& 0.1000
	& 0.1001 & 0.0098 & 0.0098 & 0.0252 & 0.0267
	& 0.1000 & 0.0069 & 0.0069 & 0.0261 & 0.0252 \\
	
	$\Xi_{33}$
	& 0.8000
	& 0.8000 & 0.0154 & 0.0153 & 0.0244 & 0.0270
	& 0.8000 & 0.0108 & 0.0108 & 0.0231 & 0.0269 \\
	
	\midrule
	
	$m_1$
	& 0.4500
	& 0.4493 & 0.0509 & 0.0507 & 0.0288 & 0.0224
	& 0.4499 & 0.0359 & 0.0358 & 0.0262 & 0.0244 \\
	
	$\delta_1$
	& 0.5803
	& 0.5781 & 0.0517 & 0.0513 & 0.0297 & 0.0204
	& 0.5794 & 0.0364 & 0.0363 & 0.0272 & 0.0224 \\
	
	$\nu_1^{-1}$
	& 0.2500
	& 0.2498 & 0.0165 & 0.0165 & 0.0265 & 0.0241
	& 0.2497 & 0.0117 & 0.0116 & 0.0276 & 0.0245 \\
	
	\midrule
	
	$m_{2,1}$
	& 0.4000
	& 0.3988 & 0.0556 & 0.0555 & 0.0297 & 0.0197
	& 0.3993 & 0.0394 & 0.0392 & 0.0298 & 0.0226 \\
	
	$m_{2,2}$
	& -0.3000
	& -0.2997 & 0.0588 & 0.0584 & 0.0212 & 0.0288
	& -0.2997 & 0.0414 & 0.0413 & 0.0229 & 0.0272 \\
	
	$\delta_{2,1}$
	& 0.5052
	& 0.5021 & 0.0600 & 0.0597 & 0.0300 & 0.0187
	& 0.5036 & 0.0425 & 0.0422 & 0.0300 & 0.0219 \\
	
	$\delta_{2,2}$
	& -0.3909
	& -0.3888 & 0.0699 & 0.0694 & 0.0210 & 0.0291
	& -0.3896 & 0.0492 & 0.0491 & 0.0218 & 0.0272 \\
	
	$\nu_2^{-1}$
	& 0.2000
	& 0.1997 & 0.0112 & 0.0111 & 0.0275 & 0.0246
	& 0.1999 & 0.0079 & 0.0079 & 0.0269 & 0.0240 \\
	
	\bottomrule
\end{tabularx}%
}
}

\begin{tablenotes}[flushleft]
	\scriptsize
	\setlength{\itemsep}{0pt}
	\setlength{\parsep}{0pt}
	\setlength{\topsep}{2pt}
	\item \textit{Note.}
	The table reports finite-sample properties of the maximum likelihood
	estimator for the standardized CST distribution with known subvector dimensions
	\(\bm n=(1,2)^{\prime}\).
	Mean and Std are the Monte Carlo mean and standard deviation, while aSD is
	the asymptotic standard deviation implied by the Fisher information
	evaluated at the true parameter.
	For the transformed skewness parameters \(\delta_k\), aSD is obtained by
	the delta method.
	The columns $\alpha_L^{0.025}$ and $\alpha_R^{0.025}$ report empirical rejection
	frequencies in the left and right 2.5\% tails of nominal 95\% intervals.
	Intervals for \(\delta_k\) are obtained by mapping the joint asymptotic
	distribution of \((m_k,\nu_k)\) through
	\(\delta_k=\delta_k(m_k,\nu_k)\); all other intervals use the normal
	approximation.
	Only unique loading entries are reported, with
	\(\Xi_{23}=\Xi_{32}\) omitted.
	The simulations use \(R=50{,}000\) replications.
\end{tablenotes}

%% file: tables/ct_vs_cst_T8000_sameDGP_table.tex
{\footnotesize
\renewcommand{\arraystretch}{1.02}
\setlength{\tabcolsep}{2pt}

\setlength{\aboverulesep}{0.35ex}
\setlength{\belowrulesep}{0.35ex}
\setlength{\cmidrulesep}{0.25ex}

\resizebox{\textwidth}{!}{%
\begin{tabularx}{\textwidth}{@{}p{1.2cm}*{8}{Y}@{}}
	\toprule
				$\theta$
		& True
	& \shortstack{CT\\Mean}
	& \shortstack{CST\\Mean}
	& \shortstack{CT\\Bias}
	& \shortstack{CST\\Bias}
	& \shortstack{CT\\RMSE}
	& \shortstack{CST\\RMSE}
	& \shortstack{CT/CST\\RMSE} \\
	\midrule

				\multicolumn{9}{l}{\textit{Panel A: Location and tail parameters}} \\
				
				$\mu_1$
				& 0.1000 & 0.0524 & 0.1000
				& -0.0476 & 0.0000
				& 0.0480 & 0.0075 & 6.4182 \\
				
				$\mu_2$
				& 0.2000 & 0.1441 & 0.2000
				& -0.0559 & 0.0000
				& 0.0566 & 0.0097 & 5.8150 \\
				
				$\mu_3$
				& 0.3000 & 0.2930 & 0.3000
				& -0.0070 & 0.0000
				& 0.0115 & 0.0101 & 1.1310 \\
				
				$1/\nu_1$
				& 0.2500 & 0.2598 & 0.2498
				& 0.0098 & -0.0002
				& 0.0153 & 0.0116 & 1.3185 \\
				
				$1/\nu_2$
				& 0.2000 & 0.2044 & 0.1998
				& 0.0044 & -0.0002
				& 0.0091 & 0.0079 & 1.1523 \\
				
				\midrule
				\multicolumn{9}{l}{\textit{Panel B: Loading-matrix parameters}} \\
				
				$\Xi_{11}$
				& 0.6000 & 0.5990 & 0.6001
				& -0.0010 & 0.0001
				& 0.0121 & 0.0120 & 1.0080 \\
				
				$\Xi_{12}$
				& 0.3000 & 0.2998 & 0.3000
				& -0.0002 & 0.0000
				& 0.0085 & 0.0083 & 1.0283 \\
				
				$\Xi_{13}$
				& 0.1000 & 0.1014 & 0.1000
				& 0.0014 & 0.0000
				& 0.0073 & 0.0069 & 1.0509 \\
				
				$\Xi_{21}$
				& 0.5000 & 0.4991 & 0.5000
				& -0.0009 & 0.0000
				& 0.0140 & 0.0137 & 1.0156 \\
				
				$\Xi_{22}$
				& 0.7000 & 0.6993 & 0.6999
				& -0.0007 & -0.0001
				& 0.0106 & 0.0105 & 1.0075 \\
				
				$\Xi_{31}$
				& 0.4000 & 0.3993 & 0.4000
				& -0.0007 & 0.0000
				& 0.0139 & 0.0137 & 1.0169 \\
				
				$\Xi_{32}$
				& 0.2000 & 0.2032 & 0.2000
				& 0.0032 & 0.0000
				& 0.0075 & 0.0067 & 1.1210 \\
				
				$\Xi_{33}$
				& 0.8000 & 0.8012 & 0.8000
				& 0.0012 & 0.0000
				& 0.0110 & 0.0108 & 1.0165 \\
				
				\bottomrule
			\end{tabularx}%
		}}
		\begin{tablenotes}[flushleft]
		\scriptsize
	\setlength{\itemsep}{0pt}
	\setlength{\parsep}{0pt}
	\setlength{\topsep}{2pt}
	\item \textit{Note.}
			The table compares CT and CST estimates when the data are generated
			from the standardized CST model with subvector dimensions
			$\bm n=(1,2)^{\prime}$, sample size $T=8{,}000$, and $R=50{,}000$
			Monte Carlo replications. In each replication, CT and CST are fitted to
			the same simulated sample. The CT model is misspecified because it
			imposes zero skewness. Bias is computed relative to the true CST
			parameter, RMSE denotes root mean squared error, and CT/CST RMSE is
			the ratio of the CT RMSE to the CST RMSE. The loading matrix is
			reported under the symmetric positive-definite diagonal loading-submatrix
			normalization; the repeated entry $\Xi_{23}=\Xi_{32}$ is therefore
			omitted. The CST-only parameters $m_k$ and $\delta_k$ are not shown
			because they have no CT counterparts.
		\end{tablenotes}

%% file: tables/empirical_RK_HAR_coefficients_table.tex
{\footnotesize
\renewcommand{\arraystretch}{1.08}
\setlength{\tabcolsep}{2pt}

\resizebox{\textwidth}{!}{%
\begin{tabularx}{\textwidth}{@{}l*{10}{Y}@{}}
\toprule
& Market
& \multicolumn{3}{c}{Health Care}
& \multicolumn{3}{c}{Industrials}
& \multicolumn{3}{c}{Information Tech.} \\
\cmidrule(lr){2-2}
\cmidrule(lr){3-5}
\cmidrule(lr){6-8}
\cmidrule(lr){9-11}
& SPX & LLY & MRK & JNJ & MMM & GE & HON & AAPL & IBM & INTC \\
\midrule

$\beta_0$
& \shortstack{0.629\\(0.113)}
& \shortstack{1.175\\(0.237)}
& \shortstack{1.306\\(0.229)}
& \shortstack{1.211\\(0.196)}
& \shortstack{1.485\\(0.222)}
& \shortstack{1.537\\(0.246)}
& \shortstack{1.518\\(0.230)}
& \shortstack{1.062\\(0.222)}
& \shortstack{1.176\\(0.230)}
& \shortstack{1.231\\(0.206)}
\\

$\beta_d$
& \shortstack{0.552\\(0.013)}
& \shortstack{0.462\\(0.014)}
& \shortstack{0.442\\(0.014)}
& \shortstack{0.425\\(0.014)}
& \shortstack{0.402\\(0.014)}
& \shortstack{0.400\\(0.014)}
& \shortstack{0.440\\(0.014)}
& \shortstack{0.439\\(0.014)}
& \shortstack{0.451\\(0.014)}
& \shortstack{0.460\\(0.014)}
\\

$\beta_w$
& \shortstack{0.345\\(0.017)}
& \shortstack{0.336\\(0.019)}
& \shortstack{0.376\\(0.019)}
& \shortstack{0.385\\(0.019)}
& \shortstack{0.417\\(0.019)}
& \shortstack{0.371\\(0.019)}
& \shortstack{0.346\\(0.019)}
& \shortstack{0.359\\(0.019)}
& \shortstack{0.355\\(0.019)}
& \shortstack{0.371\\(0.018)}
\\

$\beta_m$
& \shortstack{0.053\\(0.013)}
& \shortstack{0.148\\(0.017)}
& \shortstack{0.118\\(0.016)}
& \shortstack{0.115\\(0.017)}
& \shortstack{0.098\\(0.017)}
& \shortstack{0.145\\(0.018)}
& \shortstack{0.130\\(0.018)}
& \shortstack{0.154\\(0.017)}
& \shortstack{0.142\\(0.017)}
& \shortstack{0.104\\(0.016)}
\\

\midrule

$\mathbb{E}(Y)$
& 12.44
& 21.69
& 20.04
& 16.29
& 17.87
& 18.31
& 18.00
& 21.81
& 22.83
& 19.02
\\

$\pi$
& 0.949
& 0.946
& 0.935
& 0.926
& 0.917
& 0.916
& 0.916
& 0.951
& 0.949
& 0.935
\\

\midrule

$\sigma_\varepsilon$
& 4.213
& 8.719
& 7.588
& 5.877
& 6.059
& 7.073
& 6.124
& 7.402
& 7.049
& 6.104
\\

AR(1)
& -0.047
& -0.024
& -0.032
& -0.025
& -0.014
& -0.012
& -0.022
& -0.028
& -0.018
& -0.028
\\

Skew$^{\dagger}$
& 1.820
& 3.400
& 2.677
& 2.043
& 1.966
& 3.769
& 2.551
& 1.780
& 1.543
& 1.811
\\

Kurt$^{\dagger}$
& 16.786
& 30.551
& 25.854
& 15.850
& 12.771
& 35.212
& 20.069
& 13.636
& 11.197
& 13.759
\\

\bottomrule
\end{tabularx}%
}
}

\begin{tablenotes}[flushleft]
\scriptsize
\item \textit{Note.}
The table reports OLS estimates from asset-specific non-overlapping HAR
regressions for
\(Y_{i,t}=100\sqrt{252\,RV_{i,t}}\), together with summary statistics for
the corresponding HAR residuals.
Parameter estimates are reported in the first line of each cell, with the
usual OLS standard errors in parentheses below.
The persistence measure is
\(\pi=\beta_d+\beta_w+\beta_m\), and
\(\mathbb{E}(Y)=\beta_0/(1-\pi)\) is the HAR-implied unconditional mean.
Rounding may prevent these statistics from being reproduced exactly from
the displayed coefficients.
The quantity \(\sigma_\varepsilon\) denotes the standard deviation of the HAR residuals,
while AR(1) denotes their first-order sample autocorrelation.
Statistics marked by \(\dagger\) are computed after winsorizing each residual
series at its empirical 0.1\% and 99.9\% quantiles.
All other statistics, the HAR estimation, and the second-stage distributional
estimation use the untrimmed observations.
Kurtosis is the raw fourth standardized moment.
\end{tablenotes}

%% file: tables/table_empirical_second_stage_likelihood_v1.tex
{\footnotesize
\renewcommand{\arraystretch}{1.08}
\setlength{\tabcolsep}{2pt}

\begin{tabularx}{\textwidth}{@{}l*{7}{Y}@{}}
			\toprule
			& Gaussian
			& Multi-$t$
			& \begin{tabular}{@{}c@{}}Skew-$t$\end{tabular}
			& \multicolumn{2}{c}{Cluster}
			& \multicolumn{2}{c}{Hetero} \\
			\cmidrule(lr){5-6}
			\cmidrule(lr){7-8}
			& & &
			& $t$
			& \begin{tabular}{@{}c@{}}Skew-$t$\end{tabular}
			& $t$
			& \begin{tabular}{@{}c@{}}Skew-$t$\end{tabular} \\
			\midrule
			
			$\nu_{\mathrm{all}}$
			& $\infty$
			& \begin{tabular}{@{}c@{}}
				3.192\\[-2pt]
				{\scriptsize (0.115)}
			\end{tabular}
			& \begin{tabular}{@{}c@{}}
				3.024\\[-2pt]
				{\scriptsize (0.105)}
			\end{tabular}
			& -- & -- & -- & -- \\
			
			
			$\nu_{\mathrm{Market}}$
			& -- & -- & --
			& \begin{tabular}{@{}c@{}}
				2.606\\[-2pt]
				{\scriptsize (0.157)}
			\end{tabular}
			& \begin{tabular}{@{}c@{}}
				2.461\\[-2pt]
				{\scriptsize (0.143)}
			\end{tabular}
			& \begin{tabular}{@{}c@{}}
				2.606\\[-2pt]
				{\scriptsize (0.158)}
			\end{tabular}
			& \begin{tabular}{@{}c@{}}
				2.451\\[-2pt]
				{\scriptsize (0.139)}
			\end{tabular}
			\\[2pt]
			
			$\nu_{\mathrm{Health}}$
			& -- & -- & --
			& \begin{tabular}{@{}c@{}}
				2.494\\[-2pt]
				{\scriptsize (0.099)}
			\end{tabular}
			& \begin{tabular}{@{}c@{}}
				2.334\\[-2pt]
				{\scriptsize (0.080)}
			\end{tabular}
			& \begin{tabular}{@{}c@{}}
				2.201\\[-2pt]
				{\scriptsize (0.107)}\\[2pt]
				2.288\\[-2pt]
				{\scriptsize (0.130)}\\[2pt]
				2.879\\[-2pt]
				{\scriptsize (0.152)}
			\end{tabular}
			& \begin{tabular}{@{}c@{}}
				2.090\\[-2pt]
				{\scriptsize (0.094)}\\[2pt]
				2.218\\[-2pt]
				{\scriptsize (0.113)}\\[2pt]
				2.796\\[-2pt]
				{\scriptsize (0.146)}
			\end{tabular}
			\\[2pt]
			
			$\nu_{\mathrm{Industrials}}$
			& -- & -- & --
			& \begin{tabular}{@{}c@{}}
				2.961\\[-2pt]
				{\scriptsize (0.113)}
			\end{tabular}
			& \begin{tabular}{@{}c@{}}
				2.744\\[-2pt]
				{\scriptsize (0.093)}
			\end{tabular}
			& \begin{tabular}{@{}c@{}}
				2.987\\[-2pt]
				{\scriptsize (0.157)}\\[2pt]
				2.688\\[-2pt]
				{\scriptsize (0.146)}\\[2pt]
				2.948\\[-2pt]
				{\scriptsize (0.155)}
			\end{tabular}
			& \begin{tabular}{@{}c@{}}
				2.982\\[-2pt]
				{\scriptsize (0.157)}\\[2pt]
				2.565\\[-2pt]
				{\scriptsize (0.133)}\\[2pt]
				2.852\\[-2pt]
				{\scriptsize (0.153)}
			\end{tabular}
			\\[2pt]
			
			$\nu_{\mathrm{InfoTech}}$
			& -- & -- & --
			& \begin{tabular}{@{}c@{}}
				3.204\\[-2pt]
				{\scriptsize (0.125)}
			\end{tabular}
			& \begin{tabular}{@{}c@{}}
				3.041\\[-2pt]
				{\scriptsize (0.110)}
			\end{tabular}
			& \begin{tabular}{@{}c@{}}
				2.703\\[-2pt]
				{\scriptsize (0.139)}\\[2pt]
				3.397\\[-2pt]
				{\scriptsize (0.191)}\\[2pt]
				3.412\\[-2pt]
				{\scriptsize (0.193)}
			\end{tabular}
			& \begin{tabular}{@{}c@{}}
				2.641\\[-2pt]
				{\scriptsize (0.133)}\\[2pt]
				3.242\\[-2pt]
				{\scriptsize (0.169)}\\[2pt]
				3.442\\[-2pt]
				{\scriptsize (0.206)}
			\end{tabular}
			\\
			
			\midrule
			\multicolumn{8}{l}{\textit{Model-specific skewness components}} \\
			
			$\delta_{\mathrm{Market}}$
			& -- & -- & 0.613 (0.034)
			& -- & 0.481 (0.043)
			& -- & 0.491 (0.046) \\
			
			$\delta_{\mathrm{Health}}$
			& -- & --
			& \begin{tabular}{@{}c@{}}
				0.557 (0.026)\\
				0.578 (0.026)\\
				0.553 (0.029)
			\end{tabular}
			& --
			& \begin{tabular}{@{}c@{}}
				0.354 (0.026)\\
				0.334 (0.028)\\
				0.359 (0.031)
			\end{tabular}
			& --
			& \begin{tabular}{@{}c@{}}
				0.411 (0.033)\\
				0.400 (0.039)\\
				0.586 (0.040)
			\end{tabular}
			\\
			
			$\delta_{\mathrm{Industrials}}$
			& -- & --
			& \begin{tabular}{@{}c@{}}
				0.507 (0.026)\\
				0.528 (0.023)\\
				0.528 (0.023)
			\end{tabular}
			& --
			& \begin{tabular}{@{}c@{}}
				0.446 (0.031)\\
				0.451 (0.027)\\
				0.430 (0.029)
			\end{tabular}
			& --
			& \begin{tabular}{@{}c@{}}
				0.635 (0.036)\\
				0.549 (0.037)\\
				0.612 (0.034)
			\end{tabular}
			\\
			
			$\delta_{\mathrm{InfoTech}}$
			& -- & --
			& \begin{tabular}{@{}c@{}}
				0.566 (0.028)\\
				0.546 (0.028)\\
				0.564 (0.026)
			\end{tabular}
			& --
			& \begin{tabular}{@{}c@{}}
				0.333 (0.035)\\
				0.417 (0.038)\\
				0.387 (0.038)
			\end{tabular}
			& --
			& \begin{tabular}{@{}c@{}}
				0.433 (0.043)\\
				0.631 (0.035)\\
				0.648 (0.039)
			\end{tabular}
			\\
			
			\midrule
			$p$       & 55 & 56 & 66 & 95 & 105 & 110 & 120 \\
			$\ell$    & -151962 & -140715 & -140313 & -140152
			& \textbf{-139399} & -140493 & -139563 \\
			$\ell_m$  & -162956 & -150833 & -149460 & -150836
			& -149445 & -150828 & \textbf{-149441} \\
			$\ell_c$  & \textbf{10993} & 10118 & 9148 & 10684
			& 10045 & 10336 & 9878 \\
			BIC       & 304392 & 281907 & 281187 & 281113
			& \textbf{279692} & 281922 & 280148 \\
			\bottomrule
		\end{tabularx}%
	}

\begin{tablenotes}[flushleft]
	\scriptsize
	\item \textit{Note.} The table reports second-stage maximum likelihood estimates for the centered HAR residuals obtained from $Y_{i,t}=100\sqrt{252\,RV_{i,t}}$. For the Cluster and Hetero specifications, the residual is modeled as $\widehat\varepsilon_t=\Lambda V_t$, where $V_t$ has a standardized CT or CST distribution with $\operatorname{diag}(\bar\Xi\bar\Xi^{\prime})=\bm 1$. Parentheses report Bartlett HAC standard errors (bandwidth 5), conditional on the fitted and centered HAR residuals and excluding first-stage estimation uncertainty. The joint residual log-likelihood is denoted by $\ell$, the sum of univariate marginal log-likelihoods by $\ell_m$, and the residual dependence contribution by $\ell_c=\ell-\ell_m$. The BIC is computed as $-2\ell+p\log T$.
\end{tablenotes}

%% file: tables/table_empirical_all_Xi_estimates_v1.tex
{\footnotesize
\renewcommand{\arraystretch}{1.02}
\setlength{\tabcolsep}{2pt}

\setlength{\aboverulesep}{0.35ex}
\setlength{\belowrulesep}{0.35ex}
\setlength{\cmidrulesep}{0.25ex}

\resizebox{\textwidth}{!}{%
\begin{tabularx}{\textwidth}{@{}l*{10}{Y}@{}}
		\toprule
		& \multicolumn{1}{c}{Market}
		& \multicolumn{3}{c}{Health Care}
		& \multicolumn{3}{c}{Industrials}
		& \multicolumn{3}{c}{Information Tech.} \\
		\cmidrule(lr){2-2}
		\cmidrule(lr){3-5}
		\cmidrule(lr){6-8}
		\cmidrule(lr){9-11}
		& SPX & LLY & MRK & JNJ & MMM & GE & HON & AAPL & IBM & INTC \\
		\midrule
		
		\multicolumn{11}{l}{\textit{Panel A: Cluster-\(t\)}} \\
		
		SPX
		& \bc{\textbf{0.985}}
		& 0.044 & 0.054 & 0.060
		& \bc{0.008} & \bc{0.007} & \bc{0.002}
		& 0.066 & 0.067 & 0.111 \\
		
		LLY
		& \textbf{0.401}
		& \bc{\textbf{0.888}} & \bc{0.138} & \bc{0.121}
		& 0.028 & 0.015 & 0.013
		& \bc{0.051} & \bc{0.060} & \bc{0.094} \\
		
		MRK
		& \textbf{0.439}
		& \bc{0.138} & \bc{\textbf{0.860}} & \bc{0.163}
		& 0.010 & 0.009 & 0.005
		& \bc{0.049} & \bc{0.063} & \bc{0.129} \\
		
		JNJ
		& \textbf{0.441}
		& \bc{0.121} & \bc{0.163} & \bc{\textbf{0.868}}
		& 0.020 & 0.005 & 0.014
		& \bc{0.061} & \bc{0.045} & \bc{0.068} \\
		
		MMM
		& \bc{\textbf{0.380}}
		& 0.047 & 0.093 & 0.163
		& \bc{\textbf{0.872}} & \bc{0.131} & \bc{0.121}
		& 0.092 & 0.070 & 0.111 \\
		
		GE
		& \bc{\textbf{0.416}}
		& 0.074 & 0.102 & 0.171
		& \bc{0.131} & \bc{\textbf{0.851}} & \bc{0.140}
		& 0.093 & 0.060 & 0.093 \\
		
		HON
		& \bc{\textbf{0.431}}
		& 0.090 & 0.120 & 0.137
		& \bc{0.121} & \bc{0.140} & \bc{\textbf{0.844}}
		& 0.087 & 0.082 & 0.111 \\
		
		AAPL
		& \textbf{0.476}
		& \bc{0.079} & \bc{0.128} & \bc{0.142}
		& 0.009 & 0.011 & 0.015
		& \bc{\textbf{0.827}} & \bc{0.156} & \bc{0.147} \\
		
		IBM
		& \textbf{0.434}
		& \bc{0.088} & \bc{0.128} & \bc{0.144}
		& -0.002 & 0.020 & -0.009
		& \bc{0.156} & \bc{\textbf{0.849}} & \bc{0.143} \\
		
		INTC
		& \textbf{0.470}
		& \bc{0.086} & \bc{0.079} & \bc{0.154}
		& 0.027 & 0.021 & -0.013
		& \bc{0.147} & \bc{0.143} & \bc{\textbf{0.836}} \\
		
		\midrule
		\multicolumn{11}{l}{\textit{Panel B: Cluster-skew-\(t\)}} \\
		
		SPX
		& \bc{\textbf{0.979}}
		& 0.074 & 0.063 & 0.102
		& \bc{0.028} & \bc{0.018} & \bc{0.012}
		& 0.069 & 0.069 & 0.105 \\
		
		LLY
		& \textbf{0.340}
		& \bc{\textbf{0.910}} & \bc{0.159} & \bc{0.147}
		& 0.033 & 0.016 & 0.013
		& \bc{0.047} & \bc{0.051} & \bc{0.059} \\
		
		MRK
		& \textbf{0.387}
		& \bc{0.159} & \bc{\textbf{0.881}} & \bc{0.185}
		& 0.024 & 0.012 & 0.011
		& \bc{0.047} & \bc{0.054} & \bc{0.093} \\
		
		JNJ
		& \textbf{0.370}
		& \bc{0.147} & \bc{0.185} & \bc{\textbf{0.894}}
		& 0.032 & 0.007 & 0.018
		& \bc{0.055} & \bc{0.039} & \bc{0.040} \\
		
		MMM
		& \bc{\textbf{0.331}}
		& 0.063 & 0.085 & 0.162
		& \bc{\textbf{0.891}} & \bc{0.154} & \bc{0.144}
		& 0.070 & 0.057 & 0.082 \\
		
		GE
		& \bc{\textbf{0.368}}
		& 0.090 & 0.102 & 0.185
		& \bc{0.154} & \bc{\textbf{0.866}} & \bc{0.158}
		& 0.077 & 0.052 & 0.070 \\
		
		HON
		& \bc{\textbf{0.389}}
		& 0.110 & 0.123 & 0.146
		& \bc{0.144} & \bc{0.158} & \bc{\textbf{0.858}}
		& 0.074 & 0.072 & 0.082 \\
		
		AAPL
		& \textbf{0.456}
		& \bc{0.103} & \bc{0.141} & \bc{0.175}
		& 0.036 & 0.020 & 0.025
		& \bc{\textbf{0.824}} & \bc{0.163} & \bc{0.149} \\
		
		IBM
		& \textbf{0.412}
		& \bc{0.115} & \bc{0.146} & \bc{0.178}
		& 0.020 & 0.026 & 0.000
		& \bc{0.163} & \bc{\textbf{0.845}} & \bc{0.147} \\
		
		INTC
		& \textbf{0.451}
		& \bc{0.129} & \bc{0.111} & \bc{0.208}
		& 0.053 & 0.031 & -0.001
		& \bc{0.149} & \bc{0.147} & \bc{\textbf{0.823}} \\
		
		\midrule
		\multicolumn{11}{l}{\textit{Panel C: Hetero-\(t\)}} \\
		
		SPX
		& \bc{\textbf{0.986}}
		& 0.053 & 0.054 & 0.047
		& \bc{0.002} & \bc{0.008} & \bc{0.005}
		& 0.064 & 0.046 & 0.118 \\
		
		LLY
		& \textbf{0.313}
		& \bc{\textbf{0.926}} & \bc{0.148} & \bc{0.100}
		& 0.022 & 0.013 & 0.015
		& \bc{0.048} & \bc{0.033} & \bc{0.096} \\
		
		MRK
		& \textbf{0.400}
		& \bc{0.084} & \bc{\textbf{0.886}} & \bc{0.159}
		& 0.013 & 0.025 & 0.014
		& \bc{0.040} & \bc{0.043} & \bc{0.136} \\
		
		JNJ
		& \textbf{0.522}
		& \bc{0.099} & \bc{0.111} & \bc{\textbf{0.828}}
		& 0.022 & 0.012 & 0.035
		& \bc{0.076} & \bc{0.033} & \bc{0.106} \\
		
		MMM
		& \bc{\textbf{0.407}}
		& 0.049 & 0.079 & 0.127
		& \bc{\textbf{0.856}} & \bc{0.116} & \bc{0.188}
		& 0.089 & 0.046 & 0.135 \\
		
		GE
		& \bc{\textbf{0.407}}
		& 0.080 & 0.073 & 0.131
		& \bc{0.101} & \bc{\textbf{0.863}} & \bc{0.176}
		& 0.085 & 0.045 & 0.106 \\
		
		HON
		& \bc{\textbf{0.443}}
		& 0.120 & 0.117 & 0.105
		& \bc{0.046} & \bc{0.107} & \bc{\textbf{0.852}}
		& 0.073 & 0.059 & 0.130 \\
		
		AAPL
		& \textbf{0.450}
		& \bc{0.071} & \bc{0.135} & \bc{0.117}
		& 0.007 & 0.019 & 0.029
		& \bc{\textbf{0.845}} & \bc{0.056} & \bc{0.205} \\
		
		IBM
		& \textbf{0.451}
		& \bc{0.114} & \bc{0.140} & \bc{0.141}
		& 0.002 & 0.028 & 0.004
		& \bc{0.240} & \bc{\textbf{0.795}} & \bc{0.231} \\
		
		INTC
		& \textbf{0.492}
		& \bc{0.078} & \bc{0.071} & \bc{0.111}
		& 0.017 & 0.028 & -0.007
		& \bc{0.094} & \bc{0.058} & \bc{\textbf{0.849}} \\
		
		\midrule
		\multicolumn{11}{l}{\textit{Panel D: Hetero-skew-\(t\)}} \\
		
		SPX
		& \bc{\textbf{0.984}}
		& 0.092 & 0.042 & 0.063
		& \bc{0.009} & \bc{0.015} & \bc{0.003}
		& 0.061 & 0.051 & 0.102 \\
		
		LLY
		& \textbf{0.229}
		& \bc{\textbf{0.963}} & \bc{0.099} & \bc{0.064}
		& 0.015 & 0.010 & 0.011
		& \bc{0.037} & \bc{0.030} & \bc{0.054} \\
		
		MRK
		& \textbf{0.381}
		& \bc{0.132} & \bc{\textbf{0.896}} & \bc{0.136}
		& 0.017 & 0.020 & 0.013
		& \bc{0.043} & \bc{0.047} & \bc{0.103} \\
		
		JNJ
		& \textbf{0.497}
		& \bc{0.178} & \bc{0.126} & \bc{\textbf{0.830}}
		& 0.028 & 0.011 & 0.036
		& \bc{0.080} & \bc{0.046} & \bc{0.079} \\
		
		MMM
		& \bc{\textbf{0.408}}
		& 0.105 & 0.065 & 0.120
		& \bc{\textbf{0.862}} & \bc{0.115} & \bc{0.163}
		& 0.080 & 0.050 & 0.109 \\
		
		GE
		& \bc{\textbf{0.385}}
		& 0.125 & 0.069 & 0.125
		& \bc{0.101} & \bc{\textbf{0.877}} & \bc{0.153}
		& 0.067 & 0.052 & 0.076 \\
		
		HON
		& \bc{\textbf{0.446}}
		& 0.170 & 0.115 & 0.095
		& \bc{0.058} & \bc{0.113} & \bc{\textbf{0.847}}
		& 0.069 & 0.065 & 0.092 \\
		
		AAPL
		& \textbf{0.451}
		& \bc{0.113} & \bc{0.119} & \bc{0.116}
		& 0.017 & 0.028 & 0.029
		& \bc{\textbf{0.844}} & \bc{0.085} & \bc{0.183} \\
		
		IBM
		& \textbf{0.451}
		& \bc{0.172} & \bc{0.129} & \bc{0.132}
		& 0.015 & 0.029 & 0.005
		& \bc{0.206} & \bc{\textbf{0.811}} & \bc{0.176} \\
		
		INTC
		& \textbf{0.499}
		& \bc{0.151} & \bc{0.090} & \bc{0.127}
		& 0.028 & 0.038 & 0.003
		& \bc{0.093} & \bc{0.096} & \bc{\textbf{0.827}} \\
		
		\bottomrule
	\end{tabularx}
}}
\begin{tablenotes}[flushleft]
	\scriptsize
	\setlength{\itemsep}{0pt}
	\setlength{\parsep}{0pt}
	\setlength{\topsep}{2pt}
	\item \textit{Note.}
The table reports the estimated standardized loading matrices
\(\hat{\bar{\Xi}}\) for the unrestricted cluster and hetero specifications.
The HAR residuals satisfy
\(\varepsilon_t=\Lambda V_t\), where \(V_t\) follows a standardized CT or CST
distribution. The normalization
\(\operatorname{diag}(\hat{\bar{\Xi}}\hat{\bar{\Xi}}^{\prime})=\bm 1\)
implies that every row has unit Euclidean norm.
The residual scales \(\hat{\Lambda}\) are not reported.
Shaded cells indicate the relevant block regions.
\end{tablenotes}

%% file: tables/empirical_RK_HAR_residual_correlation_table.tex
{\footnotesize
\renewcommand{\arraystretch}{1.08}
\setlength{\tabcolsep}{2pt}

\begin{tabularx}{\textwidth}{@{}l*{10}{Y}@{}}
\toprule
& Market & \multicolumn{3}{c}{Health Care} & \multicolumn{3}{c}{Industrials} & \multicolumn{3}{c}{Information Tech.} \\
\cmidrule(lr){2-2}\cmidrule(lr){3-5}\cmidrule(lr){6-8}\cmidrule(lr){9-11}
& SPX & LLY & MRK & JNJ & MMM & GE & HON & AAPL & IBM & INTC \\
\midrule
SPX & 1.000 & 0.557 & 0.591 & 0.630 & 0.487 & 0.476 & 0.502 & 0.596 & 0.577 & 0.634 \\
LLY & 0.557 & 1.000 & 0.500 & 0.496 & 0.341 & 0.357 & 0.400 & 0.417 & 0.430 & 0.466 \\
MRK & 0.591 & 0.500 & 1.000 & 0.548 & 0.352 & 0.362 & 0.396 & 0.475 & 0.469 & 0.494 \\
JNJ & 0.630 & 0.496 & 0.548 & 1.000 & 0.409 & 0.403 & 0.429 & 0.511 & 0.492 & 0.501 \\
MMM & 0.487 & 0.341 & 0.352 & 0.409 & 1.000 & 0.471 & 0.476 & 0.367 & 0.335 & 0.406 \\
GE & 0.476 & 0.357 & 0.362 & 0.403 & 0.471 & 1.000 & 0.534 & 0.359 & 0.348 & 0.385 \\
HON & 0.502 & 0.400 & 0.396 & 0.429 & 0.476 & 0.534 & 1.000 & 0.385 & 0.359 & 0.394 \\
AAPL & 0.596 & 0.417 & 0.475 & 0.511 & 0.367 & 0.359 & 0.385 & 1.000 & 0.564 & 0.540 \\
IBM & 0.577 & 0.430 & 0.469 & 0.492 & 0.335 & 0.348 & 0.359 & 0.564 & 1.000 & 0.546 \\
INTC & 0.634 & 0.466 & 0.494 & 0.501 & 0.406 & 0.385 & 0.394 & 0.540 & 0.546 & 1.000 \\
\bottomrule
\end{tabularx}%
}

\begin{tablenotes}[flushleft]
\scriptsize
\item \textit{Note.} The table reports the empirical correlation matrix of the first-stage HAR residuals. The residuals are obtained from the non-overlapping HAR model fitted to $Y_{i,t}=100\sqrt{252\,RV_{i,t}}$.
\end{tablenotes}

%% file: tables/table_empirical_second_stage_blockXi_likelihood_v1.tex
{\footnotesize
\renewcommand{\arraystretch}{1.08}
\setlength{\tabcolsep}{2pt}

\resizebox{\textwidth}{!}{%
\begin{tabularx}{\textwidth}{@{}l*{7}{Y}@{}}
			\toprule
			& \multicolumn{3}{c}{Block matrix benchmarks}
			& \multicolumn{2}{c}{Block-$\bar{\Xi}$ Cluster}
			& \multicolumn{2}{c}{Block-$\bar{\Xi}$ Hetero} \\
			\cmidrule(lr){2-4}\cmidrule(lr){5-6}\cmidrule(lr){7-8}
			& Gaussian & Multi-$t$ & \begin{tabular}{@{}c@{}}Skew-$t$\end{tabular}
			& $t$ & \begin{tabular}{@{}c@{}}Skew-$t$\end{tabular}
			& $t$ & \begin{tabular}{@{}c@{}}Skew-$t$\end{tabular} \\
			\midrule
			
			$\nu_{\mathrm{all}}$
			& $\infty$
			& \begin{tabular}{@{}c@{}}3.141\\[-2pt]{\scriptsize (0.113)}\end{tabular}
			& \begin{tabular}{@{}c@{}}2.966\\[-2pt]{\scriptsize (0.113)}\end{tabular}
			& -- & -- & -- & -- \\
			
			
			$\nu_{\mathrm{Market}}$
			& -- & -- & --
			& \begin{tabular}{@{}c@{}}2.615\\[-2pt]{\scriptsize (0.157)}\end{tabular}
			& \begin{tabular}{@{}c@{}}2.462\\[-2pt]{\scriptsize (0.142)}\end{tabular}
			& \begin{tabular}{@{}c@{}}2.615\\[-2pt]{\scriptsize (0.157)}\end{tabular}
			& \begin{tabular}{@{}c@{}}2.449\\[-2pt]{\scriptsize (0.140)}\end{tabular}
			\\[2pt]
			
			$\nu_{\mathrm{Health}}$
			& -- & -- & --
			& \begin{tabular}{@{}c@{}}2.494\\[-2pt]{\scriptsize (0.098)}\end{tabular}
			& \begin{tabular}{@{}c@{}}2.334\\[-2pt]{\scriptsize (0.079)}\end{tabular}
			& \begin{tabular}{@{}c@{}}2.504\\[-2pt]{\scriptsize (0.096)}\\[2pt]2.347\\[-2pt]{\scriptsize (0.075)}\\[2pt]2.457\\[-2pt]{\scriptsize (0.114)}\end{tabular}
			& \begin{tabular}{@{}c@{}}2.392\\[-2pt]{\scriptsize (0.083)}\\[2pt]2.267\\[-2pt]{\scriptsize (0.063)}\\[2pt]2.390\\[-2pt]{\scriptsize (0.109)}\end{tabular}
			\\[2pt]
			
			$\nu_{\mathrm{Industrials}}$
			& -- & -- & --
			& \begin{tabular}{@{}c@{}}2.976\\[-2pt]{\scriptsize (0.113)}\end{tabular}
			& \begin{tabular}{@{}c@{}}2.756\\[-2pt]{\scriptsize (0.093)}\end{tabular}
			& \begin{tabular}{@{}c@{}}3.092\\[-2pt]{\scriptsize (0.151)}\\[2pt]2.758\\[-2pt]{\scriptsize (0.097)}\\[2pt]2.822\\[-2pt]{\scriptsize (0.112)}\end{tabular}
			& \begin{tabular}{@{}c@{}}3.031\\[-2pt]{\scriptsize (0.147)}\\[2pt]2.680\\[-2pt]{\scriptsize (0.089)}\\[2pt]2.711\\[-2pt]{\scriptsize (0.103)}\end{tabular}
			\\[2pt]
			
			$\nu_{\mathrm{InfoTech}}$
			& -- & -- & --
			& \begin{tabular}{@{}c@{}}3.211\\[-2pt]{\scriptsize (0.126)}\end{tabular}
			& \begin{tabular}{@{}c@{}}3.047\\[-2pt]{\scriptsize (0.111)}\end{tabular}
			& \begin{tabular}{@{}c@{}}2.878\\[-2pt]{\scriptsize (0.116)}\\[2pt]3.486\\[-2pt]{\scriptsize (0.195)}\\[2pt]3.158\\[-2pt]{\scriptsize (0.229)}\end{tabular}
			& \begin{tabular}{@{}c@{}}2.794\\[-2pt]{\scriptsize (0.115)}\\[2pt]3.338\\[-2pt]{\scriptsize (0.192)}\\[2pt]3.199\\[-2pt]{\scriptsize (0.253)}\end{tabular}
			\\
			
			\midrule
			\multicolumn{8}{l}{\textit{Model-specific skewness components}} \\
			
			$\delta_{\mathrm{Market}}$
			& -- & -- & 0.602 (0.037)
			& -- & 0.483 (0.043)
			& -- & 0.495 (0.042) \\
			
			$\delta_{\mathrm{Health}}$
			& -- & --
			& \begin{tabular}{@{}c@{}}0.555 (0.027)\\0.547 (0.029)\\0.535 (0.052)\end{tabular}
			& --
			& \begin{tabular}{@{}c@{}}0.348 (0.025)\\0.333 (0.028)\\0.362 (0.029)\end{tabular}
			& --
			& \begin{tabular}{@{}c@{}}0.472 (0.028)\\0.418 (0.035)\\0.526 (0.045)\end{tabular}
			\\
			
			$\delta_{\mathrm{Industrials}}$
			& -- & --
			& \begin{tabular}{@{}c@{}}0.512 (0.024)\\0.514 (0.021)\\0.511 (0.024)\end{tabular}
			& --
			& \begin{tabular}{@{}c@{}}0.446 (0.028)\\0.447 (0.027)\\0.442 (0.028)\end{tabular}
			& --
			& \begin{tabular}{@{}c@{}}0.652 (0.031)\\0.571 (0.032)\\0.582 (0.033)\end{tabular}
			\\
			
			$\delta_{\mathrm{InfoTech}}$
			& -- & --
			& \begin{tabular}{@{}c@{}}0.550 (0.027)\\0.549 (0.026)\\0.547 (0.029)\end{tabular}
			& --
			& \begin{tabular}{@{}c@{}}0.331 (0.035)\\0.386 (0.037)\\0.417 (0.036)\end{tabular}
			& --
			& \begin{tabular}{@{}c@{}}0.456 (0.040)\\0.649 (0.032)\\0.624 (0.046)\end{tabular}
			\\
			
			\midrule
			$p$       & 13 & 14 & 24 & 29 & 39 & 35 & 45 \\
			$\ell$    & -152961 & -141237 & -140845 & -140280 & \textbf{-139523} & -140693 & -139739 \\
			$\ell_m$  & -163505 & -150989 & -149650 & -150828 & -149436 & -150820 & \textbf{-149435} \\
			$\ell_c$  & 10544 & 9752 & 8805 & \textbf{10548} & 9913 & 10127 & 9696 \\
			BIC       & 306033 & 282593 & 281894 & 280807 & \textbf{279378} & 281685 & 279860 \\
			\bottomrule
		\end{tabularx}%
	
}}
\begin{tablenotes}[flushleft]
	\scriptsize
	\item \textit{Note.} The table reports second-stage maximum likelihood estimates for the centered HAR residuals obtained from $Y_{i,t}=100\sqrt{252\,RV_{i,t}}$ under the matrix restrictions in \eqref{eq:empirical-block-restriction}. For Gaussian and Multi-$t$, the block restriction is imposed on the covariance matrix; for Skew-$t$, it is imposed on the raw dispersion matrix. For the convolution specifications, the residuals are written as $\widehat\varepsilon_t=\Lambda V_t$, where $V_t$ follows a standardized CT or CST distribution and the block restriction is imposed on $\bar{\Xi}$, subject to $\operatorname{diag}(\bar{\Xi}\bar{\Xi}^{\prime})=\bm 1$. Parentheses report Bartlett HAC standard errors (bandwidth 5), conditional on the fitted and centered HAR residuals and excluding first-stage estimation uncertainty. The joint residual log-likelihood is denoted by $\ell$, the sum of univariate marginal log-likelihoods by $\ell_m$, and the residual dependence contribution by $\ell_c=\ell-\ell_m$. The BIC is computed as $-2\ell+p\log T$.
\end{tablenotes}

%% file: tables/table_empirical_blockXi_Xi_estimates_v1.tex
{\footnotesize
\renewcommand{\arraystretch}{1.00}
\setlength{\tabcolsep}{2pt}

\setlength{\aboverulesep}{0.35ex}
\setlength{\belowrulesep}{0.35ex}
\setlength{\cmidrulesep}{0.25ex}

\resizebox{\textwidth}{!}{%
\begin{tabularx}{\textwidth}{@{}l*{10}{Y}@{}}
	\toprule
	& \multicolumn{1}{c}{Market}
	& \multicolumn{3}{c}{Health Care}
	& \multicolumn{3}{c}{Industrials}
	& \multicolumn{3}{c}{Information Tech.} \\
	\cmidrule(lr){2-2}
	\cmidrule(lr){3-5}
	\cmidrule(lr){6-8}
	\cmidrule(lr){9-11}
	& SPX & LLY & MRK & JNJ & MMM & GE & HON & AAPL & IBM & INTC \\
	\midrule
	
	\multicolumn{11}{l}{\textit{Panel A: Cluster-\(t\)}} \\
	
	SPX
	& \bc{\textbf{0.988}}
	& 0.051 & 0.051 & 0.051
	& \bc{0.005} & \bc{0.005} & \bc{0.005}
	& 0.072 & 0.072 & 0.072 \\
	
	LLY
	& \textbf{0.430}
	& \bc{\textbf{0.873}} & \bc{0.139} & \bc{0.139}
	& 0.016 & 0.016 & 0.016
	& \bc{0.065} & \bc{0.065} & \bc{0.065} \\
	
	MRK
	& \textbf{0.430}
	& \bc{0.139} & \bc{\textbf{0.873}} & \bc{0.139}
	& 0.016 & 0.016 & 0.016
	& \bc{0.065} & \bc{0.065} & \bc{0.065} \\
	
	JNJ
	& \textbf{0.430}
	& \bc{0.139} & \bc{0.139} & \bc{\textbf{0.873}}
	& 0.016 & 0.016 & 0.016
	& \bc{0.065} & \bc{0.065} & \bc{0.065} \\
	
	MMM
	& \bc{\textbf{0.416}}
	& 0.101 & 0.101 & 0.101
	& \bc{\textbf{0.860}} & \bc{0.132} & \bc{0.132}
	& 0.086 & 0.086 & 0.086 \\
	
	GE
	& \bc{\textbf{0.416}}
	& 0.101 & 0.101 & 0.101
	& \bc{0.132} & \bc{\textbf{0.860}} & \bc{0.132}
	& 0.086 & 0.086 & 0.086 \\
	
	HON
	& \bc{\textbf{0.416}}
	& 0.101 & 0.101 & 0.101
	& \bc{0.132} & \bc{0.132} & \bc{\textbf{0.860}}
	& 0.086 & 0.086 & 0.086 \\
	
	AAPL
	& \textbf{0.473}
	& \bc{0.109} & \bc{0.109} & \bc{0.109}
	& 0.010 & 0.010 & 0.010
	& \bc{\textbf{0.835}} & \bc{0.145} & \bc{0.145} \\
	
	IBM
	& \textbf{0.473}
	& \bc{0.109} & \bc{0.109} & \bc{0.109}
	& 0.010 & 0.010 & 0.010
	& \bc{0.145} & \bc{\textbf{0.835}} & \bc{0.145} \\
	
	INTC
	& \textbf{0.473}
	& \bc{0.109} & \bc{0.109} & \bc{0.109}
	& 0.010 & 0.010 & 0.010
	& \bc{0.145} & \bc{0.145} & \bc{\textbf{0.835}} \\
	
	\midrule
	\multicolumn{11}{l}{\textit{Panel B: Cluster-skew-\(t\)}} \\
	
	SPX
	& \bc{\textbf{0.983}}
	& 0.073 & 0.073 & 0.073
	& \bc{0.018} & \bc{0.018} & \bc{0.018}
	& 0.075 & 0.075 & 0.075 \\
	
	LLY
	& \textbf{0.372}
	& \bc{\textbf{0.895}} & \bc{0.160} & \bc{0.160}
	& 0.019 & 0.019 & 0.019
	& \bc{0.053} & \bc{0.053} & \bc{0.053} \\
	
	MRK
	& \textbf{0.372}
	& \bc{0.160} & \bc{\textbf{0.895}} & \bc{0.160}
	& 0.019 & 0.019 & 0.019
	& \bc{0.053} & \bc{0.053} & \bc{0.053} \\
	
	JNJ
	& \textbf{0.372}
	& \bc{0.160} & \bc{0.160} & \bc{\textbf{0.895}}
	& 0.019 & 0.019 & 0.019
	& \bc{0.053} & \bc{0.053} & \bc{0.053} \\
	
	MMM
	& \bc{\textbf{0.370}}
	& 0.108 & 0.108 & 0.108
	& \bc{\textbf{0.875}} & \bc{0.153} & \bc{0.153}
	& 0.070 & 0.070 & 0.070 \\
	
	GE
	& \bc{\textbf{0.370}}
	& 0.108 & 0.108 & 0.108
	& \bc{0.153} & \bc{\textbf{0.875}} & \bc{0.153}
	& 0.070 & 0.070 & 0.070 \\
	
	HON
	& \bc{\textbf{0.370}}
	& 0.108 & 0.108 & 0.108
	& \bc{0.153} & \bc{0.153} & \bc{\textbf{0.875}}
	& 0.070 & 0.070 & 0.070 \\
	
	AAPL
	& \textbf{0.453}
	& \bc{0.135} & \bc{0.135} & \bc{0.135}
	& 0.022 & 0.022 & 0.022
	& \bc{\textbf{0.832}} & \bc{0.151} & \bc{0.151} \\
	
	IBM
	& \textbf{0.453}
	& \bc{0.135} & \bc{0.135} & \bc{0.135}
	& 0.022 & 0.022 & 0.022
	& \bc{0.151} & \bc{\textbf{0.832}} & \bc{0.151} \\
	
	INTC
	& \textbf{0.453}
	& \bc{0.135} & \bc{0.135} & \bc{0.135}
	& 0.022 & 0.022 & 0.022
	& \bc{0.151} & \bc{0.151} & \bc{\textbf{0.832}} \\
	
	\midrule
	\multicolumn{11}{l}{\textit{Panel C: Hetero-\(t\)}} \\
	
	SPX
	& \bc{\textbf{0.991}}
	& 0.041 & 0.041 & 0.041
	& \bc{0.004} & \bc{0.004} & \bc{0.004}
	& 0.063 & 0.063 & 0.063 \\
	
	LLY
	& \textbf{0.442}
	& \bc{\textbf{0.871}} & \bc{0.118} & \bc{0.118}
	& 0.023 & 0.023 & 0.023
	& \bc{0.074} & \bc{0.074} & \bc{0.074} \\
	
	MRK
	& \textbf{0.442}
	& \bc{0.118} & \bc{\textbf{0.871}} & \bc{0.118}
	& 0.023 & 0.023 & 0.023
	& \bc{0.074} & \bc{0.074} & \bc{0.074} \\
	
	JNJ
	& \textbf{0.442}
	& \bc{0.118} & \bc{0.118} & \bc{\textbf{0.871}}
	& 0.023 & 0.023 & 0.023
	& \bc{0.074} & \bc{0.074} & \bc{0.074} \\
	
	MMM
	& \bc{\textbf{0.431}}
	& 0.087 & 0.087 & 0.087
	& \bc{\textbf{0.862}} & \bc{0.123} & \bc{0.123}
	& 0.076 & 0.076 & 0.076 \\
	
	GE
	& \bc{\textbf{0.431}}
	& 0.087 & 0.087 & 0.087
	& \bc{0.123} & \bc{\textbf{0.862}} & \bc{0.123}
	& 0.076 & 0.076 & 0.076 \\
	
	HON
	& \bc{\textbf{0.431}}
	& 0.087 & 0.087 & 0.087
	& \bc{0.123} & \bc{0.123} & \bc{\textbf{0.862}}
	& 0.076 & 0.076 & 0.076 \\
	
	AAPL
	& \textbf{0.495}
	& \bc{0.094} & \bc{0.094} & \bc{0.094}
	& 0.021 & 0.021 & 0.021
	& \bc{\textbf{0.830}} & \bc{0.139} & \bc{0.139} \\
	
	IBM
	& \textbf{0.495}
	& \bc{0.094} & \bc{0.094} & \bc{0.094}
	& 0.021 & 0.021 & 0.021
	& \bc{0.139} & \bc{\textbf{0.830}} & \bc{0.139} \\
	
	INTC
	& \textbf{0.495}
	& \bc{0.094} & \bc{0.094} & \bc{0.094}
	& 0.021 & 0.021 & 0.021
	& \bc{0.139} & \bc{0.139} & \bc{\textbf{0.830}} \\
	
	\midrule
	\multicolumn{11}{l}{\textit{Panel D: Hetero-skew-\(t\)}} \\
	
	SPX
	& \bc{\textbf{0.990}}
	& 0.049 & 0.049 & 0.049
	& \bc{0.009} & \bc{0.009} & \bc{0.009}
	& 0.066 & 0.066 & 0.066 \\
	
	LLY
	& \textbf{0.419}
	& \bc{\textbf{0.885}} & \bc{0.116} & \bc{0.116}
	& 0.020 & 0.020 & 0.020
	& \bc{0.067} & \bc{0.067} & \bc{0.067} \\
	
	MRK
	& \textbf{0.419}
	& \bc{0.116} & \bc{\textbf{0.885}} & \bc{0.116}
	& 0.020 & 0.020 & 0.020
	& \bc{0.067} & \bc{0.067} & \bc{0.067} \\
	
	JNJ
	& \textbf{0.419}
	& \bc{0.116} & \bc{0.116} & \bc{\textbf{0.885}}
	& 0.020 & 0.020 & 0.020
	& \bc{0.067} & \bc{0.067} & \bc{0.067} \\
	
	MMM
	& \bc{\textbf{0.425}}
	& 0.093 & 0.093 & 0.093
	& \bc{\textbf{0.867}} & \bc{0.117} & \bc{0.117}
	& 0.069 & 0.069 & 0.069 \\
	
	GE
	& \bc{\textbf{0.425}}
	& 0.093 & 0.093 & 0.093
	& \bc{0.117} & \bc{\textbf{0.867}} & \bc{0.117}
	& 0.069 & 0.069 & 0.069 \\
	
	HON
	& \bc{\textbf{0.425}}
	& 0.093 & 0.093 & 0.093
	& \bc{0.117} & \bc{0.117} & \bc{\textbf{0.867}}
	& 0.069 & 0.069 & 0.069 \\
	
	AAPL
	& \textbf{0.488}
	& \bc{0.103} & \bc{0.103} & \bc{0.103}
	& 0.025 & 0.025 & 0.025
	& \bc{\textbf{0.831}} & \bc{0.137} & \bc{0.137} \\
	
	IBM
	& \textbf{0.488}
	& \bc{0.103} & \bc{0.103} & \bc{0.103}
	& 0.025 & 0.025 & 0.025
	& \bc{0.137} & \bc{\textbf{0.831}} & \bc{0.137} \\
	
	INTC
	& \textbf{0.488}
	& \bc{0.103} & \bc{0.103} & \bc{0.103}
	& 0.025 & 0.025 & 0.025
	& \bc{0.137} & \bc{0.137} & \bc{\textbf{0.831}} \\
	
	\bottomrule
\end{tabularx}%
}
}

\begin{tablenotes}[flushleft]
	\scriptsize
	\setlength{\itemsep}{0pt}
	\setlength{\parsep}{0pt}
	\setlength{\topsep}{2pt}
	\item \textit{Note.}
	The table reports the estimated standardized loading matrices
	\(\hat{\bar{\Xi}}\) under the block-\(\bar{\Xi}\) restrictions.
	The centered HAR residuals satisfy
	\(\varepsilon_t=\Lambda V_t\), where \(V_t\) follows a standardized CT or
	CST distribution. The normalization
	\(\operatorname{diag}(\hat{\bar{\Xi}}\hat{\bar{\Xi}}^{\prime})=\bm 1\)
	implies that every row has unit Euclidean norm.
	Repeated coefficients within each sector are imposed by the block
	specification rather than generated by numerical rounding.
	The estimated residual-scale matrix \(\hat{\Lambda}\) is not reported.
	Shaded cells indicate the relevant block regions.
\end{tablenotes}

%% file: tables/table_empirical_PIT_CT_CST_diagnostics.tex
\begin{tabularx}{\textwidth}{@{}ll*{5}{Y}@{}}
	\toprule
	Specification & Distribution & $\bar u$ & $\mathrm{Var}(u)$ & KS stat.
	& Mean $|\mathrm{Skew}(z)|$ & Mean $|\mathrm{Ex.Kurt}(z)|$ \\
	\midrule
	Unrestricted & Cluster-$t$      & 0.4689 & 0.0762 & 0.0810 & 0.5496 & 0.6076 \\
	Unrestricted & Cluster-skew-$t$ & 0.4932 & 0.0770 & 0.0306 & 0.1639 & 0.3125 \\
	Unrestricted & Hetero-$t$       & 0.4689 & 0.0766 & 0.0811 & 0.5491 & 0.6184 \\
	Unrestricted & Hetero-skew-$t$  & 0.4928 & 0.0777 & 0.0312 & 0.1509 & 0.3540 \\
	\midrule
	Block-$\bar\Xi$ & Cluster-$t$      & 0.4688 & 0.0764 & 0.0809 & 0.5491 & 0.6016 \\
	Block-$\bar\Xi$ & Cluster-skew-$t$ & 0.4931 & 0.0772 & 0.0305 & 0.1643 & 0.3066 \\
	Block-$\bar\Xi$ & Hetero-$t$       & 0.4688 & 0.0767 & 0.0810 & 0.5476 & 0.6051 \\
	Block-$\bar\Xi$ & Hetero-skew-$t$  & 0.4931 & 0.0778 & 0.0305 & 0.1466 & 0.3512 \\
	\bottomrule
\end{tabularx}
\begin{tablenotes}[flushleft]
	\footnotesize
	\item Note. The table reports probability integral transform diagnostics for the fitted univariate marginal distributions of the centered HAR residuals. The transforms are $u_{j,t}=\hat F_j(\widehat\varepsilon_{j,t})$ and $z_{j,t}=\Phi^{-1}(u_{j,t})$. Under correct marginal specification, the PIT is approximately uniform on $[0,1]$, with mean $0.5$ and variance $1/12\simeq0.0833$. The KS column reports the average marginal Kolmogorov--Smirnov statistic; the last two columns report mean absolute skewness and excess kurtosis of the normal scores across series. These are descriptive in-sample diagnostics.
\end{tablenotes}